\documentclass[aps,twocolumn,showpacs,amsmath,amssymb,prc]{revtex4-2}
\usepackage{graphicx,here}
\usepackage{multirow}
\usepackage{tikz}
\usepackage{epstopdf}
\usepackage[percent]{overpic}
\usepackage{scrextend}
\usepackage{xcolor,xspace}
\usepackage[ulem=normalem]{changes}
\usepackage{hyperref}
\hypersetup{colorlinks=True,urlcolor=blue,linkcolor=blue,citecolor=blue,filecolor=black}

\colorlet{Changes@Color}{red}
\usepackage{dcolumn,color,footnote,bm,braket}
\usepackage{url,longtable,tabularx}
\usepackage{threeparttable}

\usepackage{fancybox}
\usetikzlibrary{matrix}

\newcommand\+{\dagger}
\newcommand\jr{j_{\rho}}
\newcommand\mr{m_{\rho}}

\newcommand\mgt{M^{\mathrm{GT}}_{2\nu}}
\newcommand\mf{M^{\mathrm{F}}_{2\nu}}

\newcommand\mbb{M_{2\nu}}
\newcommand\mbbe{M_{2\nu}^{\mathrm{eff}}}

\newcommand\hb{\hat{H}_{\mathrm{B}}}
\newcommand\hf{\hat{H}_{\mathrm{F}}}
\newcommand\hbf{\hat{V}_{\mathrm{BF}}}
\newcommand\hff{\hat{V}_{\nu\pi}}

\newcommand\db{\beta\beta}
\newcommand\tnbb{2\nu\beta\beta}
\newcommand\znbb{0\nu\beta\beta}
\newcommand\gae{g_{\mathrm{A}}^{\mathrm{eff}}}

\newcommand\ga{g_{\mathrm{A}}}
\newcommand\gv{g_{\mathrm{V}}}

\newcommand\trgg{0^+_1\to0^+_1}
\newcommand\trge{0^+_1\to0^+_2}

\newcommand\ft{\log{ft}}

\newcommand\qbb{Q_{\beta\beta}}

\newcommand\taubb{\tau_{1/2}^{(2\nu)}}
\newcommand\vd{v_{\mathrm{d}}}
\newcommand\vssd{v_{\mathrm{ssd}}}
\newcommand\vsss{v_{\mathrm{ss}}}
\newcommand\vt{v_{\mathrm{t}}}

\begin{document}

\title{Effects of pairing strength on the nuclear structure and double-$\beta$ decay predictions within the mapped interacting boson model}

\author{Kosuke Nomura}
\email{nomura@sci.hokudai.ac.jp}
\affiliation{Department of Physics, 
Hokkaido University, Sapporo 060-0810, Japan}
\affiliation{Nuclear Reaction Data Center, 
Hokkaido University, Sapporo 060-0810, Japan}

\date{\today}

\begin{abstract}
The low-energy nuclear structure and two-neutrino 
double-$\beta$ ($2\nu\beta\beta$) decay are studied within 
the interacting boson model (IBM) that 
is based on the nuclear energy density functional (EDF). 
The IBM Hamiltonian describing the 
initial and final even-even nuclei, and the 
interacting boson fermion-fermion Hamiltonian 
producing the intermediate states of the 
neighboring odd-odd nuclei are determined by the 
microscopic inputs provided by 
the self-consistent mean-field (SCMF) 
calculations employing a relativistic EDF 
and a separable pairing force. 
Sensitivities of the low-lying structure and 
$2\nu\beta\beta$-decay properties to the 
pairing strength are specifically analyzed. 
It is shown that the SCMF calculations with 
decreased and increased pairing strengths 
lead to quadrupole-quadrupole interaction 
strengths in the IBM that are, respectively, 
significantly enhanced and reduced in magnitude. 
When the increased pairing is adopted, 
in particular, the energy levels of 
the excited $0^+$ states are lowered, 
and the predicted $2\nu\beta\beta$-decay 
nuclear matrix elements (NMEs) 
increase in magnitude systematically. 
The mapped IBM employing the increased pairing force 
generates effective NMEs and half-lives 
that are in a reasonable agreement 
with the experimental data for the 
$^{76}$Ge$\to^{76}$Se, 
$^{82}$Se$\to^{82}$Kr, and 
$^{100}$Mo$\to^{100}$Ru decays in particular, 
whereas the calculation with the standard 
pairing strength is adequate to provide 
an overall good description of the 
effective NMEs in agreement with data. 
\end{abstract}

\maketitle

\section{Introduction}

The double-$\beta$ ($\db$) decay is a rare nuclear process 
by which the neutron $N$ and proton $Z$ numbers 
decrease (or increase) and increase (or decrease) by two, 
emitting two electrons (positrons) and some 
light particles such as neutrinos. 
Since this nuclear decay process, especially the one that 
does not emit neutrinos (neutrinoless $\db$ decay: $\znbb$) 
concerns several conservation laws required for the electroweak 
fundamental interactions in the standard model, 
and the nature and masses of neutrinos, 
a number of underground experiments 
aimed to detect the $\db$ decay have been running 
and proposed all over the world 
\cite{primakoff1959,haxton1984,doi1985,tomoda1991,suhonen1998,faessler1998,vogel2012,vergados2012,engel2017}. 
See also, e.g., Refs.~\cite{avignone2008,ejiri2019,agostini2023} 
for a review on the related experimental investigations.

Theoretical studies on the $\db$ decay in the 
context of low-energy nuclear physics mainly consist in 
the calculations of the corresponding 
nuclear matrix elements (NMEs). 
The predicted $\db$ NMEs 
currently available are, however, largely 
at variance with different theoretical approaches 
by a factor of 2 to 3. 
Reducing the theoretical uncertainties arising in 
a given nuclear model is, therefore, a crucial  
step toward the consistent understanding of the 
$\db$ decay. Accurate computations of the 
NMEs would be, in turn, a stringent 
test for the model, as 
the nuclear wave functions used to compute 
the NMEs should be sensitive to the 
model's assumptions, parameters, etc. 
The two-neutrino $\db$ ($\tnbb$) decay, in particular, 
is an allowed decay, and a number of experimental 
data are available 
(see, e.g., \cite{barabash2020,agostini2023-76Ge,augier2023-100Mo-Letter,augier2023-100Mo,adams2021-130Te}) to compare 
with theoretical calculations. 
For the calculations of the $\tnbb$ decay, 
the so-called closure approximation, 
which is considered valid for the $0\nu$ modes, is not 
a good approximation, but the  
intermediate states of the neighboring odd-odd nuclei 
should be explicitly taken into account. 
Theoretical predictions on the $\tnbb$-decay 
NMEs that do not assume the closure approximation 
have been reported, such as in the quasiparticle 
random phase approximation (QRPA) 
\cite{suhonen1998,pirinen2015,simkovic2018}, 
nuclear shell model (NSM) 
\cite{caurier2007,yoshinaga2018,caurier1990,caurier2012,senkov2016,coraggio2019}, and interacting boson model (IBM) \cite{yoshida2013}.

Recently, a calculation of the two-neutrino $\db$ 
decay ($\tnbb$) NMEs of a number of candidate nuclei 
was reported \cite{nomura2022bb}, employing the 
neutron-proton IBM (IBM-2) \cite{OAIT,OAI} that is 
based on the self-consistent mean-field (SCMF) 
calculation within the framework of the nuclear 
energy density functional (EDF) 
\cite{RS,bender2003,vretenar2005,niksic2011,robledo2019}. 
In that study, the IBM-2 Hamiltonians producing 
the low-lying states of the initial even-even nuclei 
including $^{48}$Ca to $^{198}$Pt, and those of the final 
ones including $^{48}$Ti to $^{198}$Hg, 
were completely determined so that the 
triaxial quadrupole 
potential energy surface (PES), 
which is obtained 
from the constrained relativistic Hartree-Bogoliubov 
(RHB) \cite{vretenar2005,niksic2011} 
SCMF calculation employing the density-dependent 
point-coupling (DD-PC1) \cite{DDPC1} EDF 
and the separable pairing force of 
finite range \cite{tian2009}, is mapped 
onto that of the boson system. 
The calculation for the $\tnbb$-decay NMEs 
was made without the closure approximation, 
and the intermediate 
odd-odd nuclei were treated in terms 
of the particle-core coupling 
scheme within the neutron-proton interacting 
boson-fermion-fermion model (IBFFM-2) 
\cite{brant1984,IBFM}, 
with the building blocks being also 
determined by the same SCMF calculation.

The mapped IBM-2 study of Ref.~\cite{nomura2022bb} 
has shown that the calculated $\tnbb$-decay 
NMEs with mass-dependent  
quenching factors generally fell into 
the spectrum of various theoretical values 
available in the 
literature, and were more less consistent 
with the experimental systematic \cite{barabash2020}. 
The amounts of the quenching were, however, 
shown to be also different among the decay processes. 
For instance, the NME for the 
$^{76}$Ge$\to^{76}$Se decay was 
calculated to be substantially small to such 
an extent that does not require a quenching, 
whereas a too large NME was obtained for the 
$^{150}$Nd$\to^{150}$Sm decay, for which 
a much larger quenching than the former, 
approximately by a factor of 5, was needed. 
The fact that the quenching of the NMEs was 
required, and that it was at variance with the 
decay processes indicated a need for further 
investigating possible uncertainties in 
the mapped IBM-2 descriptions. 
Indeed, dependencies of the 
$\tnbb$-decay NMEs on several 
model assumptions 
and parameters within this framework 
were studied in Ref.~\cite{nomura2022bb}, 
and it was suggested that a possible 
refinement of the model could be made, 
for instance, at the 
level of the SCMF calculations and/or the 
employed EDF, on which the IBM-2 and IBFFM-2 
Hamiltonians and the $\tnbb$-decay 
operators were built.

It is the aim of the present article to 
pursue further the last point, that is, 
to explore the sensitivities of the mapped IBM-2 
predictions on the $\tnbb$-decay NMEs, 
along with the properties of the low-lying states 
of the relevant even-even and odd-odd nuclei, 
to the underlying SCMF calculations. 
Among those controllable parameters 
in the SCMF model, in the 
present study the effects of the 
pairing strength in the RHB-SCMF calculations on 
the mapped IBM-2 predictions 
are specifically analyzed for those
candidate nuclei,  
$^{48}$Ca, $^{76}$Ge, $^{82}$Se, $^{96}$Zr, $^{100}$Mo, 
$^{116}$Cd, $^{128}$Te, $^{130}$Te, $^{136}$Xe, and $^{150}$Nd, 
where experimental data are available.

In previous applications of the 
mapped IBM-2 to a number of nuclear structure 
phenomena, there has been a problem that 
the microscopically derived 
quadrupole-quadrupole boson interaction strength 
in the IBM-2 Hamiltonian, responsible for deformation, 
is considerably larger in magnitude than those 
used in the conventional IBM fits, and this leads 
to substantial deviations from the observed low-lying 
energy spectra, such as that of the excited $0^+$ 
states, which are generally predicted to 
be too high as compared to the experimental data. 
On one hand, this has been handled 
on the IBM's side, that is, either 
by incorporating the effects of configuration 
mixing \cite{duval1981}, i.e., the mixing between 
several configurations associated with 
particle-hole excitations that are different 
in boson numbers 
(see, e.g., Refs.~\cite{nomura2016sc,nomura2016zr,nomura2022octcm}), 
or by introducing dynamical pairing degree of freedom 
as additional collective coordinate to the quadrupole 
deformations \cite{nomura2020pv,nomura2021pv}. 
On the other hand, the discrepancy in the 
calculation of the excited $0^+$ states 
has been, in many cases, attributed 
to the properties of the underlying SCMF calculations also, 
since any of the representative relativistic and 
nonrelativistic EDFs appears to produce 
PESs that are steep in both $\beta$ and $\gamma$ 
deformations, and exhibit a too pronounced 
energy minimum to be used as an input 
to the IBM.

Increasing the 
strength of the pairing correlations would 
soften the PES, as the stronger pairing generally 
favors a less deformed configuration, so that the 
quadrupole-quadrupole strength in the 
IBM-2 is expected to be reasonably reduced. 
The increased pairing strength in both the 
relativistic and nonrelativistic EDF frameworks 
has been shown to provide a better agreement 
with the experimental 
energy spectrum of the deformed nucleus 
$^{168}$Er in the mapped IBM \cite{nomura2021pds}. 
It was shown more recently that 
the reduction of the bosonic quadrupole-quadrupole 
interaction strength 
allows one to reproduce the measured $\ft$ values for the 
single-$\beta$ decays in the neutron-rich even-even 
Zr isotopes \cite{homma2024}.

The paper is organized as follows. 
Section~\ref{sec:theory} describes the theoretical 
procedure. The results of the nuclear structure 
calculations for each even-even and odd-odd nucleus, 
excitation spectra, and electromagnetic transition 
properties are presented in Sec.~\ref{sec:str}. 
Section~\ref{sec:db} presents results of the 
calculated $\tnbb$-decay NMEs and half-lives 
resulting from the different pairing strengths 
in comparison to the experimental data. 
A summary and concluding remarks are 
given in Sec.~\ref{sec:summary}.

\section{Theoretical framework\label{sec:theory}}

\subsection{Self-consistent mean-field calculations}

To obtain the  
microscopic inputs to the IBM-2 and IBFFM-2 Hamiltonians, 
the triaxial quadrupole constrained SCMF calculations 
are carried out employing the 
RHB method \cite{vretenar2005,niksic2011}
with the particle-hole channel given by 
the DD-PC1 interaction. 
The particle-particle part is modeled by
the separable pairing force of finite range \cite{tian2009}, 
with the pairing matrix element defined in the 
coordinate space
\begin{align}
\label{eq:pair1}
 V({\bf r}_{1},{\bf{r}_2},{\bf{r}'_1},{\bf{r}'_2})
=-V\delta({\bf R}-{\bf{R}'})P({\bf{r}})P({\bf{r}'})\frac{1}{2}(1-P^{\sigma}),
\end{align}
where ${\bf R}=({\bf r}_{1}+{\bf r}_{2})/2$ 
and ${\bf r}={\bf r}_{2} - {\bf r}_{2}$ are 
the center-of-mass and relative coordinates, respectively, 
and the factor $P({\bf{r}})$
a Gaussian function 
\begin{align}
\label{eq:pair2}
 P({\bf{r}}) = \frac{1}{(4{\pi}a^2)^{3/2}}e^{-{\bf r}^{2}/4a^{2}}. 
\end{align}
The strength $V=728$ MeV fm$^{3}$ 
and the parameter $a=0.644$ fm are taken to be the
same for protons and neutrons, and these values 
were determined in \cite{tian2009} so that the
$^{1}S_{0}$ pairing gap of infinite nuclear matter 
resulting from the 
Hartree-Fock-Bogoliubov (HFB) model calculation 
using the Gogny-D1S EDF \cite{D1S} should 
be reproduced. 
In the present study, in addition to the 
default value $V=728$ MeV fm$^{3}$ 
two other values are employed for the 
RHB-SCMF calculations: 
$655$ and 837 MeV fm$^{3}$, which 
correspond to the decrease and 
increase of the original value 
$V$ by 10 \% and 15 \%, respectively. 
The other parameter, $a$, is here kept constant 
for the sake of simplicity.

The particular choices of the 
pairing strength, i.e., scaling it with factors 
0.9 and 1.15, are here considered as 
two illustrative cases in which quantitative changes 
in various calculated results on the low-lying 
states and $\tnbb$ decay are clearly observed. 
Use of the above two scaling factors 
is also inspired by a global systematic 
study of the separable pairing strength within the 
relativistic EDF, that was reported in Ref.~\cite{teeti2021}. 
In that study, global fits of the pairing 
interaction strength to the empirical odd-even mass 
staggering over the entire mass table suggest 
that the strength does depend on nucleon numbers, 
and for majority of the studied nuclei 
the modified pairing strengths with the scaling 
factor $f$, being typically within the range 
$0.9 \lesssim f \lesssim 1.2$, were considered for 
medium-mass and heavy regions. 
In addition, the earlier mapped IBM 
study on the $^{168}$Er energy spectrum 
\cite{nomura2021pds}  
reported a generally more reasonable description 
of the low-lying non-yrast levels 
including that of the excited $0^+_2$ 
state with the increased pairing by 15 \% 
than with the default strength.

A set of the RHB-SCMF calculations are performed for each 
even-even nucleus with constraints on the mass 
quadrupole moments that are associated with the 
triaxial quadrupole deformations $\beta$ and $\gamma$ 
in the geometrical model \cite{BM}. 
The RHB-SCMF calculations yield the potential energy surface (PES), 
that is, total mean-field energy defined as a 
function of the $\beta$ and $\gamma$ deformations, 
and then it is used as the input to construct 
the IBM-2 Hamiltonian. 
The RHB-SCMF calculations further provide single-particle 
energies, and occupation probabilities at spherical 
configuration for the neighboring 
odd-odd nuclei. These quantities are to be 
used to construct the IBFFM-2 Hamiltonian, and 
are obtained from the 
standard RHB calculations without blocking, 
with constraints to zero deformation 
and with the particle number constrained to odd numbers 
(see Refs.~\cite{nomura2016odd,nomura2017odd-2} 
for details).

\subsection{IBM-2 Hamiltonian}

In order to calculate low-lying 
states and transition properties 
starting from the SCMF calculations, 
one should go beyond the mean-field approximation 
by restoring symmetries and including 
quantum fluctuations around the mean-field solution 
\cite{RS,bender2003,niksic2011,robledo2019}. 
These so-called beyond-mean-field 
effects are here taken into account by 
mapping the SCMF results onto the IBM-2. 
The IBM-2 comprises the neutron $s_{\nu}$ 
and proton $s_{\pi}$ monopole bosons, and 
neutron $d_{\nu}$ and proton $d_{\pi}$ 
quadrupole bosons. 
The $s_{\nu}$ ($s_{\pi}$) and $d_{\nu}$ ($d_{\pi}$) 
bosons are connected to  
the collective monopole $S_\nu$ ($S_\pi$) 
and quadrupole $D_\nu$ ($D_\pi$) 
pairs of valence neutrons (protons) 
with spin and parity values $J=0^{+}$ and $J=2^{+}$, 
respectively \cite{OAI}.

The IBM-2 Hamiltonian employed in this study takes the form
\begin{align}
\label{eq:hb}
 \hb = 
\epsilon_{d}(\hat{n}_{d_{\nu}}+\hat{n}_{d_{\pi}})
+\kappa\hat{Q}_{\nu}\cdot\hat{Q}_{\pi}
+ \kappa'\hat{L}\cdot\hat{L} \; .
\end{align}
$\hat{n}_{d_\rho}=d^\+_\rho\cdot\tilde d_{\rho}$ 
($\rho=\nu,\pi$) is the $d$-boson number operator, 
with $\epsilon_{d}$ the single $d$-boson
energy relative to the $s$-boson one, and 
$\tilde d_{\rho\mu}=(-1)^\mu d_{\rho-\mu}$. 
The second term is the quadrupole-quadrupole 
interaction between neutron and proton bosons, 
with 
$\hat Q_{\rho}=d_{\rho}^\+ s_{\rho} + s_{\rho}^\+\tilde d_{\rho} + \chi_{\rho}(d^\+_{\rho}\times\tilde{d}_{\rho})^{(2)}$ being 
the quadrupole operator in the boson system. 
The last term in Eq.~(\ref{eq:hb}) is a rotational 
term with 
$\hat{L}=\sqrt{10}\sum_{\rho}(d^{\+}_{\rho}\times\tilde{d}_{\rho})^{(1)}$ 
being the bosonic angular momentum operator.

Since there appears no interaction 
between unlike-bosons for those nuclei 
corresponding either to $N_{\pi}=0$ 
or $N_{\nu}=0$, the following Hamiltonian 
is considered: 
\begin{align}
\label{eq:hb-semimagic}
 \hb = 
\epsilon_{d\rho} \hat{n}_{d_{\rho}} 
+\kappa_\rho \hat{Q}_{\rho}\cdot\hat{Q}_{\rho} \; , 
\end{align}
which is nothing but the Hamiltonian in the IBM-1, 
where no distinction is made between neutron and 
proton bosons. 
The IBM-1-like Hamiltonian (\ref{eq:hb-semimagic}) 
is here employed specifically 
for $^{48}$Ca, $^{116}$Sn, and $^{136}$Xe, 
having $N_\pi=0$, $N_\pi=0$, and $N_{\nu}=0$, 
respectively.

The strength parameters for the Hamiltonian (\ref{eq:hb}), 
or (\ref{eq:hb-semimagic}), are determined by 
using the SCMF-to-IBM mapping procedure 
\cite{nomura2008,nomura2010}, so that the 
following approximate equality is satisfied 
in the vicinity of the global mean-field minimum. 
\begin{align}
\label{eq:map}
 E_{\mathrm{SCMF}}(\beta,\gamma)
\approx
E_{\mathrm{IBM}}(\beta,\gamma) \; .
\end{align}
Here $E_{\mathrm{SCMF}}(\beta,\gamma)$ represents 
the SCMF PES, and $E_{\mathrm{IBM}}(\beta,\gamma)$ 
on the right-hand side the corresponding 
PES in the boson system, which is given as 
the energy expectation value 
$\braket{\Phi|\hb|\Phi}/\braket{\Phi|\Phi}$, 
with the wave function $\ket{\Phi}$ being 
a boson coherent state 
\cite{dieperink1980,ginocchio1980} that is 
defined as 
\begin{align}
\label{eq:coherent}
 \ket{\Phi}=\prod_{\rho=\nu,\pi}
\left[
s_{\rho}^{\+}+\sum_{\mu=-2}^{+2}\alpha_{\rho\mu}d_{\rho\mu}^{\+}
\right]^{N_{\rho}}\ket{0} \; ,
\end{align}
up to the normalization factor. 
The amplitudes $\alpha_{\rho\mu}$ are given 
as $\alpha_{\rho0}=\beta_{\rho}\cos{\gamma_{\rho}}$, 
$\alpha_{\rho\pm1}=0$, and 
$\alpha_{\rho\pm2}=\beta_{\rho}\sin{\gamma_{\rho}}/\sqrt{2}$, 
where $\beta_{\rho}$ and $\gamma_{\rho}$ are 
boson analogs of the deformation variables. 
$\ket{0}$ represents the boson vacuum, i.e., the inert core. 
$N_{\nu}$ ($N_{\pi}$) is the number of neutron 
(proton) bosons, and is counted as half the number 
of valence neutron (proton) particles/holes 
with respect to the nearest doubly magic nucleus 
\cite{OAIT,OAI}. 
Only, for the $^{48}$Ca and $^{48}$Ti nuclei, 
the inert core is taken to be $^{40}$Ca, in order to 
have the number of bosons be enough to 
produce boson-boson interactions. 
Furthermore, 
both the neutron and proton $\beta_\rho$ 
and $\gamma_{\rho}$ deformations 
are assumed to be equal to each other, 
$\beta_{\nu}=\beta_{\pi}$ and $\gamma_{\nu}=\gamma_{\pi}$. 
As in Ref.~\cite{nomura2022bb} it is also assumed that 
the $\beta_\rho$ deformation is 
proportional to the geometrical one, 
$\beta_{\nu}=\beta_{\pi}\propto\beta$, while  
the $\gamma_\rho$ is identical to the 
geometrical counterpart, 
$\gamma_{\nu}=\gamma_{\pi}\equiv\gamma$ 
\cite{ginocchio1980,nomura2008}.

The parameter $\kappa'$ for the third term 
of Eq.~(\ref{eq:hb}), 
$\hat L \cdot \hat L$, is determined \cite{nomura2011rot}
separately from the other parameters, 
so that the cranking moment of inertia 
calculated in the intrinsic frame of the boson system 
\cite{Schaaser86} at the global minimum 
is equal to the corresponding Inglis-Belyaev 
\cite{inglis1956,belyaev1961}
moment of inertia obtained from the RHB-SCMF  
calculation. 
This term is, however, neglected for most of 
the studied even-even nuclei, since it turns out 
to have only minor effects on the low-lying energy levels, 
except for a few nuclei with specific choices of 
the pairing strength. 
Details are given in Sec.~\ref{sec:para}.

\subsection{IBFFM-2 Hamiltonian}

The extension to the IBFFM-2 is made 
by introducing unpaired nucleon degrees of freedom and 
their couplings to the even-even IBM-2 space. 
The IBFFM-2 Hamiltonian is written as
\begin{align}
\label{eq:ham-ibffm2}
 \hat{H}=\hb + \hf^{\nu} + \hf^{\pi} + \hbf^{\nu} + \hbf^{\pi} + \hff,
\end{align}
The first term $\hb$ is the 
IBM-2 Hamiltonian (\ref{eq:hb}) [or (\ref{eq:hb-semimagic})]. 
The second and third terms represent 
the single-nucleon Hamiltonians 
of the form 
\begin{align}
\label{eq:hf}
 \hf^{\rho} = -\sum_{\jr}\epsilon_{\jr}\sqrt{2\jr+1}
  (a_{\jr}^\+\times\tilde a_{\jr})^{(0)}
\equiv
\sum_{\jr}\epsilon_{\jr}\hat{n}_{\jr},
\end{align}
where $\epsilon_{\jr}$ stands for the 
single-particle energy of the odd neutron ($\rho=\nu$)
or proton ($\rho=\pi$) orbital $\jr$. 
$a_{\jr}^{(\+)}$ represents 
a particle annihilation (or creation) operator, 
with $\tilde{a}_{\jr}$ defined by 
$\tilde{a}_{\jr\mr}=(-1)^{\jr -\mr}a_{\jr-\mr}$. 
The operator $\hat{n}_{\jr}$ stands for the number operator 
for the unpaired particle. 
Within the present formalism, 
the single-particle energy 
$\epsilon_{\jr}$ in Eq.~(\ref{eq:hf}) 
is replaced with the quasiparticle energy 
$\tilde\epsilon_{\jr}$.

The fourth (fifth) term of Eq.~(\ref{eq:ham-ibffm2}) 
stands for the interaction between the odd 
neutron (proton) and the IBM-2 core, 
and has a specific form \cite{IBFM}
\begin{equation}
\label{eq:hbf}
 \hbf^{\rho}
=\Gamma_{\rho}\hat{V}_{\mathrm{dyn}}^{\rho}
+\Lambda_{\rho}\hat{V}_{\mathrm{exc}}^{\rho}
+A_{\rho}\hat{V}_{\mathrm{mon}}^{\rho} \; ,
\end{equation}
where the first, second, and third 
terms are dynamical quadrupole, 
exchange, and monopole interactions, respectively. 
Expressions for the terms in Eq.~(\ref{eq:hbf}) 
are given in the generalized seniority scheme as
\cite{IBFM,scholten1985}
\begin{widetext}
\begin{align}
\label{eq:dyn}
&\hat{V}_{\mathrm{dyn}}^{\rho}
=\sum_{\jr\jr'}\gamma_{\jr\jr'}
(a^{\+}_{\jr}\times\tilde{a}_{\jr'})^{(2)}
\cdot\hat{Q}_{\rho'},\\
\label{eq:exc}
&\hat{V}^{\rho}_{\mathrm{exc}}
=-\left(
s_{\rho'}^\+\times\tilde{d}_{\rho'}
\right)^{(2)}
\cdot
\sum_{\jr\jr'\jr''}
\sqrt{\frac{10}{N_{\rho}(2\jr+1)}}
\beta_{\jr\jr'}\beta_{\jr''\jr}
:\left[
(d_{\rho}^{\+}\times\tilde{a}_{\jr''})^{(\jr)}\times
(a_{\jr'}^{\+}\times\tilde{s}_{\rho})^{(\jr')}
\right]^{(2)}:
+ (\text{H.c.}) \; ,\\
\label{eq:mon}
&\hat{V}_{\mathrm{mon}}^{\rho}
=\hat{n}_{d_{\rho}}\hat{n}_{\jr} \; ,
\end{align}
\end{widetext}
Here the factors 
$\gamma_{\jr\jr'}=(u_{\jr}u_{\jr'}-v_{\jr}v_{\jr'})Q_{\jr\jr'}$, 
and $\beta_{\jr\jr'}=(u_{\jr}v_{\jr'}+v_{\jr}u_{\jr'})Q_{\jr\jr'}$, 
with 
$Q_{\jr\jr'}=\braket{\ell_{\rho}\frac{1}{2}\jr\|Y^{(2)}\|\ell'_\rho\frac{1}{2}\jr'}$ 
the matrix element of the fermion 
quadrupole operator in the single-particle basis. 
$\hat{Q}_{\rho'}$ in (\ref{eq:dyn}) is the same boson 
quadrupole operator as in the boson 
Hamiltonian (\ref{eq:hb}). 
The notation $:(\cdots):$ in (\ref{eq:exc}) 
means normal ordering.

The last term of Eq.~(\ref{eq:ham-ibffm2}) 
corresponds to the odd neutron-proton interaction 
that is given as
\begin{align}
\label{eq:hff}
\hff
=& 4\pi({\vd} + \vssd {\bm{\sigma}}_{\nu}
\cdot{\bm{\sigma}}_{\pi} )
\delta(\bm{r})
\delta(\bm{r}_{\nu}-r_0)
\delta(\bm{r}_{\pi}-r_0)
\nonumber\\
& - \frac{1}{\sqrt{3}} \vsss {\bm{\sigma}}_{\nu}
\cdot{\bm{\sigma}}_{\pi}
+ \vt
\left[
\frac{3({\bm\sigma}_{\nu}\cdot{\bf r})
({\bm\sigma}_{\pi}\cdot{\bf r})}{r^2}
-{\bm{\sigma}}_{\nu}
\cdot{\bm{\sigma}}_{\pi}
\right] \; .
\end{align}
The first term consists of the $\delta$, and 
spin-spin $\delta$ interactions, and the second, 
and third terms represent the spin-spin 
and tensor interactions, respectively. 
$\vd$, $\vssd$, $\vsss$, and $\vt$
are strength parameters. 
Note that $\bm{r}=\bm{r}_{\nu}-\bm{r}_{\pi}$ 
is the relative coordinate of the 
neutron and proton, and $r_0=1.2A^{1/3}$ fm.

The strength parameters for the IBFFM-2 Hamiltonian 
(\ref{eq:ham-ibffm2}) are obtained by using the 
procedure developed in 
Refs.~\cite{nomura2016odd,nomura2019dodd}: 
(i) The quasiparticle energies $\tilde\epsilon_{\jr}$, 
occupation $v_{\jr}$, and unoccupation $u_{\jr}$ 
amplitudes provided by the RHB-SCMF calculations 
are input to $\hf^\rho$ (\ref{eq:hf}) 
and $\hbf^\rho$ (\ref{eq:hbf}); 
(ii) the coupling constants $\Gamma_\rho$, $\Lambda_\rho$, 
and $A_\rho$ are fit to reproduce a few low-lying 
levels of each of the neighboring odd-$N$ 
and odd-$Z$ nuclei, 
separately for positive- and negative-parity states; 
(iii) the parameters for $\hff$ (\ref{eq:hff}) 
are determined so 
as to reproduce the ground-state spin and 
a few energy levels of each odd-odd nucleus. 
The employed parameters for the IBFFM-2 Hamiltonian 
are given in Tables~\ref{tab:parabfn}, \ref{tab:parabfp}, 
and \ref{tab:paraff} in the Appendix~\ref{sec:paraoo}. 
Single-particle spaces taken for the odd nucleons 
are given in Tables~\ref{tab:parabfn}, and \ref{tab:parabfp}. 
The even-even boson core nuclei and neighboring 
odd-$N$ and odd-$Z$ nuclei are summarized in 
Table~I of Ref.~\cite{nomura2022bb}.

\subsection{$\tnbb$ decay operators}

The $\tnbb$-decay NME requires to calculate 
the Gamow-Teller (GT) and Fermi (F) transitions 
for the single-$\beta$ decay 
from the initial even-even to intermediate odd-odd, 
and that from the intermediate odd-odd, 
to final even-even nuclei. 
The corresponding GT and F operators take the forms
\begin{align}
\label{eq:ogt}
&\hat{T}^{\rm GT}
=\sum_{j_{\nu}j_{\pi}}
\eta_{j_{\nu}j_{\pi}}^{\mathrm{GT}}
\left(\hat P_{j_{\nu}}\times\hat P_{j_{\pi}}\right)^{(1)}, \\
\label{eq:ofe}
&\hat{T}^{\rm F}
=\sum_{j_{\nu}j_{\pi}}
\eta_{j_{\nu}j_{\pi}}^{\mathrm{F}}
\left(\hat P_{j_{\nu}}\times\hat P_{j_{\pi}}\right)^{(0)} \; ,
\end{align}
where the coefficients 
$\eta_{j_{\nu}j_{\pi}}^{\mathrm{GT}}$ and 
$\eta_{j_{\nu}j_{\pi}}^{\mathrm{F}}$  are, 
to the lowest order,  
\begin{align}
\label{eq:eta}
\eta_{j_{\nu}j_{\pi}}^{\mathrm{GT}}
&= - \frac{1}{\sqrt{3}}
\left\langle
\ell_{\nu}\frac{1}{2};j_{\nu}
\bigg\|{\bm\sigma}\bigg\|
\ell_{\pi}\frac{1}{2};j_{\pi}
\right\rangle
\delta_{\ell_{\nu}\ell_{\pi}} \\
\eta_{j_{\nu}j_{\pi}}^{\mathrm{F}}
&=-\sqrt{2j_{\nu}+1}
\delta_{j_{\nu}j_{\pi}} \; .
\end{align}
$\hat P_{\jr}$ is here given by one of the 
one-particle creation operators
\begin{subequations}
 \begin{align}
\label{eq:creation1}
&A^{\+}_{\jr\mr} = \zeta_{\jr} a_{{\jr}\mr}^{\+}
 + \sum_{\jr'} \zeta_{\jr\jr'} s^{\+}_\rho (\tilde{d}_{\rho}\times a_{\jr'}^{\+})^{(\jr)}_{\mr}
\\
\label{eq:creation2}
&B^{\+}_{\jr\mr}
=\theta_{\jr} s^{\+}_\rho\tilde{a}_{\jr\mr}
 + \sum_{\jr'} \theta_{\jr\jr'} (d^{\+}_{\rho}\times\tilde{a}_{\jr'})^{(\jr)}_{\mr},
\end{align}
and annihilation operators
\begin{align}
\label{eq:annihilation1}
&\tilde{A}_{\jr\mr}=(-1)^{\jr-\mr}A_{\jr-\mr}\\ 
\label{eq:annihilation2}
&\tilde{B}_{\jr\mr}=(-1)^{\jr-\mr}B_{\jr-\mr} \; .
\end{align}
\end{subequations}
The operators in Eqs.~(\ref{eq:creation1}) 
and (\ref{eq:annihilation1}) 
conserve the boson number, whereas those 
in Eqs.~(\ref{eq:creation2}) and (\ref{eq:annihilation2}) 
do not. 
The GT (\ref{eq:ogt}) and Fermi (\ref{eq:ofe}) 
operators are formed as a pair of the above operators, 
depending on the particle or hole nature of bosons in 
the even-even IBM-2 core. 
Coefficients $\zeta_{j}$, $\zeta_{jj'}$, 
$\theta_{j}$, and $\theta_{jj'}$ 
are dependent on the $v_{\jr}$ and $u_{\jr}$ 
amplitudes, for which the same values as those 
used in the IBFFM-2 calculations for 
the odd-odd nuclei are employed. 
The expressions for these coefficients are 
found, e.g., in Appendix D of Ref.~\cite{nomura2022bb}. 
A more detailed description of the derivation of 
the one-particle transfer operator within 
the generalized seniority scheme is found in 
Refs.~\cite{dellagiacoma1988phdthesis,DELLAGIACOMA1989,IBFM}.

The GT and F matrix elements that enter the 
$\tnbb$ NME are calculated via the formulas
\begin{align}
\label{eq:mgt}
 &\mgt=\sum_{N}
\frac{\braket{0^{+}_{F}\|\hat{T}^{\rm GT}\|1^{+}_{N}}\braket{1^{+}_{N}\|\hat{T}^{\rm GT}\|0^{+}_{1,I}}}
{E_{N}-E_{I}+\frac{1}{2}(Q_{\beta\beta}+2m_{e}c^{2})}
\\
\label{eq:mf}
&\mf=\sum_{N}
\frac{\braket{0^{+}_{F}\|\hat{T}^{\rm F}\|0^{+}_{N}}\braket{0^{+}_{N}\|\hat{T}^{\rm F}\|0^{+}_{1,I}}}
{E_{N}-E_{I}+\frac{1}{2}(Q_{\beta\beta}+2m_{e}c^{2})}\; ,
\end{align}
respectively. 
In the denominators $E_{I}$ ($E_{N}$) stands for the energy 
of the initial (intermediate) state, 
$\qbb$ is the $Q$ value for 
the $\tnbb$ decay, and $m_e$ is the electron mass, 
i.e., $m_e=0.511$ MeV/$c^2$. 
The sums in Eqs.~(\ref{eq:mgt}) and (\ref{eq:mf}) are 
taken over all the intermediate states $1^{+}_{N}$ and 
$0^{+}_{N}$ obtained from the IBFFM-2 Hamiltonian 
below the excitation energy of 30 MeV. 
For the $\qbb$ value, experimental data available at NNDC 
database \cite{data} are adopted. 
Using $\mgt$ (\ref{eq:mgt}) and $\mf$ (\ref{eq:mf}) 
transition matrix elements, 
the $\tnbb$-decay NME is obtained through
\begin{align}
\label{eq:mbb}
 M_{2\nu}=\ga^{2} {m_{e}c^{2}}
\left[
\mgt - \left(\frac{g_\mathrm{V}}{\ga}\right)^{2}\mf
\right],
\end{align}
with $\gv=1$ and $\ga=1.269$ the vector and 
axial vector coupling constants, respectively. 
The corresponding half-life, 
denoted $\tau_{1/2}^{(2\nu)}$, 
can be readily calculated by \cite{tomoda1991}
\begin{align}
\label{eq:taubb}
 \left[\tau_{1/2}^{(2\nu)}\right]^{-1}
=G_{2\nu}|M_{2\nu}|^{2} \; ,
\end{align}
where $G_{2\nu}$ is the phase-space factor  
in year$^{-1}$ for the $0^+\to0^+$ $\tnbb$ decay. 
The $G_{2\nu}$ values calculated in Ref.~\cite{kotila2012} 
are used.

%
%
\begin{figure*}[ht]
\begin{center}
\includegraphics[width=\linewidth]{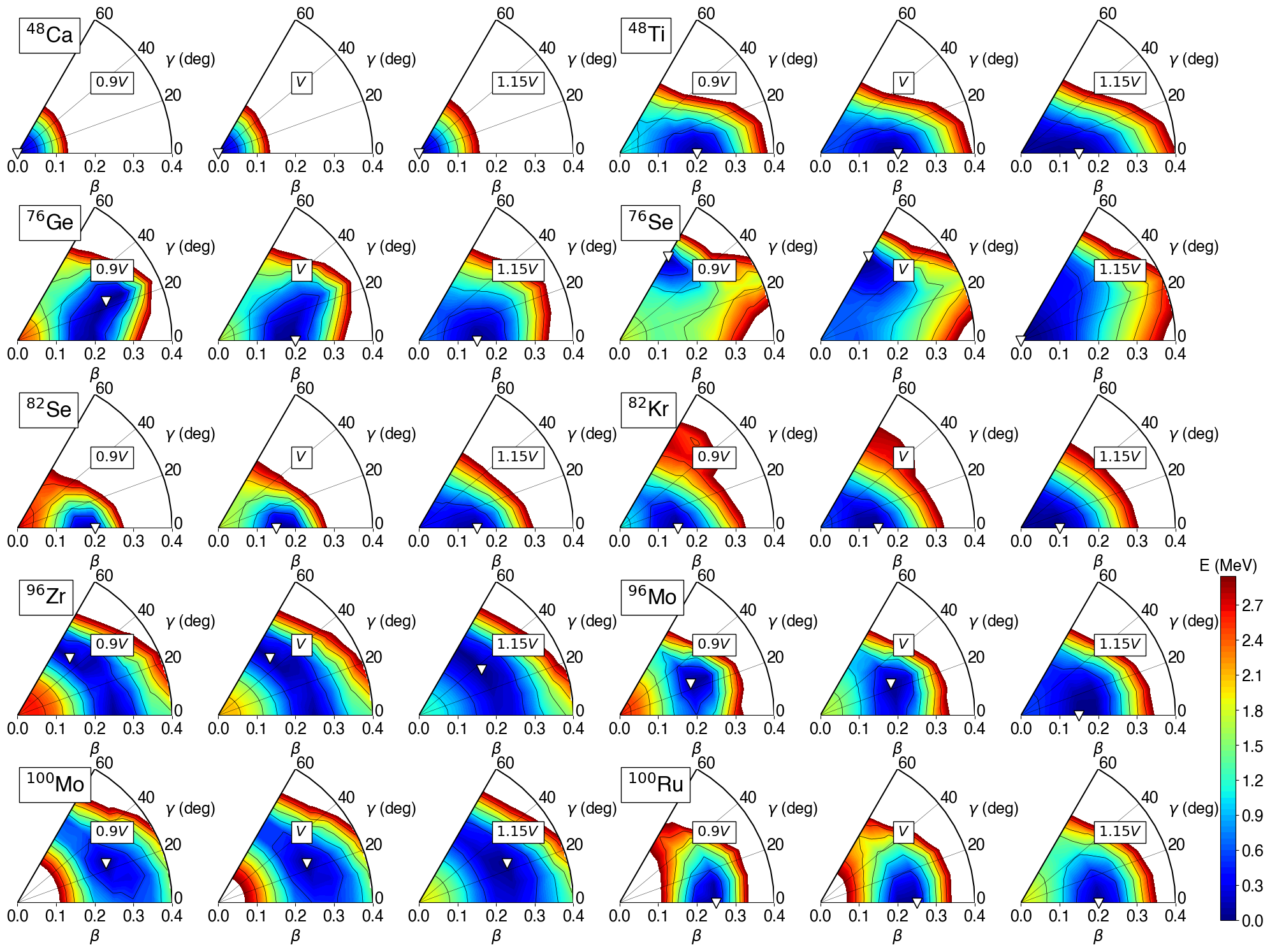}
\caption{
Triaxial quadrupole ($\beta,\gamma$) 
potential energy surfaces for the 
even-even nuclei from $^{48}$Ca to $^{100}$Ru 
obtained from the constrained SCMF calculations 
within the RHB method using 
the DD-PC1 EDF and the separable pairing 
force with strengths $0.9V=655$ MeV fm$^3$ 
(first and fourth columns), 
$V=728$ MeV fm$^3$ (second and fifth columns), 
and $1.15V=837$ MeV fm$^3$ (third and sixth columns). 
The energy difference between neighboring contours 
is 0.3 MeV, and the global minimum is 
indicated by the open triangle. 
}
\label{fig:pes1}
\end{center}
\end{figure*}

%
%
\begin{figure*}[ht]
\begin{center}
\includegraphics[width=\linewidth]{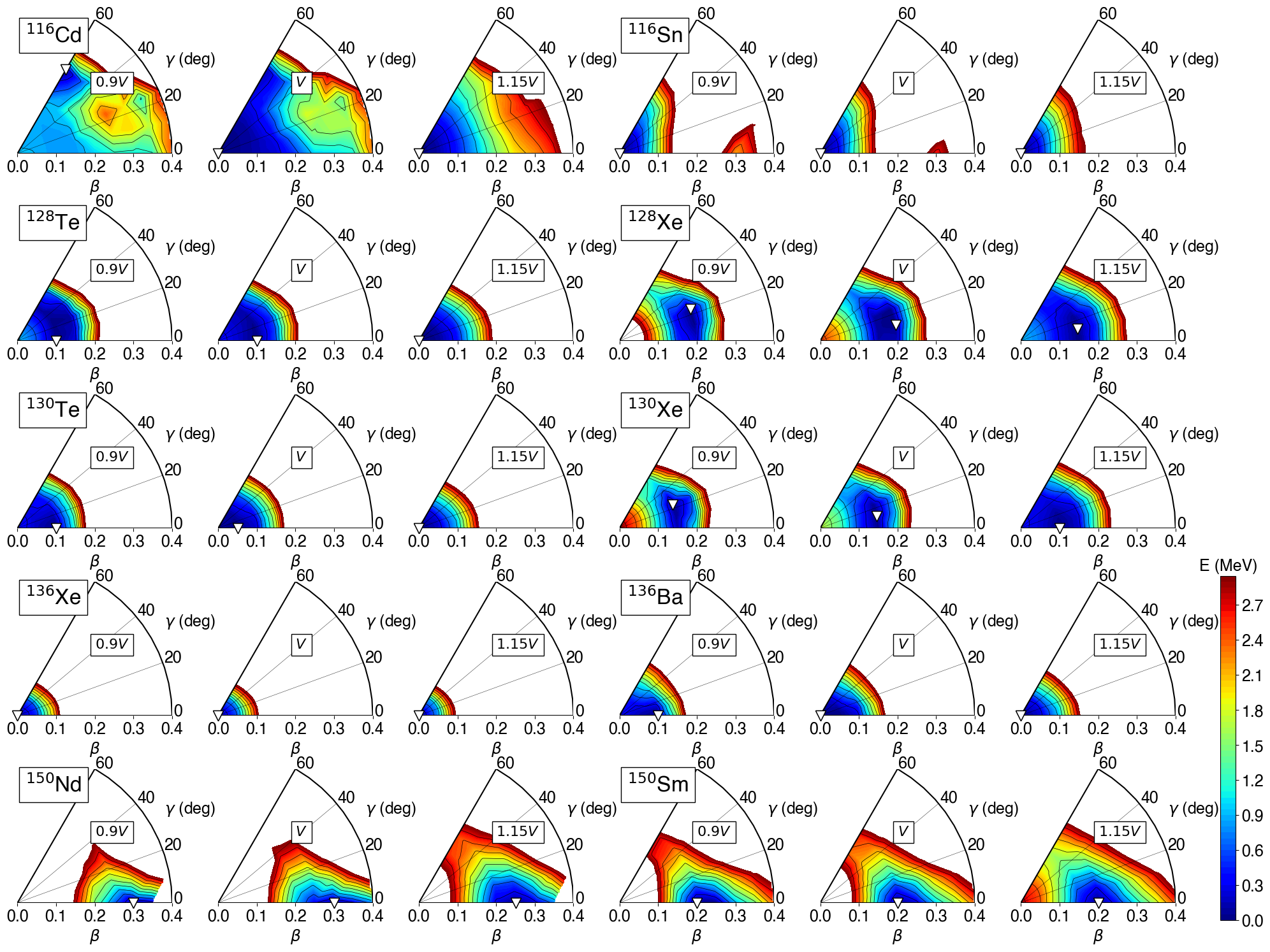}
\caption{Same as Fig.~\ref{fig:pes1}, but for the 
even-even nuclei from $^{116}$Cd to $^{150}$Sm.}
\label{fig:pes2}
\end{center}
\end{figure*}

%
%
\begin{figure}[ht]
\begin{center}
\includegraphics[width=\linewidth]{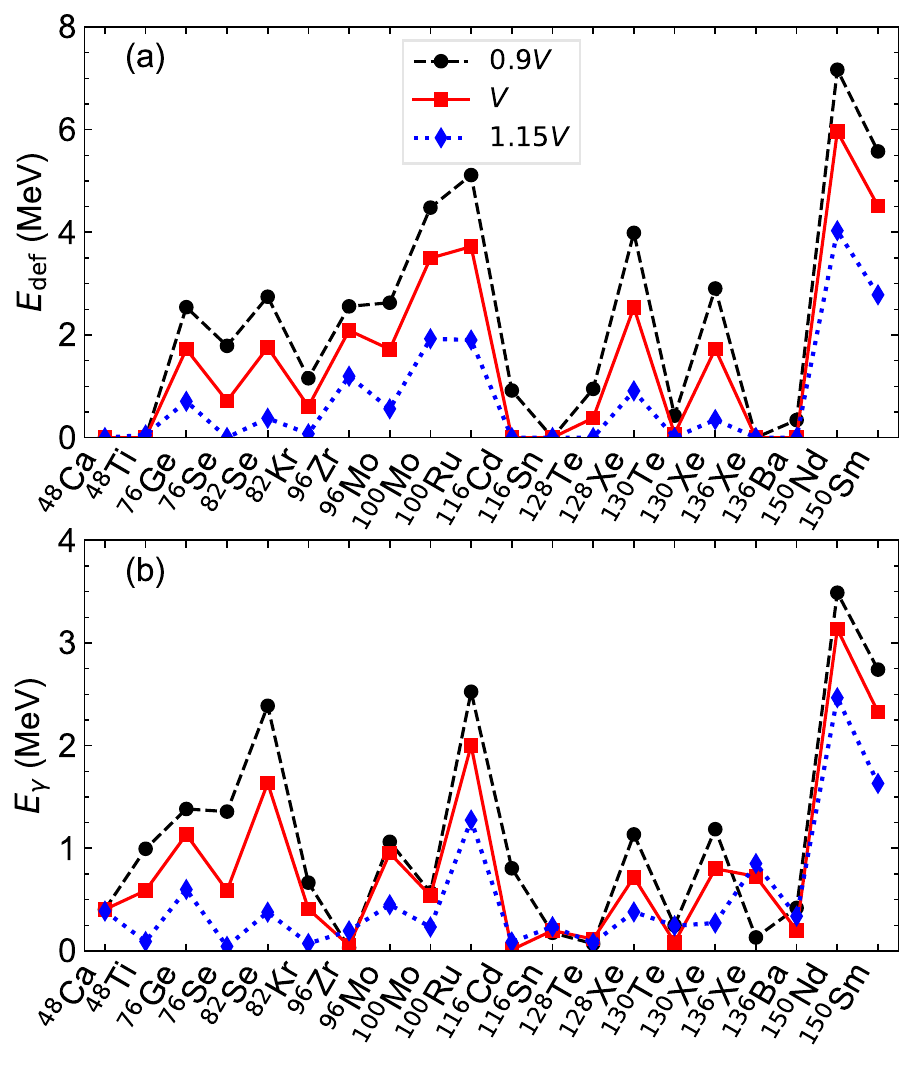}
\caption{Energies $E_{\rm def}$ (a) and $E_{\gamma}$ 
extracted from the SCMF PESs for the different pairing 
strengths. See the main text for the definitions 
of the above quantities. 
}
\label{fig:edef}
\end{center}
\end{figure}

%
%
\begin{figure}
\begin{center}
\includegraphics[width=\linewidth]{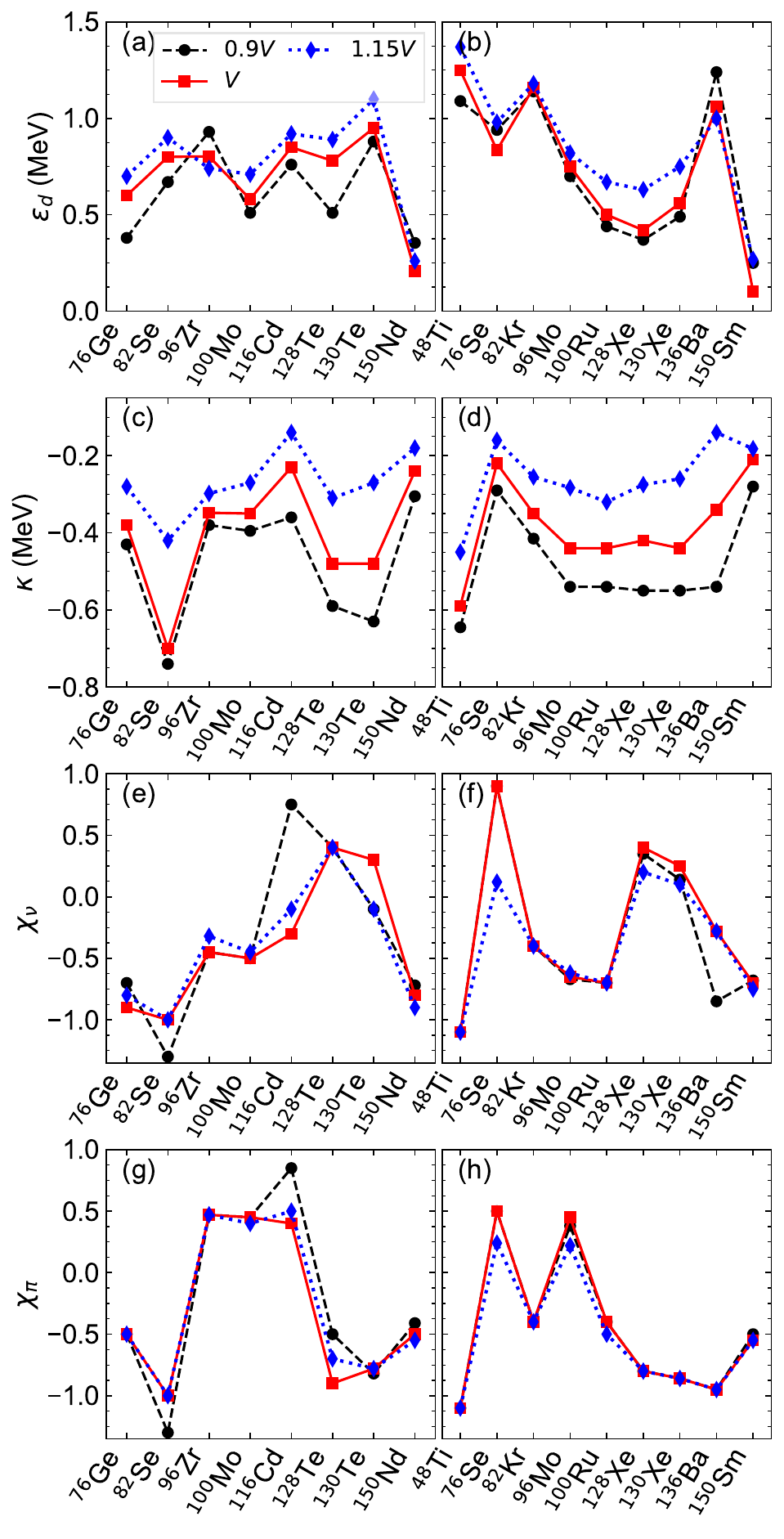}
\caption{Derived parameters for the IBM-2 
Hamiltonian (\ref{eq:hb}) 
for the even-even nuclei with the 
reduced ($0.9V$), default ($V$), and increased ($1.15V$) 
strengths of the separable pairing force.}
\label{fig:para}
\end{center}
\end{figure}

%
\begin{figure}
\begin{center}
\includegraphics[width=\linewidth]{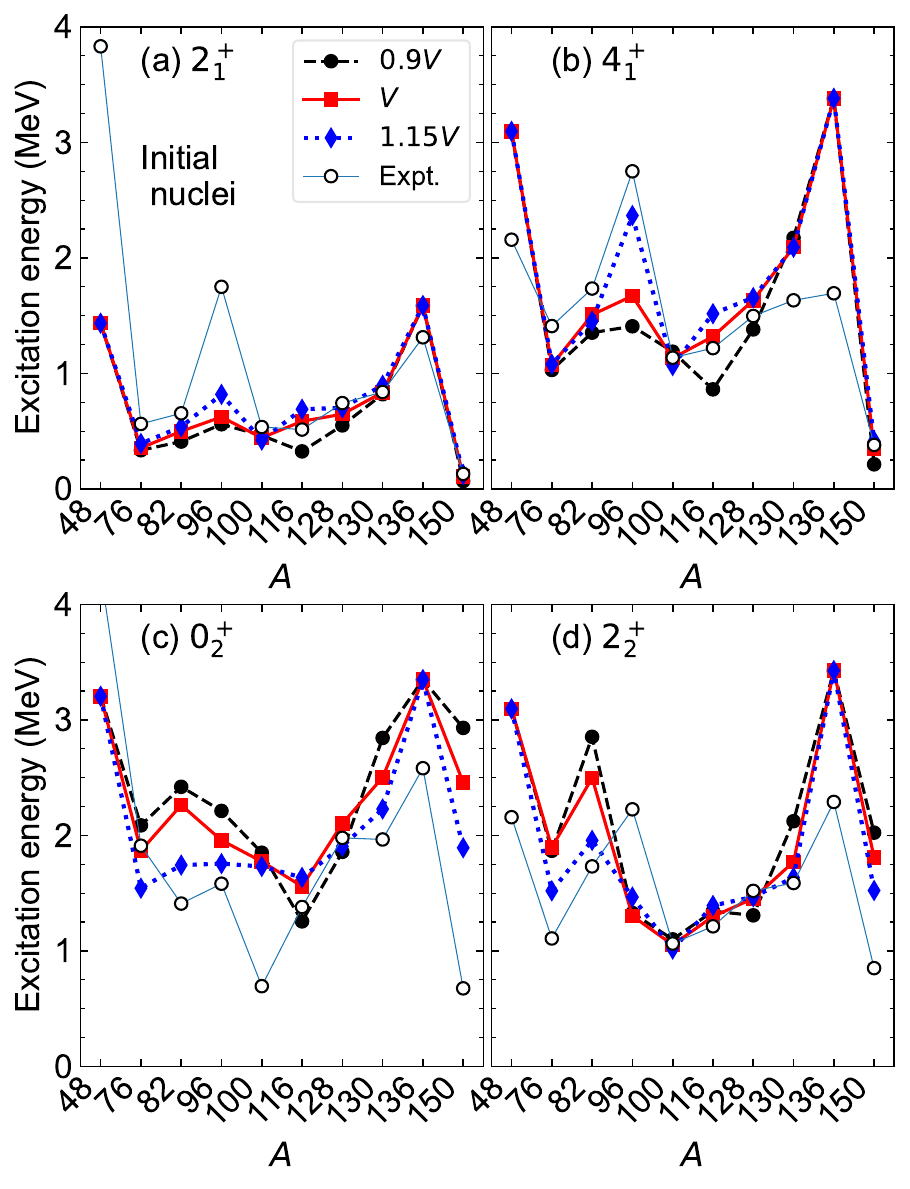}
\caption{Excitation energies of the 
$2^+_1$, $4^+_1$, $0^+_2$, and $2^+_2$ states 
calculated with the mapped IBM-2 for the initial 
even-even nuclei with the reduced ($0.9V$), 
default ($V$), and increased ($1.15V$) pairing 
strengths in the RHB-SCMF SCMF calculations. 
Experimental values are taken from the 
NNDC database \cite{data}.}
\label{fig:ee1}
\end{center}
\end{figure}

%
\begin{figure}
\begin{center}
\includegraphics[width=\linewidth]{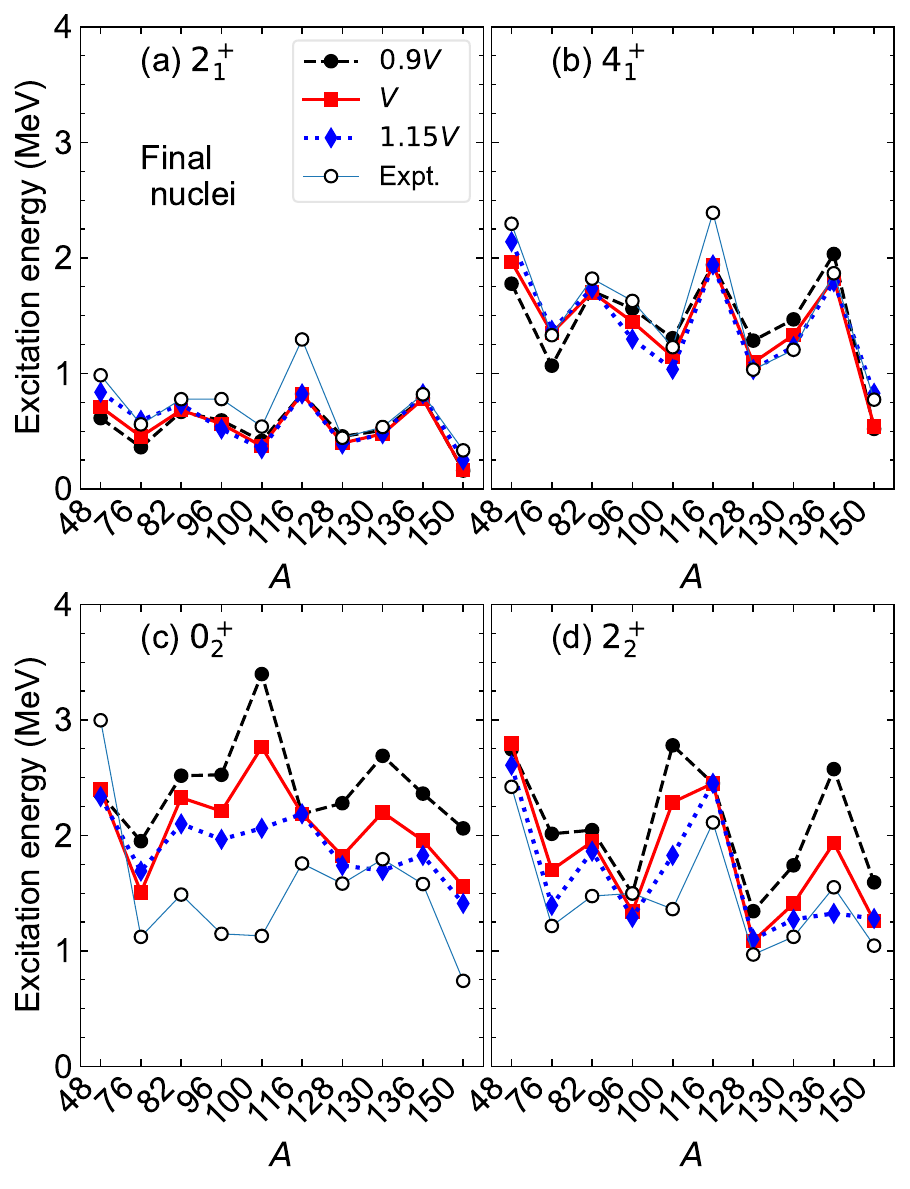}
\caption{Same as Fig.~\ref{fig:ee1}, but for the 
final even-even nuclei.}
\label{fig:ee2}
\end{center}
\end{figure}

%
\begin{figure}
\begin{center}
\includegraphics[width=\linewidth]{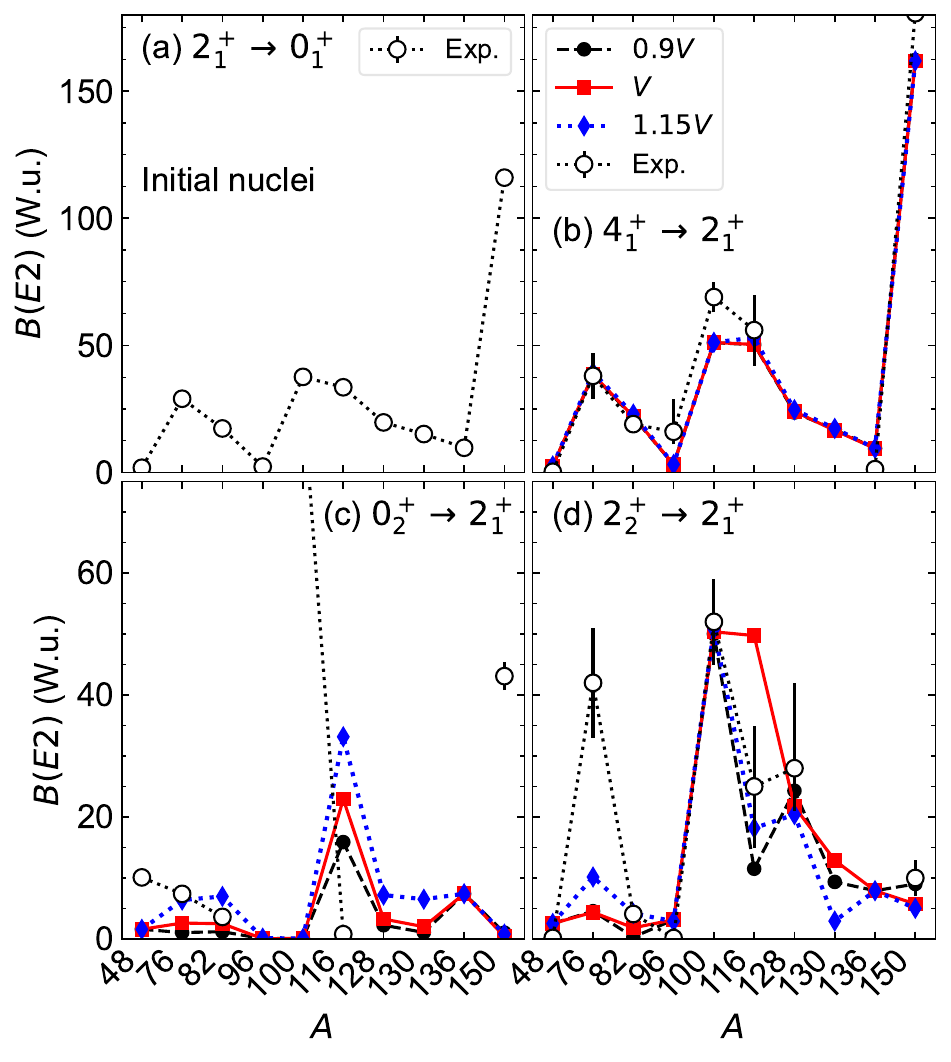}
\caption{$B(E2)$ values 
[in Weisskopf units (W.u.)] for the $E2$ transitions 
(a) $2^+_1 \to 0^+_1$, (b) $4^+_1 \to 2^+_1$, 
(c) $0^+_2 \to 2^+_1$, and (d) $2^+_2 \to 2^+_1$ 
calculated for the initial even-even nuclei 
by using the reduced ($0.9V$), default ($V$), 
and increased ($1.15V$) pairing strengths. 
Experimental data are adopted from the NNDC database. 
Note that since the effective boson charges are fit to the 
experimental $B(E2;2^+_1 \to 0^+_1)$ values \cite{data}, 
the calculated values for this transition are 
not included in the plot.}
\label{fig:ee1-em}
\end{center}
\end{figure}

%
\begin{figure}
\begin{center}
\includegraphics[width=\linewidth]{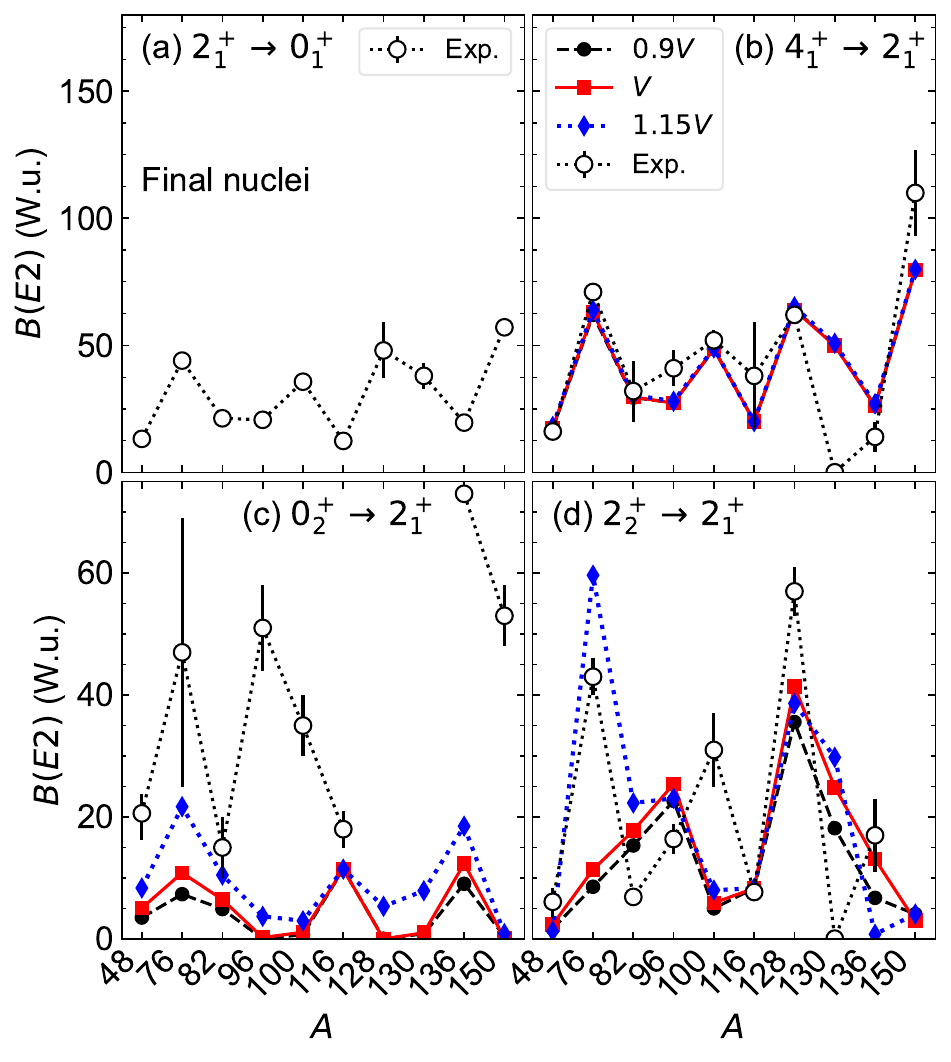}
\caption{Same as Fig.~\ref{fig:ee1-em}, but for the 
final even-even nuclei.}
\label{fig:ee2-em}
\end{center}
\end{figure}

%
\begin{figure}
\begin{center}
\includegraphics[width=\linewidth]{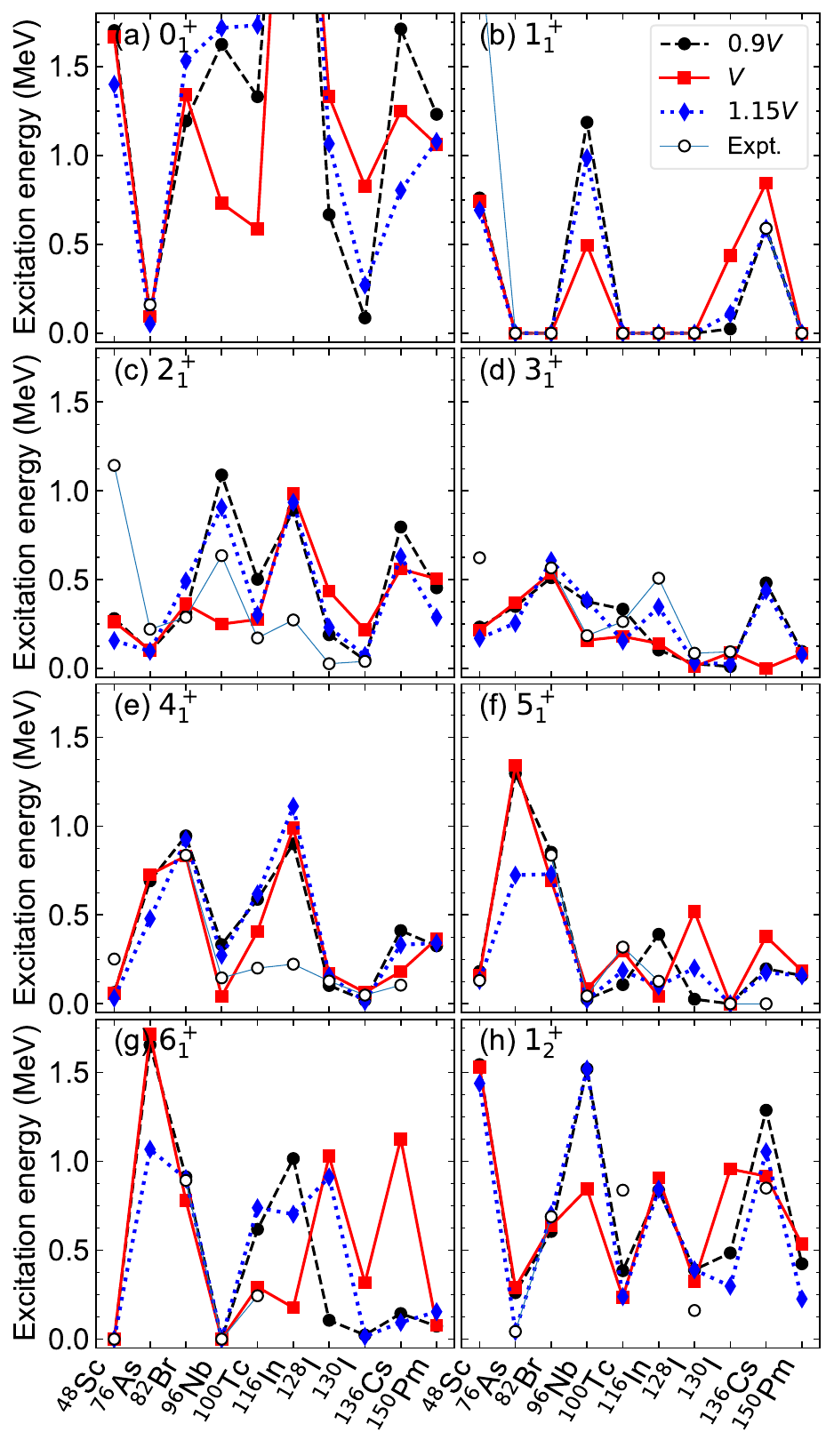}
\caption{Excitation energies of the 
$0^+_1$, $1^+_1$, $2^+_1$, $3^+_1$, 
$4^+_1$, $5^+_1$, $6^+_1$, and $1^+_2$ states 
calculated with the IBFFM-2 for the intermediate 
odd-odd nuclei with the pairing strengths of 
$0.9V$, $V$, and $1.15V$. Experimental data 
are taken from the NNDC database \cite{data}}
\label{fig:oo}
\end{center}
\end{figure}

\section{Low-energy nuclear structures\label{sec:str}}

\subsection{Even-even nuclei }

\subsubsection{Potential energy surfaces\label{sec:pes}}

Figures~\ref{fig:pes1} and \ref{fig:pes2} display 
the PESs in terms of the quadrupole triaxial 
($\beta,\gamma$) deformations for the even-even initial and 
final nuclei, obtained from 
the constrained RHB-SCMF calculations employing 
the DD-PC1 EDF, 
combined with the reduced ($0.9V$), default ($V$), and 
increased ($1.15V$) strengths of the separable 
pairing force (\ref{eq:pair1}). 
The PESs with the default pairing strength are 
taken from Ref.~\cite{nomura2022bb} without any change, 
and their properties were discussed there. 
A general effect of reducing the pairing strength 
is that the potential becomes steeper in both 
$\beta$ and $\gamma$ deformations, and 
in some cases the global minimum occurs at 
a larger $\beta$ deformation. 
If one increases the pairing strength with respect 
to the default one, on the other hand, 
the PES generally looks even softer in $\beta$ 
and $\gamma$ deformations, and the location 
of the global minimum shifts in the direction to 
the origin, hence disfavoring the strong deformation.

Figure~\ref{fig:edef} shows the energies 
(a) $E_{\rm def}$, defined as the difference between 
the mean-field energies at the origin 
and at the global minimum, 
and (b) $E_\gamma$, defined as the difference in energy 
between the global minimum and saddle point along 
axial deformation, 
i.e., the $\gamma=0^{\circ}$ and $60^{\circ}$ axes. 
The former quantity represents an energy gained 
by the deformation, while the latter can be used 
as a measure of the $\gamma$ softness. 
It is seen from Fig.~\ref{fig:edef} that 
the quantity $E_{\rm def}$ is reduced (increased) by a few MeV 
when the increased $1.15V$ (reduced $0.9V$) 
pairing strength is used. 
The increased pairing strength generally leads to 
a lower $E_{\gamma}$ energy, implying that 
the PES becomes softer in $\gamma$ deformation.


The corresponding IBM-2 PES in the case of the 
default pairing strength $V$ were presented in 
Ref.~\cite{nomura2022bb}. 
It was shown \cite{nomura2022bb} that 
the basic features of the SCMF PESs in the neighborhood 
of the global minimum are reproduced 
by the IBM-2. Discrepancies between the 
SCMF and IBM-2 PESs in their topology 
were shown to arise such that the latter is 
in most cases flatter in the region of 
large $\beta$ deformations, and that triaxial minima 
found in the SCMF PESs for 
$^{96}$Zr, $^{96}$Mo, $^{100}$Mo, $^{128}$Xe, 
and $^{130}$Xe cannot be reproduced in the IBM-2 
surfaces. These discrepancies can be attributed to 
the limited degrees of freedom and form of the 
Hamiltonian in the IBM-2. 
Similar observations hold for the mapped IBM-2 PESs 
both with the reduced and increased pairing strengths.

\subsubsection{Derived IBM-2 parameters\label{sec:para}}

Figure~\ref{fig:para} shows the derived parameters 
for the IBM-2 Hamiltonian (\ref{eq:hb}) 
for the even-even nuclei. 
What is worth noticing are the facts that the derived 
single-$d$ boson energy $\epsilon_d$ 
[Figs.~\ref{fig:para}(a) and \ref{fig:para}(b)] 
is basically larger when a larger pairing 
strength is considered, and that the quadrupole-quadrupole 
interaction strength $\kappa$ derived in the 
case of a stronger pairing force is systematically 
reduced in magnitude, most notably, 
by approximately a factor of 4 for $^{136}$Ba 
[Figs.~\ref{fig:para}(c) and \ref{fig:para}(c)].

In addition, the ratio of these parameters, 
$|\kappa|/\epsilon_d$, is systematically lowered for 
the increased pairing strength, representative cases 
being $^{76}$Ge, $^{128}$Te, $^{100}$Ru, 
$^{128}$Xe, and $^{130}$Xe. 
In the case of $^{76}$Ge, for instance, 
the ratios corresponding to $0.9V$, 
$V$, and $1.15V$ pairing strengths are 
1.1, 0.88, and 0.47, respectively. 
Since the $\hat n_{d}$ term favors a spherical shape and 
the $\hat Q_{\nu}\cdot \hat Q_{\pi}$ term 
gives rise to deformation, 
the ratio of their strength parameters 
$|\kappa|/\epsilon_d$ provides 
an implication for 
the degree of deformation and collectivity. 
The reduction of the ratio with the increased pairing 
strength is reasonable, as the pairing correlations 
generally prefer a less deformed shape, and indeed 
the SCMF PES tends to be softer with enhanced  
pairing strength (cf. Fig.~\ref{fig:edef}).

From Figs.~\ref{fig:para}(e) to \ref{fig:para}(h)
the derived parameters $\chi_\nu$ and $\chi_\pi$  
do not depend much on the pairing strengths. 
A few exceptions are perhaps the $\chi_\nu$ values for 
$^{116}$Cd, and $^{76}$Se, for which the values 
corresponding to the pairing strengths of 
$0.9V$ and $1.15V$ are quite large 
and small, respectively. 
This reflects the fact that the SCMF PESs for 
these nuclei exhibit a more pronounced potential 
valley on the oblate side with the reduced pairing strength 
(see Figs.~\ref{fig:pes1} and \ref{fig:pes2}).

As noted, in the present 
IBM-2 calculations the $\hat L \cdot \hat L$ 
term is considered only for 
$^{150}$Nd, $^{150}$Sm, $^{96}$Zr, and $^{76}$Se 
when particular pairing strengths are considered. 
For these nuclei, this term has certain 
influences on the energy spectra, 
and the corresponding strength parameter $\kappa'$ 
are also appreciable: 
for $^{150}$Nd the values $\kappa'=-0.0082$ MeV 
(with $V$) and $-0.024$ MeV (with $0.9V$); 
for $^{150}$Sm $\kappa'=0.0095$ MeV
(with $V$) and $0.022$ MeV (with $1.15V$); 
for $^{96}$Zr $\kappa'=0.021$ MeV (with $V$)
and $0.063$ MeV (with $1.15V$); 
and for $^{76}$Se $\kappa'=0.021$ MeV (with $V$)
are employed.

The parameters for the like-boson interactions 
in Eq.~(\ref{eq:hb-semimagic}) 
specifically considered for the semimagic nuclei 
are as follows. 
$\epsilon_{d\nu}=1.5$ MeV and 
$\kappa_\nu=-0.057$ MeV ($^{48}$Ca, and $^{116}$Sn); 
$\chi_\nu=0.8$ ($^{116}$Sn) and 0 ($^{48}$Ca); 
while $\epsilon_{d\pi}=1.5$ MeV, 
$\kappa_\pi=-0.057$ MeV, and 
$\chi_\pi=-0.8$ for $^{138}$Xe. 
These values are taken to be the same between 
the IBM-2 calculations based on the 
different pairing strengths. 

Concerning the $\tnbb$ decays of $^{48}$Ca, $^{116}$Cd, 
and $^{150}$Nd, the boson core nuclei for the odd-odd 
intermediate nuclei $^{48}$Sc, $^{116}$In, and $^{150}$Pm 
are taken to be $^{46}$Ca, $^{118}$Sn, and $^{148}$Nd, 
respectively, which are different from 
either of the initial and final nuclei. 
The IBM-2 parameters used for these boson core 
nuclei are shown in Table~IX of Ref.~\cite{nomura2022bb} 
in the case of the default pairing strength, 
and the same parameters are here employed. 
As for the $^{46}$Ca and $^{118}$Sn nuclei, 
the same IBM-2 parameters 
are used for the calculations with the 
modified pairing strengths $0.9V$ and $1.15V$. 
Regarding $^{148}$Nd, however, $\epsilon_d$ and $\kappa$ 
parameters are here changed with respect to those 
with the default $V$: 
$\epsilon_d=0.21$ (0.48) MeV 
and $\kappa=-0.265$ ($-0.21$) MeV, for the 
strength of $0.9V$ ($1.15V$).

\subsubsection{Low-lying states\label{sec:energy}}

Figures~\ref{fig:ee1} and \ref{fig:ee2} show 
the excitation energies of the $2^+_1$, 
$4^+_1$, $0^+_2$, and $2^+_2$ states of the initial 
and final even-even nuclei resulting from the 
mapped IBM-2, respectively. 
One sees that the description of the energies for the 
yrast states $2^+_1$ and $4^+_1$ is not strongly 
affected by changing the pairing strength in the 
underlying RHB-SCMF calculations, 
except perhaps for the $^{96}$Zr and $^{116}$Cd nuclei. 
For the $^{96}$Zr nucleus, in particular, 
there is observed a 
certain improvement of the description 
of the $4^+_1$ excitation energy. 
Also for $^{96}$Zr, the measured $2^+_1$ energy level 
is particularly high, which is due to the filling of the 
neutron $d_{5/2}$ subshell. 
The present IBM-2 cannot reproduce it, since 
the SCMF PESs for this nucleus 
with the three pairing choices all suggest a well 
deformed minimum (see Fig.~\ref{fig:pes1}).

As one can see in Figs.~\ref{fig:ee1}(c), \ref{fig:ee1}(d), 
\ref{fig:ee2}(c), and \ref{fig:ee2}(d), 
dependence of the calculated excitation energies 
on the choice of the pairing strength is even more 
visible for the non-yrast states $0^+_2$ and $2^+_2$. 
For almost all the even-even nuclei considered, 
by the increase of the separable pairing force, 
both the $0^+_2$ and $2^+_2$ energy levels 
are generally lowered, and are in some cases 
in a better agreement with the experimental 
data \cite{data}. 
This result is an immediate consequence of the 
reduced quadrupole-quadrupole interaction strength 
in the IBM-2 
[cf. Figs.~\ref{fig:para}(c) and \ref{fig:para}(d)], 
and further confirms the effect of increasing  
the pairing strength in the SCMF calculations, 
which produce the PESs with a potential valley that is 
much less pronounced. 

Significant deviations of the calculated $0^+_2$ 
energy levels from the experimental data 
are still present, e.g., for $^{100}$Mo and $^{150}$Nd 
[Fig.~\ref{fig:ee1}(c)], and $^{96}$Mo, $^{100}$Ru, 
and $^{150}$Sm [Fig.~\ref{fig:ee2}(c)], even though 
the increased pairing strength is considered. 
Given the fact that these are all predicted 
to have a well deformed ground 
state (cf. Figs.~\ref{fig:pes1} and \ref{fig:pes2}), 
characterized by the large $E_{\rm def}$ energies, 
perhaps an even larger pairing strength 
would be required so the PES becomes much more softer, 
leading to a much weaker quadrupole-quadrupole 
interaction strength $\kappa$ for the IBM-2 
Hamiltonian. 
The low-energy $0^+_2$ levels, which are supposed 
to play a part especially in the $^{76}$Ge$\to^{76}$Se,
$^{100}$Mo$\to^{100}$Mo, and $^{150}$Nd$\to^{150}$Sm 
decays, could be adequately handled by 
the version of the IBM-2 that includes 
configuration mixing between the normal and 
intruder states \cite{duval1981}.  
Within the version of the mapped IBM-2 
with a single (normal) configuration as 
considered here, increasing the pairing strength 
does not lower the $0^+_2$ energy level 
as dramatically as with the configuration-mixing 
IBM-2. 
A more realistic calculation would, therefore, 
require one to incorporate in the mapping 
procedure effects of the intruder states 
by using the configuration-mixing IBM-2, 
which would be of 
particular importance for the 
$^{100}$Mo$\to^{100}$Mo$(0^+_2)$ 
and $^{150}$Nd$\to^{150}$Sm$(0^+_2)$ decays.

\subsubsection{$E2$ transitions}

The $B(E2)$ transition rates for the even-even nuclei 
are computed in the IBM-2 by using the $E2$ operator 
\begin{eqnarray}
\label{eq:e2b}
 \hat T^{(E2)}_{\rm B} = 
e^{\rm B}_\nu \hat Q_{\nu} + e^{\rm B}_\pi \hat Q_{\pi} \; ,
\end{eqnarray}
where $\hat Q_{\rho}$ are the same quadrupole operators 
used in the Hamiltonian (\ref{eq:hb}) with the same 
value of the $\chi_\rho$ parameter. 
The effective boson charges $e^{\rm B}_\rho$ are here 
assumed to be the same between neutron and proton 
systems, $e^{\rm B}_\nu = e^{\rm B}_\pi \equiv e^{\rm B}$, 
which is then determined so as to reproduce the 
experimental $B(E2; 2^+_1 \to 0^+_1)$ value \cite{data} 
for each nucleus. 
Figures~\ref{fig:ee1-em} and \ref{fig:ee2-em} display 
the calculated and experimental $B(E2;2^+_1 \to 0^+_1)$, 
$B(E2;4^+_1 \to 2^+_1)$, $B(E2;0^+_2 \to 2^+_1)$, 
and $B(E2;2^+_2 \to 2^+_1)$ transition strengths 
for the even-even nuclei. 
The calculated $B(E2;4^+_1 \to 2^+_1)$ values 
are, in general, in a good agreement with 
experiment, and the results from the 
different pairing strengths are strikingly similar 
to each other. 
The $B(E2;0^+_2 \to 2^+_1)$, 
and $B(E2;2^+_2 \to 2^+_1)$ transition rates 
are, however, at variance between the calculations 
with different pairing strengths. 
A significant improvement of the description of the
$B(E2;2^+_2 \to 2^+_1)$ transition rate 
due to the increase of the 
pairing force is observed for $^{76}$Se. 
The modification of the separable pairing strength 
thus appears to affect the wave functions for the 
even-even nuclei, especially those of the 
non-yrast states.

\subsection{Intermediate odd-odd nuclei}

Figure~\ref{fig:oo} depicts 
the calculated excitation energies 
of low-spin positive-parity 
states of the intermediate odd-odd nuclei 
resulting from the IBFFM-2 with the three different 
pairing strengths. 
The correct ground-state 
spin is reproduced by any of the three IBFFM-2 
calculations, except for the $^{136}$Cs nucleus. 
One can see that 
the IBFFM-2 descriptions based on the 
three choices of the pairing strength 
are rather different from each other. 
There appears to be, however, no general 
tendency of reaching a better agreement 
with the experimental data by changing the 
pairing strength in either way. 
The differences in the calculated energy levels 
due to the choices of the pairing strength 
are primarily attributed to the differences between 
the respective IBFFM-2 parameters, 
which, e.g., in $^{96}$Nb, $^{128}$I, 
and $^{130}$I, differ significantly from each other 
(see Table~\ref{tab:paraff} in the Appendix~\ref{sec:paraoo}).

Furthermore, 
the $E2$ and $M1$ transition properties for the 
odd-odd nuclei are studied. 
The $E2$ operator is given by
\begin{align}
 \label{eq:e2}
\hat T^{(E2)}
= \hat T^{(E2)}_\text{B}
+ \hat T^{(E2)}_\text{F} \; ,
\end{align}
with the boson operator $\hat T^{(E2)}_\text{B}$ 
defined in Eq.~(\ref{eq:e2b}), and the fermion one
\begin{align}
 \label{eq:e2f}
\hat T^{(E2)}_\mathrm{F}
=-\frac{1}{\sqrt{5}}
&\sum_{\rho=\nu,\pi}
\sum_{\jr\jr'}
(u_{\jr}u_{\jr'}-v_{\jr}v_{\jr'})
\nonumber\\
&\times
\left\langle
\ell_\rho\frac{1}{2}\jr 
\bigg\| 
e^\mathrm{F}_\rho r^2 Y^{(2)} 
\bigg\|
\ell_\rho'\frac{1}{2}\jr'
\right\rangle
(a_{\jr}^\dagger\times\tilde a_{\jr'})^{(2)} \; .
\end{align}
The neutron and proton effective charges,  
$e^\mathrm{F}_\nu =0.5$ $e$b and
$e^\mathrm{F}_\pi =1.5$ $e$b, 
are taken from Ref.~\cite{nomura2022bb}. 
The $M1$ transition operator 
$\hat T^{(M1)}$ reads 
\begin{align}
 \label{eq:m1}
\hat T^{(M1)}
=\sqrt{\frac{3}{4\pi}}
&\sum_{\rho=\nu,\pi}
\Biggl[
g_\rho^\mathrm{B}\hat L_\rho
-\frac{1}{\sqrt{3}}
\sum_{\jr\jr'}
(u_{\jr}u_{\jr'}+v_{\jr}v_{\jr'})
\nonumber \\
&\times
\left\langle \jr \| g_l^\rho{\bf \ell}+g_s^\rho{\bf s} 
\| \jr' \right\rangle
(a_{\jr}^\+\times\tilde a_{\jr'})^{(1)}
\Biggr] \; ,
\end{align}
where $\hat L_\rho$ is the angular momentum operator 
in the boson system (\ref{eq:hb}), and 
the empirical $g$ factors for the neutron and
proton bosons, $g_\nu^\mathrm{B}=0\,\mu_N$ and 
$g_\pi^\mathrm{B}=1.0\,\mu_N$, respectively, are considered. 
For the neutron (or proton) $g$ factors, the 
free values $g_\ell^\nu=0\,\mu_N$ and $g_s^\nu=-3.82\,\mu_N$
($g_\ell^\pi=1.0\,\mu_N$ and $g_s^\pi=5.58\,\mu_N$) 
are employed, with $g_s^\rho$ quenched by 30 \%.

Table~\ref{tab:em-oo} gives the calculated 
electric quadrupole $Q(I)$ and magnetic dipole 
$\mu(I)$ moments, and $B(M1)$ transition probabilities 
in the cases of the three different 
pairing strengths in the RHB-SCMF calculations 
in comparison with the available 
experimental data \cite{data,stone2005}. 
The transition properties appear to be sensitive 
to the choice of the pairing strength. 
Notable difference is found   
in the $\mu(1^+_1)$ and $\mu(6^+_1)$ moments  
for the $^{76}$As and $^{96}$Nb nuclei, respectively, 
since not only their magnitudes but also signs 
are different between the pairing strengths considered.

The IBFFM-2 wave function for 
the $1^+_1$ ground state of $^{76}$As 
is here accounted for by the mixture of the 
neutron-proton pair components 
$[\nu p_{1/2} \otimes \pi p_{3/2}]^{(J=1^+)}$ (45 \%), 
and 
$[\nu p_{1/2} \otimes \pi f_{5/2}]^{(J=5^+)}$ (38 \%) 
when the default ($V$) and reduced ($0.9V$) 
pairing strengths are employed in the 
RHB-SCMF calculations. 
For the increased pairing ($1.15V$), 
the dominant configuration is of the type 
$[\nu p_{1/2} \otimes \pi p_{3/2}]^{(J=1^+)}$ (72 \%), 
and there are numerous minor contributions 
from other pair components. 
Since the compositions of the $1^+_1$ wave function 
and the employed boson-fermion and fermion-fermion 
interaction parameters 
are similar between the IBFFM-2 calculations with 
$V$ and $0.9V$, the difference between the $\mu(1^+_1)$ 
values from the two calculations probably 
arises from the difference between 
the even-even boson-core ($^{76}$Ge) parameters. 
Concerning the $^{96}$Nb nucleus, 
the IBFFM-2 wave function of the $6^+_1$ 
ground state is mostly (86 \%) composed 
of the pair configuration 
$[\nu d_{5/2} \otimes \pi g_{9/2}]^{(J=6^+)}$ 
in the case of the default pairing strength ($V$). 
In those calculations in which reduced ($0.9V$) 
and increased ($1.15V$) 
pairing strengths are employed, however, 
the pair component of the type 
$[\nu h_{11/2} \otimes \pi p_{1/2}]^{(J=6^+)}$ 
makes a dominant (89 \%, and 87 \%, respectively) 
contribution to the the IBFFM-2 $6^+_1$ 
wave functions.

The differences in the nature of the 
wave functions, and the calculated 
electromagnetic transition properties 
for the odd-odd nuclei 
among the three cases of the pairing strength 
arise from the differences in 
the parameters involved in the IBFFM-2 Hamiltonian, 
which are especially dependent on the 
strength parameters for the even-even 
boson core. 
One cannot draw any definite conclusion 
from Table~\ref{tab:em-oo} that 
increasing or reducing the pairing 
strength in the RHB-SCMF model is particularly good 
for describing many of the electromagnetic transition 
properties of the odd-odd nuclei studied. 
It appears to be rather reasonable to use the 
standard pairing strength $V$, since only 
in that case is a reasonable description obtained 
for the $\mu(1^+_1)$ and $\mu(1^+_6)$ moments 
of $^{76}$As, and $^{96}$Nb, respectively 
(cf. Table~\ref{tab:em-oo}).

%
%
\begin{table*}
\caption{\label{tab:em-oo}
Calculated electric quadrupole $Q(I)$ (in $e$b) and magnetic 
dipole $\mu(I)$ (in $\mu_N$) moments, and the 
$B(M1)$ transition strengths (in W.u.) 
of the intermediate 
odd-odd nuclei, obtained from the 
RHB-SCMF mapped IBFFM-2 with the reduced ($0.9V$), 
default ($V$), and increased ($1.15V$) 
pairing strengths. 
The experimental values 
are taken from Refs.~\cite{data,stone2005}.
}
 \begin{center}
 \begin{ruledtabular}
  \begin{tabular}{lccccc}
\multirow{2}{*}{nucleus} &
\multirow{2}{*}{property} &
\multicolumn{3}{c}{IBFFM-2} &
\multirow{2}{*}{Experiment} \\
\cline{3-5}
& & $0.9V$ & $V$ & $1.15V$ & \\
\hline
$^{48}$Sc
& $\mu(6^+_{1})$ & $3.100$ & $3.098$ & $3.091$ & $+3.737\pm0.012$ \\
[1.0ex]
$^{76}$As
& $\mu(1^+_{1})$ & $-0.095$ & $0.388$ & $2.235$ & $+0.559\pm0.005$ \\
[1.0ex]
$^{96}$Nb
& $\mu(6^+_{1})$ & $-0.479$ & $4.547$ & $-0.874$ & $4.976\pm0.004$ \\
& $B(M1;4^+_{1}\to5^+_{1})$ & $0.0039$ & $1.2298$ & $0.0263$ & $>0.021$ \\
& $B(M1;2^+_{1}\to3^+_{1})$ & $0.0030$ & $0.2426$ & $0.0025$ & $>0.00017$ \\
[1.0ex]
$^{116}$In
& $\mu(1^+_{1})$ & $2.958$ & $2.478$ & $2.996$ & $+2.7876\pm0.0006$ \\
& $Q(1^+_{1})$ & $0.126$ & $0.213$ & $0.110$ & $0.11$ \\
& $\mu(5^+_{1})$ & $0.870$ & $0.177$ & $0.505$ & $4.435\pm0.015$ \\
& $Q(5^+_{1})$ & $-0.764$ & $-0.813$ & $-0.747$ & $+0.802\pm0.012$ \\
& $B(M1;4^+_{1}\to5^+_{1})$ & $0.0002$ & $0.0072$ & $0.0054$ & $>0.18$ \\
& $B(M1;2^+_{1}\to1^+_{1})$ & $0.1236$ & $0.2249$ & $0.0549$ & $>0.016$ \\
& $B(M1;4^+_{2}/5^+_{2}\to4^+_{1})$ & $0.0010/0.1709$ & $0.0621/0.0060$ & $0.0616/0.0687$ & $0.00013\pm0.00006$ \\
& $B(M1;4^+_{2}/5^+_{2}\to5^+_{1})$ & $0.0013/0.0085$ & $0.0002/0.0275$ & $0.0148/0.0044$ & $0.00013\pm0.00006$ \\
& $B(M1;3^+_{1}\to4^+_{1})$ & $0.0014$ & $0.0090$ & $0.0005$ & $>0.0080$ \\
& $B(M1;3^+_{1}\to2^+_{1})$ & $0.0232$ & $0.1079$ & $0.0505$ & $>0.0066$ \\
[1.0ex]
$^{128}$I
& $B(M1;3^+_{2}\to3^+_{1})$ & $0.0002$ & $0.0068$ & $0.0034$ & $>0.0017$ \\
& $B(M1;3^+_{2}\to2^+_{1})$ & $0.0088$ & $0.0011$ & $0.0005$ & $>0.011$ \\
& $B(M1;1^+_{2}/2^+_{2}\to2^+_{1})$ & $0.6268/0.0090$ & $0.0022/0.0223$ & $0.0170/0.0166$ & $>0.0026$ \\
& $B(M1;1^+_{2}/2^+_{2}\to1^+_{1})$ & $0.1847/0.0085$ & $0.0004/0.0006$ & $0.0002/0.0044$ & $>0.0046$ \\
& $B(M1;3^+_{3}\to2^+_{2})$ & $0.0036$ & $0.0041$ & $0.0092$ & $>0.0095$ \\
& $B(M1;3^+_{3}\to3^+_{1})$ & $0.0000$ & $0.0087$ & $0.0002$ & $>0.00011$ \\
& $B(M1;3^+_{3}\to2^+_{1})$ & $0.0000$ & $0.0021$ & $0.0185$ & $>0.00051$ \\
& $B(M1;4^+_{2}\to3^+_{1})$ & $0.0267$ & $0.0542$ & $0.0105$ & $>0.0027$ \\
& $B(M1;4^+_{2}\to3^+_{2})$ & $0.4234$ & $0.0733$ & $0.1520$ & $>0.0019$ \\
& $B(M1;4^+_{2}\to4^+_{1})$ & $0.0048$ & $0.0045$ & $0.0855$ & $>0.00050$ \\
[1.0ex]
$^{130}$I
& $\mu(5^+_{1})$ & $2.606$ & $3.900$ & $2.845$ & $3.349\pm0.007$ \\
[1.0ex]
$^{136}$Cs
& $\mu(5^+_{1})$ & $2.381$ & $3.570$ & $2.379$ & $+3.711\pm0.005$ \\
& $Q(5^+_{1})$ & $0.196$ & $0.267$ & $0.191$ & $+0.225\pm0.010$ \\
  \end{tabular}
 \end{ruledtabular}
 \end{center}
\end{table*}

\section{$\tnbb$ decay\label{sec:db}}

\subsection{GT and F transitions\label{sec:gtf-dist}}

Tables~\ref{tab:gtf1} and \ref{tab:gtf2} present 
the calculated $\mgt$ (\ref{eq:mgt}) 
and $\mf$ (\ref{eq:mf}) for 
the ground-state-to-ground-state ($0^+_1 \to 0^+_1$), 
and for the ground-state-to-first-excited 
state ($0^+_1 \to 0^+_2$) decays, respectively. 
As can be seen in Tables~\ref{tab:gtf1} and \ref{tab:gtf2}, 
by the increase 
of the pairing strength the predicted $\mgt$ value 
for the $0^+_1 \to 0^+_1$ decay generally increases 
in magnitude. This is also true for the $\mf$ values. 
For some of the studied $\tnbb$ decays, the absolute value 
$|\mf|$ is so large as to be in the same order of 
magnitude as $|\mgt|$. 
This was already pointed out in the previous 
mapped IBM-2 study of Ref.~\cite{nomura2022bb}, 
and appears to occur irrespective of which 
pairing strength is considered in the present analysis. 
This is so for those $\tnbb$ decays in which 
the neutrons and protons 
are in the same major oscillator shell 
so that the Fermi transitions are allowed. 
For the $^{48}$Ca$\to^{48}$Ti decay in particular, 
the $|\mf|$ value is equal to or even larger than 
$|\mgt|$. 
The large $|\mf|/|\mgt|$ ratio indicates that there 
is a spurious isospin symmetry breaking 
that is not expected in the $\tnbb$ decay. 
Effective ways to restore the isospin 
symmetry would be, for instance, 
to simply discard $\mf$ in the calculations of $\mbb$, and to make 
some modifications to the Fermi transition 
operator (\ref{eq:ofe}) so that the Fermi matrix elements 
should vanish in the closure approximation 
(see Refs.~\cite{barea2013,barea2015} 
for detailed discussions). 
In the present study, no such treatment is made 
to restore the isospin symmetry broken 
in the employed model.

It should be mentioned that 
the results in the case of the default pairing 
strength $V$ are found in Table~III of 
Ref.~\cite{nomura2022bb}, and that one can notice 
slight deviations of the present $\mgt$ and $\mf$ values 
from those in the previous study \cite{nomura2022bb} 
in some instances. 
This is mainly due to the following differences between 
the present calculation and that of \cite{nomura2022bb}. 
First, as already mentioned, 
in some of the even-even and odd-odd 
nuclei modifications to the 
IBM-2 as well as IBFFM-2 parameters are made 
in the present calculation employing the same 
default pairing strength as in Ref.~\cite{nomura2022bb}. 
Second, in the present IBFFM-2 calculation 
the maximum number of iterations 
in the numerical (Lanczos) diagonalization of the 
IBFFM-2 Hamiltonian is set to be 200000 times 
for all the odd-odd nuclei and in all the three 
cases of the separable pairing strength, 
whereas in Ref.~\cite{nomura2022bb} the number 
of iterations was much less and also 
at variance with the nuclei. 
Third, the truncation of the maximum energy for the 
intermediate states for the calculations of $\mgt$ 
and $\mf$ is here set to be 30 MeV, while in 
\cite{nomura2022bb} it was 10 MeV. 
These modifications, especially the second and third ones, 
could have affected the predictions of the 
$\mgt$, as well as $\mf$, and hence $\mbb$ values 
since these quantities require to include contributions 
from higher-lying intermediate states, which should be 
sensitive to the convergence of the IBFFM-2 
eigenvalues and to the truncation to their energy range.

\begin{table*}
\caption{\label{tab:gtf1}
GT and F matrix elements obtained from the mapped 
IBM-2 based on the separable pairing strengths 
of $0.9V$, $V$, and $1.15V$ in the RHB-SCMF 
calculations for the $0^+_1 \to 0^+_1$ 
$\tnbb$ decays of the candidate nuclei. 
}
 \begin{center}
 \begin{ruledtabular}
  \begin{tabular}{lcccccc}
\multirow{2}{*}{$0^+_1 \to 0^+_1$ decay}
&\multicolumn{3}{c}{$\mgt$}
&\multicolumn{3}{c}{$\mf$}\\
\cline{2-4}\cline{5-7}
&{$0.9V$}&{$V$}&{$1.15V$}
&{$0.9V$}&{$V$}&{$1.15V$}\\
\hline
$^{48}$Ca$\to^{48}$Ti & $0.060$ & $0.031$ & $0.077$ & $0.029$ & $0.016$ & $-0.018$ \\ 
$^{76}$Ge$\to^{76}$Se & $0.024$ & $0.036$ & $-0.128$ & $0.000$ & $-0.002$ & $0.059$ \\ 
$^{82}$Se$\to^{82}$Kr & $0.024$ & $-0.052$ & $0.131$ & $0.000$ & $0.011$ & $-0.045$ \\ 
$^{96}$Zr$\to^{96}$Mo & $-0.087$ & $0.175$ & $-0.159$ & $-0.000$ & $0.001$ & $-0.000$ \\ 
$^{100}$Mo$\to^{100}$Ru & $0.465$ & $0.483$ & $0.574$ & $-0.005$ & $-0.000$ & $-0.000$ \\ 
$^{116}$Cd$\to^{116}$Sn & $-0.225$ & $0.275$ & $0.337$ & $-0.000$ & $0.000$ & $-0.001$ \\ 
$^{128}$Te$\to^{128}$Xe & $0.035$ & $-0.102$ & $0.073$ & $-0.007$ & $0.005$ & $-0.041$ \\ 
$^{130}$Te$\to^{130}$Xe & $0.008$ & $-0.038$ & $-0.118$ & $-0.006$ & $0.023$ & $0.076$ \\ 
$^{136}$Xe$\to^{136}$Ba & $-0.091$ & $-0.102$ & $-0.232$ & $-0.004$ & $0.029$ & $0.099$ \\ 
$^{150}$Nd$\to^{150}$Sm & $0.299$ & $-0.369$ & $-0.501$ & $-0.000$ & $0.000$ & $0.000$ \\
  \end{tabular}
 \end{ruledtabular}
 \end{center}
\end{table*}

\begin{table*}
\caption{\label{tab:gtf2}
Same as Table~\ref{tab:gtf1}, but for the 
$0^+_1 \to 0^+_2$ $\tnbb$ decay.}
 \begin{center}
 \begin{ruledtabular}
  \begin{tabular}{lcccccc}
\multirow{2}{*}{$0^+_1 \to 0^+_2$ decay}
&\multicolumn{3}{c}{$\mgt$}
&\multicolumn{3}{c}{$\mf$}\\
\cline{2-4}\cline{5-7}
&{$0.9V$}&{$V$}&{$1.15V$}
&{$0.9V$}&{$V$}&{$1.15V$}\\
\hline
$^{48}$Ca$\to^{48}$Ti & $0.037$ & $0.069$ & $0.024$ & $-0.103$ & $-0.068$ & $-0.068$ \\ 
$^{76}$Ge$\to^{76}$Se & $-0.056$ & $0.064$ & $-0.100$ & $0.032$ & $-0.036$ & $0.094$ \\ 
$^{82}$Se$\to^{82}$Kr & $-0.074$ & $0.090$ & $0.168$ & $0.039$ & $-0.055$ & $-0.092$ \\ 
$^{96}$Zr$\to^{96}$Mo & $0.038$ & $-0.068$ & $0.052$ & $0.000$ & $-0.001$ & $0.001$ \\ 
$^{100}$Mo$\to^{100}$Ru & $0.269$ & $0.154$ & $0.054$ & $-0.001$ & $0.000$ & $-0.000$ \\ 
$^{116}$Cd$\to^{116}$Sn & $0.106$ & $-0.035$ & $0.118$ & $0.001$ & $-0.001$ & $-0.003$ \\ 
$^{128}$Te$\to^{128}$Xe & $-0.027$ & $0.032$ & $-0.134$ & $0.024$ & $0.002$ & $0.112$ \\ 
$^{130}$Te$\to^{130}$Xe & $0.092$ & $0.037$ & $0.363$ & $-0.052$ & $-0.018$ & $-0.232$ \\ 
$^{136}$Xe$\to^{136}$Ba & $0.045$ & $0.009$ & $-0.027$ & $0.022$ & $-0.000$ & $-0.000$ \\ 
$^{150}$Nd$\to^{150}$Sm & $0.095$ & $-0.207$ & $0.156$ & $0.000$ & $-0.000$ & $0.000$ \\
  \end{tabular}
 \end{ruledtabular}
 \end{center}
\end{table*}

%
%
\begin{figure*}[ht]
\begin{center}
\includegraphics[width=.49\linewidth]{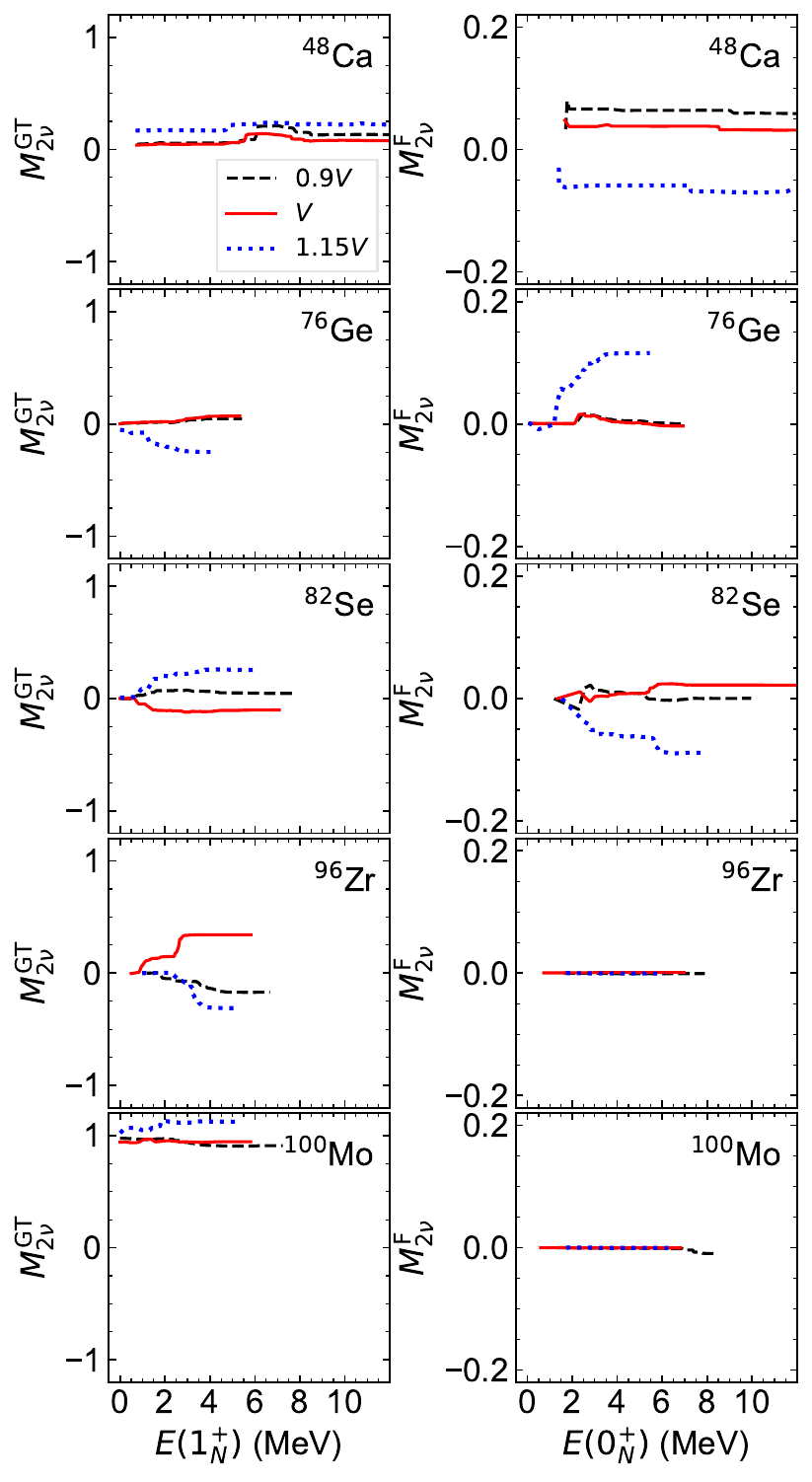} 
\includegraphics[width=.49\linewidth]{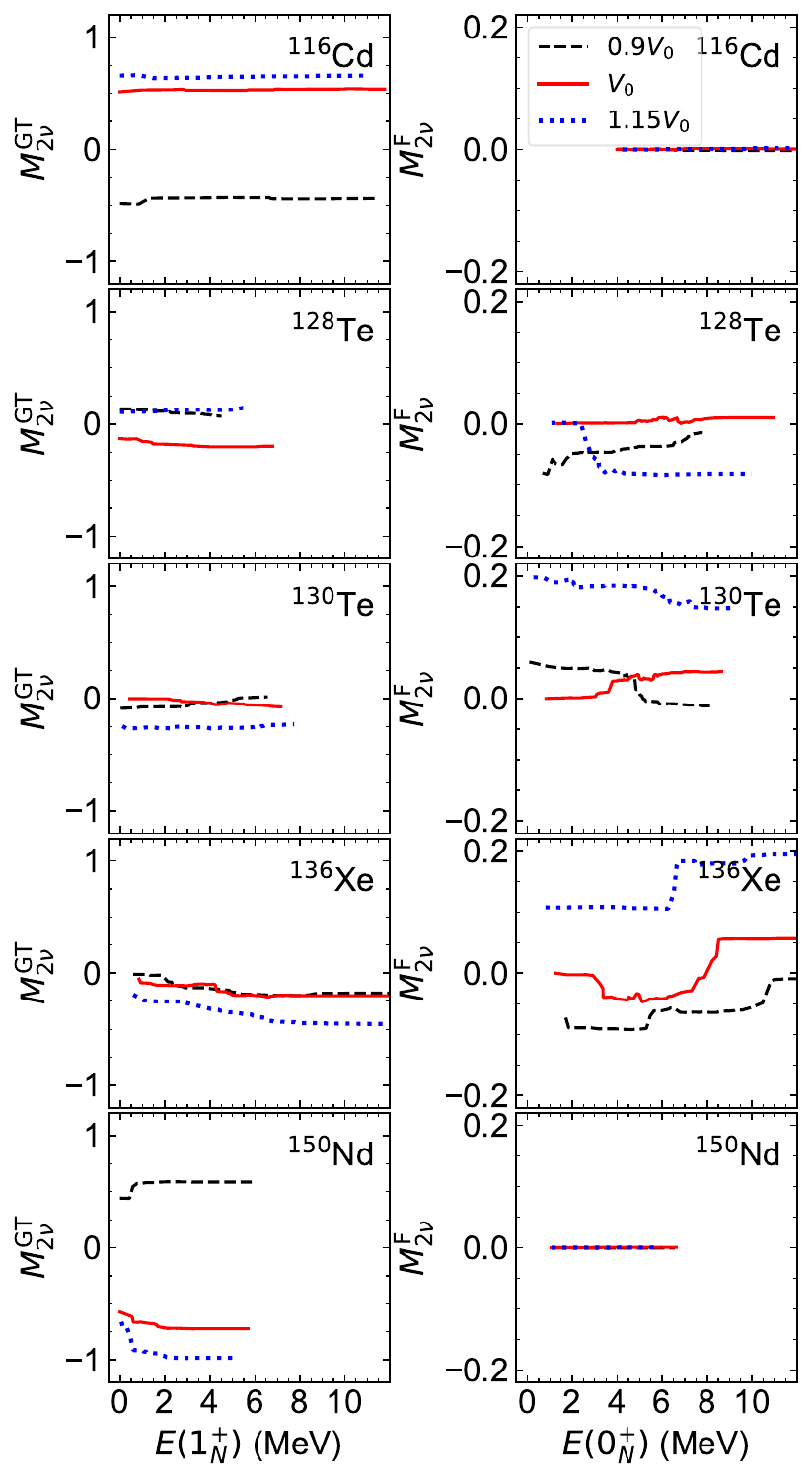} 
\caption{Running sums of the GT (\ref{eq:mgt}) 
(first and third columns) 
and F (\ref{eq:mf}) (second and fourth columns) 
transition strengths for the $\tnbb$ decay of the 
candidate even-even nuclei 
as functions of the excitation energies of the 
$1^+_{N}$ and $0^+_{N}$ intermediate states. 
The calculated results with the reduced, default, and 
increased pairing strengths employed in the 
RHB-SCMF calculations are compared.}
\label{fig:sum}
\end{center}
\end{figure*}

Figure~\ref{fig:sum} depicts the 
running sums of the $\mgt$ (\ref{eq:mgt}) 
and $\mf$ (\ref{eq:mf}) matrix elements for the 
$0^+_1 \to 0^+_1$ $\tnbb$ decays as functions of the 
excitation energies $E(1^+_N)$ and $E(0^+_N)$ of the 
intermediate states, respectively. 
The GT sums in most cases appear to be accounted for 
by the contributions from the lower-lying $1^+$ 
states, typically below $E(1^+_N)\approx 3$ MeV. 
This result is consistent with the so-called 
single-state dominance (SSD) \cite{griffiths1992,civitarese1998} 
or low-lying-state dominance (LLSD) \cite{moreno2008} 
hypotheses drawn from the pnQRPA studies 
for the $\tnbb$ decay. 
The behaviors of the GT sums are also 
at variance with the calculations employing the different 
pairing strengths in the RHB-SCMF input, 
with representative cases being the $^{116}$Cd 
and $^{150}$Nd decays. 
Among the three IBM-2 results, 
the GT running sums resulting from the increased pairing 
strength $1.15V$ exhibit the strongest dependence 
on the $1^+$ excitation energies so that they continue 
to increase in magnitude, implying that 
contributions from higher-lying intermediate states 
are more important than in the calculations with 
weaker pairing forces. 

Regarding the Fermi transitions, in the majority 
of the considered decay processes the contributions 
from the low-lying $0^+_N$ states, with typically up to 
$E(0^+_N) \approx 5$ MeV, determine most 
of the $\mf$ matrix element. 
The Fermi running sums 
seem to show a stronger dependence on the 
intermediate energies than the GT sums. 
Peculiar behaviors of the calculated Fermi sums are 
found for the $^{128}$Te, $^{130}$Te, and $^{136}$Xe decays, 
where especially the sums obtained with the reduced 
pairing strength $0.9V$ decrease in magnitude 
with $E(0^+_N)$.

%
%
\begin{figure}[ht]
\begin{center}
\includegraphics[width=\linewidth]{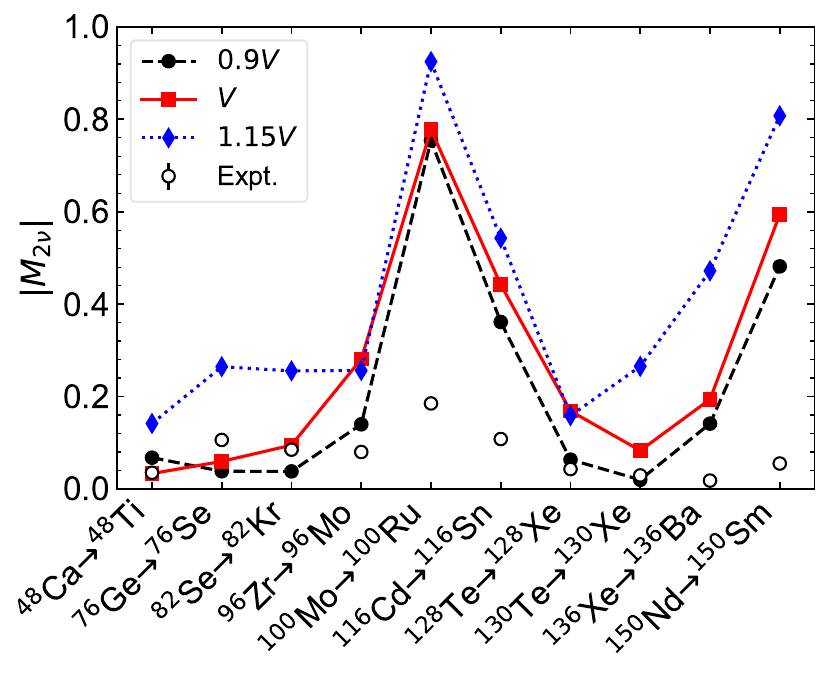}
\caption{Calculated NMEs $\mbb$ 
for the considered $\tnbb$ decays, 
obtained from the mapped IBM-2 employing 
the three different pairing strengths 
($0.9V$, $V$, and $1.15V$) in the RHB-SCMF 
calculations. The bare $\ga$ factor is employed, hence 
no quenching is made. Experimental $\mbb$, extracted 
from the measured half-lives, are taken from Ref.~\cite{barabash2020}. 
}
\label{fig:nme}
\end{center}
\end{figure}

%
%
\begin{figure}[ht]
\begin{center}
\includegraphics[width=\linewidth]{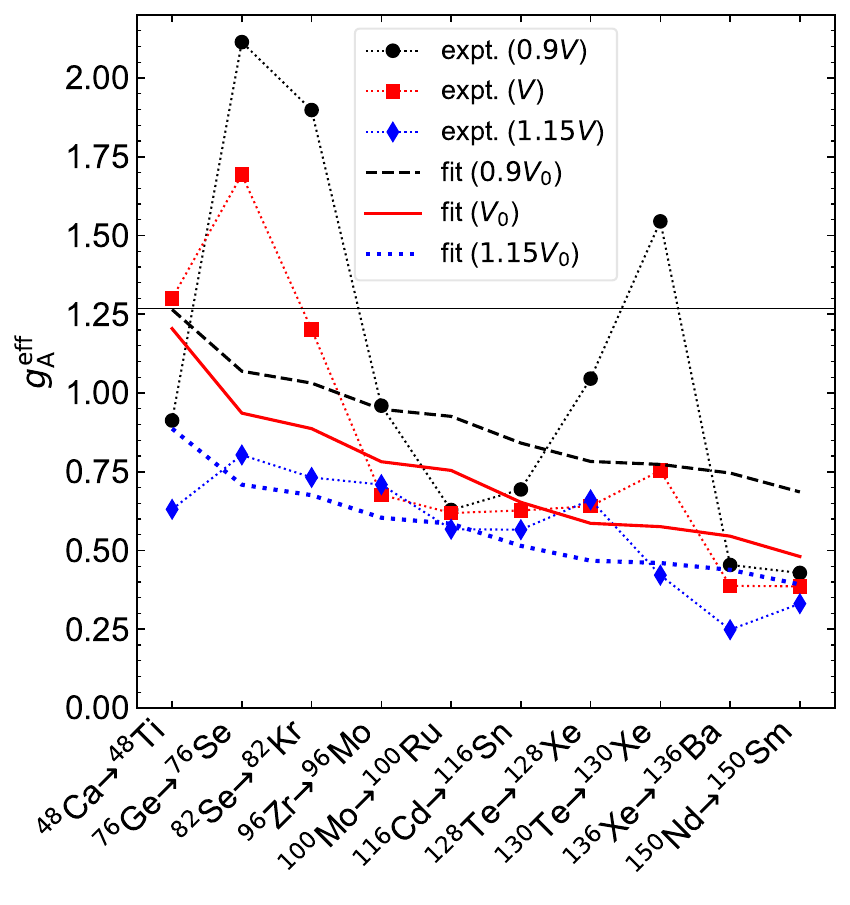}
\caption{Effective $\gae$ factors that are required to 
reproduce the experimental $|\mbbe|$ values 
(denoted by ``expt.''), and the mass-dependent 
$\gae$ factors (denoted by ``fit'') obtained by using 
the formula Eq.~(\ref{eq:gae}), for the different pairing 
strengths in the RHB-SCMF calculations. 
The free-nucleon value $\ga$ is indicated by the 
sold horizontal line.}
\label{fig:ga}
\end{center}
\end{figure}

%
%
\begin{figure}[ht]
\begin{center}
\includegraphics[width=\linewidth]{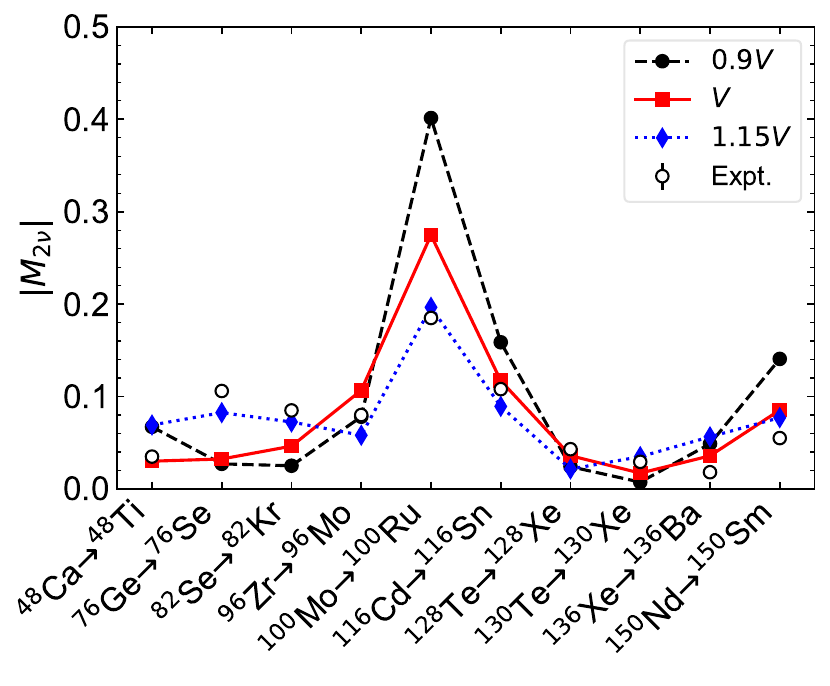}
\caption{Effective NMEs $|\mbbe|$ calculated by 
using the $\gae$ factors defined in Eq.~(\ref{eq:gae}) 
in the cases of the three different pairing strengths.}
\label{fig:nme-q}
\end{center}
\end{figure}

\subsection{$\tnbb$ NMEs}

Figure~\ref{fig:nme} displays the calculated 
$\mbb$ (\ref{eq:mbb}) for the $\trgg$ 
$\tnbb$ decay for the considered nuclei. 
The $\mbb$ values 
calculated for the $0^+_1 \to 0^+_1$ $\tnbb$ decays, and 
those for the $^{100}$Mo$(0^+_1)\to^{100}$Ru$(0^+_2)$
and $^{150}$Nd$(0^+_1)\to^{150}$Sm$(0^+_2)$ decays 
are listed from the second to fourth columns 
of Table~\ref{tab:nme}. 
The experimental data \cite{barabash2020} included 
both in the figure and table are those extracted from the 
measured half-lives with the phase-space factors 
$G_{2\nu}$ taken from Ref.~\cite{kotila2012}, 
and are referred to as ``Recommended Value'' in Table~3 
of Ref.~\cite{barabash2020}.

As is evident from Fig.~\ref{fig:nme}, the predicted 
$|\mbb|$ values with the bare (unquenched) 
$\ga$ factor are, in most cases, 
substantially larger than the experimental 
values regardless of which pairing 
strength is used in the RHB-SCMF calculations, 
illustrative cases being the 
$^{100}$Mo$\to^{100}$Ru, $^{116}$Cd$\to^{116}$Sn, 
and $^{150}$Nd$\to^{150}$Sm decays. 
For the $^{48}$Ca$\to^{48}$Ti, $^{76}$Ge$\to^{76}$Se, 
and $^{82}$Se$\to^{82}$Kr decays, in contrast, 
the predicted $|\mbb|$ with the default pairing strength $V$ 
are approximately equal to or even lower than the 
experimental $|\mbbe|$ values.  
A remarkable finding in Fig.~\ref{fig:nme} 
is that, when the increased pairing strength 
is adopted, the $|\mbb|$ values systematically 
become larger. 

In the last two rows of Table~\ref{tab:nme} 
the $0^+_1 \to 0^+_2$ decay $|\mbb|$ values
are also shown for the $^{100}$Mo and $^{150}$Nd, 
for which experimental data \cite{barabash2020} 
are available. 
As compared to the $|\mbb(0^+_1 \to 0^+_1)|$ values, 
which exhibit an increase with the enhanced pairing force, 
one can hardly see a general trend of 
the $|\mbb(0^+_1 \to 0^+_2)|$ values for the $^{100}$Mo 
and $^{150}$Nd decays due to the modification of the 
pairing strength.

\begin{table*}
\caption{\label{tab:nme}
Predicted $|\mbb|$ obtained from the mapped IBM-2 
employing the separable pairing force with the 
reduced ($0.9V$), default ($V$), and increased ($1.15V$) 
strengths. From the second to fourth columns are $|\mbb|$ 
values obtained with the bare $\ga$ factor, and 
from the fifth to seventh columns are those 
with the effective $\gae$ factor 
defined in Eq.~(\ref{eq:gae}). 
The effective $|\mbbe|$ extracted from the experimental 
$\tnbb$ half-lives \cite{barabash2020} 
are shown in the eighth column. 
}
 \begin{center}
 \begin{ruledtabular}
  \begin{tabular}{lccccccc}
\multirow{2}{*}{Decay}
&\multicolumn{3}{c}{$|\mbb|$ with $\ga$}
&\multicolumn{3}{c}{$|\mbb|$ with $\gae$}&
\multirow{2}{*}{$|\mbbe|$}\\
\cline{2-4}\cline{5-7}
&{$0.9V$}&{$V$}&{$1.15V$}
&{$0.9V$}&{$V$}&{$1.15V$}&\\
\hline
$^{48}$Ca$\to^{48}$Ti & 0.068 & 0.033 & 0.142 & 0.067 & 0.030 & 0.069 & $0.035\pm0.003$ \\ 
$^{76}$Ge$\to^{76}$Se & 0.038 & 0.060 & 0.265 & 0.027 & 0.032 & 0.083 & $0.106\pm0.004$ \\ 
$^{82}$Se$\to^{82}$Kr & 0.038 & 0.095 & 0.256 & 0.025 & 0.046 & 0.072 & $0.085\pm0.001$ \\ 
$^{96}$Zr$\to^{96}$Mo & 0.140 & 0.281 & 0.256 & 0.078 & 0.107 & 0.058 & $0.088\pm0.004$ \\ 
$^{100}$Mo$\to^{100}$Ru & 0.754 & 0.778 & 0.925 & 0.401 & 0.275 & 0.197 & $0.185\pm0.002$ \\ 
$^{116}$Cd$\to^{116}$Sn & 0.361 & 0.442 & 0.543 & 0.159 & 0.117 & 0.089 & $0.108\pm0.003$ \\ 
$^{128}$Te$\to^{128}$Xe & 0.063 & 0.169 & 0.159 & 0.024 & 0.036 & 0.022 & $0.043\pm0.003$ \\ 
$^{130}$Te$\to^{130}$Xe & 0.020 & 0.083 & 0.265 & 0.007 & 0.017 & 0.035 & $0.0293\pm0.0009$ \\ 
$^{136}$Xe$\to^{136}$Ba & 0.141 & 0.194 & 0.472 & 0.049 & 0.036 & 0.056 & $0.0181\pm0.0006$ \\ 
$^{150}$Nd$\to^{150}$Sm & 0.482 & 0.594 & 0.808 & 0.141 & 0.085 & 0.077 & $0.055\pm0.003$ \\ 
$^{100}$Mo$\to^{100}$Ru$(0^+_2)$ & 0.434 & 0.248 & 0.086 & 0.231 & 0.088 & 0.018 & $0.151\pm0.004$ \\ 
$^{150}$Nd$\to^{150}$Sm$(0^+_2)$ & 0.153 & 0.333 & 0.251 & 0.045 & 0.048 & 0.024 & $0.044\pm0.005$ \\
  \end{tabular}
 \end{ruledtabular}
 \end{center}
\end{table*}

To make a more reasonable comparison with experiment, 
effective $\ga$ factors, denoted as $\gae$, are considered. 
As in the previous mapped IBM-2 
study \cite{nomura2022bb}, 
while both $\gv$ and ratio $\gv/\ga$ in Eq.~(\ref{eq:mbb}) 
remain unchanged, only the $\ga$ factor is modified 
in such a way that 
\begin{eqnarray}
\label{eq:mbbe}
 \mbb \to \mbbe = \left(\frac{\gae}{\ga}\right)^2 \mbb \; .
\end{eqnarray}
The quenching factor $q$ is also extracted from 
the above relation, $q=\gae/\ga$. 
The $\gae$ is here assumed to be a smooth function 
of the mass number $A$, and the following functional 
form was shown \cite{nomura2022bb} 
to give an overall good description 
of the experimental NMEs: 
\begin{eqnarray}
\label{eq:gae}
 \gae = ce^{-dA} \; ,
\end{eqnarray}
with $c$ and $d$ being numerical constants that 
are fitted to the experimental $\mbbe$ 
values \cite{barabash2020}. 
Note the constant $c=\ga$ in Ref.~\cite{nomura2022bb}.

Figure~\ref{fig:ga} exhibits those $\gae$ values 
(shown as solid symbols connected by thin solid lines) 
that would be required so that 
the calculated $\mbbe$'s agree with the data. 
The $\gae$ values corresponding to the pairing 
strengths of $0.9V$ and $V$ 
appear to be significantly at variance with the 
different decay processes, and 
those for $^{76}$Ge, $^{82}$Se, and $^{130}$Te decays 
are particularly large, being close to or 
much larger than the bare value, $\ga=1.269$, 
represented by the horizontal solid line in the 
figure. 
On the other hand, the $\gae$ values that are expected 
for the calculated $\mbb$ result with 
the increased pairing strength $1.15V$ change smoothly 
as functions of the mass $A$.

The function (\ref{eq:gae}) is then fitted 
to those effective $\ga$ values extracted from 
the data for each nucleus, giving rise to 
the numerical constants $c$ and $d$ as 
$(c,d)=(1.69,0.006)$, $(1.86,0.009)$, and $(1.30,0.008)$ 
for the calculations with the pairing strengths of 
$0.9V$, $V$, and $1.15V$, respectively. 
Note that the free nucleon value $\ga=1.269$ at $A=1$ 
is included in the fit. 
The fitted $\gae$'s (depicted as the thick lines 
in Fig.~\ref{fig:ga}) do exhibit smoothly 
decreasing systematic as functions of $A$ 
for all the three pairing strengths, 
with the corresponding values for the 
masses $A=48$ to 150 changing within the 
ranges $\gae= 1.26-0.69$ (with $0.9V$), 
$1.20-0.48$ (with $V$), and 
$0.89-0.39$ (with $1.15V$). 
The quenching factors $q$ estimated 
for the masses $A=48$ to 150 also decrease monotonously 
in the intervals $q = 1.00-0.54$, $0.95-0.38$, and 
$0.70-0.31$, for the calculations with the 
reduced ($0.9V$), default ($V$), 
and increased ($1.15V$) pairing strengths, 
respectively. 
Figure~\ref{fig:nme-q} depicts the resultant $|\mbbe|$ 
values, which are computed by using the effective 
$\gae$ (\ref{eq:gae}) determined by the aforementioned procedure, 
and compare them with the experimental data 
\cite{barabash2020}. 
As one can see, while certain improvements appear to be 
made by using the increased pairing strength $1.15V$, 
particularly 
for the $^{76}$Ge, $^{82}$Se, and $^{100}$Mo decays, 
the calculations with the original  
pairing strength $V$ provide an overall good description 
of the observed $|\mbbe|$ values. 
The calculated results with the reduced pairing 
strength $0.9V$ seem to be, in many cases, rather 
far from the experimental data.

The predicted $|\mbbe|$ with the 
effective $\gae$, which are shown 
in Fig.~\ref{fig:nme-q}, are also listed from the 
fifth to seventh columns of Table~\ref{tab:nme}. 
The table also gives results for the 
$|\mbbe(0^+_1 \to 0^+_2)|$ NMEs for the 
$^{100}$Mo and $^{150}$Nd decays, for which 
the same $\gae$ values as those used for the 
$|\mbbe(0^+_1 \to 0^+_1)|$ ones are employed. 
Smaller $|\mbbe(0^+_1 \to 0^+_1)|$ are obtained 
with the stronger pairing 
interaction, reflecting that the 
larger quenching is expected 
from the systematic of the unquenched $|\mbb|$
calculated with the increased pairing strength. 
Note, however, that the use of the same $\gae$ values for 
the $0^+_1 \to 0^+_1$ and $0^+_1 \to 0^+_2$ decays 
may not be entirely justified, that is, some different 
quenching may need to be made for the decays to the 
different final states.

\subsection{Half-lives\label{sec:tau}}

The calculated $\tnbb$-decay half-lives $\taubb$ 
(\ref{eq:taubb}), with the NMEs $|\mbb|$ given in 
Table~\ref{tab:nme}, are listed in Table~\ref{tab:tau}. 
The experimental data \cite{barabash2020}, 
shown in the eighth column, 
correspond to those that are 
referred to as ``Average (or Recommended) value'' 
in Table~1 of Ref.~\cite{barabash2020}, 
which are based on the measured $\tnbb$ 
$\taubb$ from the 1990s till 2020. 
As in the case of the $|\mbbe|$ results 
in Fig.~\ref{fig:nme-q}, 
the calculations with the default pairing strength $V$ 
provide an overall good description of the 
$\taubb$ data, with $\mbb$ quenched with 
the effective $\gae$ factors of Eq.~(\ref{eq:gae}). 
Increasing the pairing 
strength to $1.15V$ leads to some improvements 
in specific cases of the $^{76}$Ge$\to^{76}$Se, 
$^{82}$Se$\to^{82}$Kr, and 
$^{100}$Mo$\to^{100}$Ru decays. 
As for the $\taubb$'s for the 
$^{136}$Xe$\to^{136}$Ba and 
$^{100}$Mo$\to^{100}$Ru$(0^+_2)$ decays, however, 
enhancing the pairing does not seem to work well, 
as the predicted $\taubb$'s are here 
by about one and two orders of magnitude 
shorter and longer, respectively, 
than the experimental ones. 
It should be worth mentioning more recent 
measurements of the $\tnbb$-decay 
$\taubb$ concerning some of the candidate nuclei: 
a GERDA experiment for the $^{76}$Ge decay 
\cite{agostini2023-76Ge} obtained 
$\taubb=(2.022\pm0.018_{\rm stat}\pm0.038_{\rm syst})\times 10^{21}$ yr, 
CUPID-Mo experiments on the $^{100}$Mo decay 
reported 
$\taubb=[7.07\pm0.02({\rm stat})\pm0.11({\rm syst})]\times 10^{18}$ yr 
for the $\trgg$ transition \cite{augier2023-100Mo-Letter}, 
and $[7.5\pm0.8({\rm stat})^{+0.4}_{-0.3}]\times 10^{20}$ yr 
for the $\trge$ transition \cite{augier2023-100Mo}, 
and a CUORE measurement on $^{130}$Te \cite{adams2021-130Te} provided 
$\taubb=[7.71^{+0.08}_{-0.06}({\rm stat})^{+0.12}_{-0.15}({\rm syst})]\times 10^{20}$ yr. 
All these new entries present crucial updates 
on the $\tnbb$-decay $\taubb$ data with high accuracy, 
and are more or less close to the average values 
of Ref.~\cite{barabash2020} listed in Table~\ref{tab:tau}.

Concerning the $^{100}$Mo and $^{150}$Nd decays, 
ratios of the measured $\taubb$ values 
for the $0^+_1 \to 0^+_1$ to $0^+_1 \to 0^+_2$ 
decays are computed as
\begin{align}
\label{eq:tauratio1}
 \frac{\taubb\left[^{100}\mathrm{Mo}(0^{+}_{1})\to^{100}\mathrm{Ru}(0^{+}_{2})\right]}
{\taubb\left[^{100}\mathrm{Mo}(0^{+}_{1})\to^{100}\mathrm{Ru}(0^{+}_{1})\right]}
= 94.9^{+7.4}_{-5.9} \; ,
\end{align}
while the predicted ratios in the present calculation 
are much larger: 165, 536, and 6279, obtained 
for the pairing strengths of 
$0.9V$, $V$, and $1.15V$, respectively. 
The experimental ratio for the $^{150}$Nd decay, 
\begin{align}
\label{eq:tauratio2}
\frac{\taubb\left[^{150}\mathrm{Nd}(0^{+}_{1})\to^{150}\mathrm{Sm}(0^{+}_{2})\right]}{\taubb\left[^{150}\mathrm{Nd}(0^{+}_{1})\to^{150}\mathrm{Sm}(0^{+}_{1})\right]} = 12.8^{+3.3}_{-2.3} \; ,
\end{align}
is reproduced reasonably well by the present calculation, 
with the predicted values being 
83, 27, and 87, for the pairing strengths of 
$0.9V$, $V$, and $1.15V$, respectively. 
These ratios are independent of the effective $\ga$ 
factors if the same $\gae$ values are used 
in the calculations of the NMEs for the $\trgg$ and 
$\trge$ decays. The description of the 
ratio for the $^{100}$Mo decays (\ref{eq:tauratio1}) 
could be improved if different $\gae$ values 
are considered between the $\trgg$ and 
$\trge$ decays.

%
%
\begin{table*}
\caption{\label{tab:tau}
Calculated $\tnbb$-decay 
$\taubb$'s (\ref{eq:taubb}) (in year) 
with the unquenched $\ga$ factor 
(from the second to fourth columns), and 
with the effective $\gae$ factors defined in Eq.~(\ref{eq:gae}) 
(from the fifth to seventh columns) obtained from the 
mapped IBM-2 with the different pairing strengths 
$0.9V$, $V$, and $1.15V$. 
The experimental $\taubb$ 
\cite{barabash2020} are included in the 
eighth column. 
}
 \begin{center}
 \begin{ruledtabular}
  \begin{tabular}{lccccccc}
\multirow{2}{*}{Decay}
&\multicolumn{3}{c}{$\tau_{1/2}^{(2\nu)}$ (yr), with $\ga$}
&\multicolumn{3}{c}{$\tau_{1/2}^{(2\nu)}$ (yr), with $\gae$}
&\multirow{2}{*}{Experiment}\\
\cline{2-4}\cline{5-7}
& $0.9V$ & $V$ & $1.15V$ & $0.9V$ & $V$ & $1.15V$ & \\
\hline
$^{48}$Ca$\to^{48}$Ti & $1.41\times 10^{19}$ & $5.79\times 10^{19}$ & $3.20\times 10^{18}$ & $1.42\times 10^{19}$ & $7.14\times 10^{19}$ & $1.34\times 10^{19}$ & $5.3^{+1.2}_{-0.8}\times10^{19}$ \\ 
$^{76}$Ge$\to^{76}$Se & $1.42\times 10^{22}$ & $5.85\times 10^{21}$ & $2.97\times 10^{20}$ & $2.82\times 10^{22}$ & $1.98\times 10^{22}$ & $3.05\times 10^{21}$ & $(1.88\pm0.08)\times10^{21}$ \\ 
$^{82}$Se$\to^{82}$Kr & $4.34\times 10^{20}$ & $6.97\times 10^{19}$ & $9.59\times 10^{18}$ & $9.95\times 10^{20}$ & $2.92\times 10^{20}$ & $1.19\times 10^{20}$ & $(0.87^{+0.02}_{-0.01})\times10^{20}$ \\ 
$^{96}$Zr$\to^{96}$Mo & $7.50\times 10^{18}$ & $1.86\times 10^{18}$ & $2.23\times 10^{18}$ & $2.40\times 10^{19}$ & $1.29\times 10^{19}$ & $4.35\times 10^{19}$ & $(2.3\pm0.2)\times10^{19}$ \\ 
$^{100}$Mo$\to^{100}$Ru & $5.32\times 10^{17}$ & $5.00\times 10^{17}$ & $3.53\times 10^{17}$ & $1.88\times 10^{18}$ & $4.01\times 10^{18}$ & $7.82\times 10^{18}$ & $(7.06^{+0.15}_{-0.13})\times10^{18}$ \\ 
$^{116}$Cd$\to^{116}$Sn & $2.77\times 10^{18}$ & $1.85\times 10^{18}$ & $1.23\times 10^{18}$ & $1.44\times 10^{19}$ & $2.64\times 10^{19}$ & $4.53\times 10^{19}$ & $(2.69\pm0.09)\times10^{19}$ \\ 
$^{128}$Te$\to^{128}$Xe & $9.28\times 10^{23}$ & $1.31\times 10^{23}$ & $1.48\times 10^{23}$ & $6.41\times 10^{24}$ & $2.87\times 10^{24}$ & $8.03\times 10^{24}$ & $(2.25\pm0.09)\times10^{24}$ \\ 
$^{130}$Te$\to^{130}$Xe & $1.67\times 10^{21}$ & $9.48\times 10^{19}$ & $9.29\times 10^{18}$ & $1.21\times 10^{22}$ & $2.24\times 10^{21}$ & $5.37\times 10^{20}$ & $(7.91\pm0.21)\times10^{20}$ \\ 
$^{136}$Xe$\to^{136}$Ba & $3.49\times 10^{19}$ & $1.86\times 10^{19}$ & $3.13\times 10^{18}$ & $2.93\times 10^{20}$ & $5.44\times 10^{20}$ & $2.20\times 10^{20}$ & $(2.18\pm0.05)\times10^{21}$ \\ 
$^{150}$Nd$\to^{150}$Sm & $1.18\times 10^{17}$ & $7.78\times 10^{16}$ & $4.21\times 10^{16}$ & $1.39\times 10^{18}$ & $3.77\times 10^{18}$ & $4.61\times 10^{18}$ & $(9.34\pm0.65)\times10^{18}$ \\ 
$^{100}$Mo$\to^{100}$Ru$(0^+_2)$ & $8.78\times 10^{19}$ & $2.68\times 10^{20}$ & $2.22\times 10^{21}$ & $3.10\times 10^{20}$ & $2.15\times 10^{21}$ & $4.91\times 10^{22}$ & $6.7^{+0.5}_{-0.4}\times10^{20}$ \\ 
$^{150}$Nd$\to^{150}$Sm$(0^+_2)$ & $9.84\times 10^{18}$ & $2.08\times 10^{18}$ & $3.66\times 10^{18}$ & $1.15\times 10^{20}$ & $1.01\times 10^{20}$ & $4.01\times 10^{20}$ & $1.2^{+0.3}_{-0.2}\times10^{20}$ \\ 
  \end{tabular}
 \end{ruledtabular}
 \end{center}
\end{table*}

\section{Concluding remarks\label{sec:summary}}

The low-energy nuclear structure and $\tnbb$-decay NMEs 
without the closure approximation 
have been investigated 
within the mapped IBM-2 that is 
based on the SCMF calculation employing the 
relativistic EDF DD-PC1 
and separable pairing force of finite range. 
The IBM-2 Hamiltonian describing the initial 
and final even-even nuclei has been completely determined 
by mapping the RHB-SCMF PES onto the bosonic counterpart. 
The particle-boson and particle-particle 
interactions in the IBFFM-2 Hamiltonian 
used to compute intermediate states 
of the neighboring odd-odd nuclei have also been 
determined by using the results 
of the RHB-SCMF calculations. 
In the present analysis, the effects of changing 
the pairing strength in the RHB calculations on 
the spectroscopic properties of low-lying states 
and $\tnbb$-decay NMEs have been specifically studied. 

When the increased pairing with respect to the 
default one by 15 \% is chosen, 
the SCMF PESs for the candidate even-even nuclei 
have been shown to be substantially 
softer in both the axial $\beta$ 
and triaxial $\gamma$ deformations, and the 
potential valley becomes less pronounced. 
With the pairing strength 
reduced by 10 \% 
from the default one, on the other hand, 
a PES with a much more pronounced potential valley 
has been obtained which is steep in the 
$\beta$ and $\gamma$ directions. 
The derived strength parameters for the pairing-like 
term ($\hat n_d$) and quadrupole-quadrupole interaction 
($\hat Q_{\nu} \cdot \hat Q_{\pi}$) 
in the IBM-2 have been shown to be significantly 
at variance with the different IBM-2 calculations 
using the reduced, default, and increased 
pairing strengths in the RHB-SCMF calculations.

The calculated energy spectra for the 
even-even nuclei indicated that with the increased 
pairing strength $1.15V$, the energy levels for the 
non-yrast states $0^+_2$ and $2^+_2$ are 
generally lowered, being in better agreement 
with experiment than in the cases of the 
weaker pairing strength $0.9V$ and $V$. 
An even more accurate description of 
the excited $0^+_2$ states in the even-even 
nuclei and its influence on the 
$\db$-decay NMEs would be better investigated 
within the version of the IBM-2 that 
incorporates the 
configuration mixing between the normal 
and intruder states. 
The energy levels for the intermediate odd-odd nuclei 
have been shown to be more or less sensitive 
to the choice of the pairing strength. 
The electromagnetic transition properties for both 
the even-even and odd-odd nuclear systems 
have exhibited certain sensitivities to the pairing strengths. 
A notable consequence is that the calculation with the 
default pairing strength has given the 
best agreement with the experimental 
transition properties for many of the 
odd-odd nuclei.

An effect of modifying the pairing strength 
in the RHB-SCMF calculation on $\tnbb$ decay is that 
with the increased strength, resultant $\tnbb$-decay 
NMEs, $|\mbb|$, become systematically larger. 
A quenching of the NMEs with the effective $\gae$ 
factor that is a smooth function of the mass $A$ 
has been introduced as in Ref.~\cite{nomura2022bb}. 
In the calculation employing 
the increased pairing strength $1.15V$, 
the effective NMEs turned out be in a fairly 
good agreement with the experimental $|\mbbe|$ values 
for the $^{76}$Ge, $^{82}$Se, and $^{100}$Mo 
decays in particular. 
In many other decays, however, the calculated results 
with the default $V$ pairing strength 
have been shown to be adequate to provide 
an overall good agreement with the data.

Of several assumptions, and approximations 
introduced in the employed theoretical approach, 
the uncertainties in the $\tnbb$ NME predictions 
could arise, to a larger extent, 
from the SCMF models and properties of the employed EDF, 
which underlie the mapped IBM-2 study. 
The present study 
indicates that the strength of the pairing interaction 
is considered among the most important 
parameters that may affect both low-lying states and 
decay processes. 
In the meantime, 
it remains an open question to investigate 
thoroughly how relevant other building blocks 
involved in the model are in the predictions of the 
NMEs, as well as the low-lying structures, 
such as those related to the parametrizations of the EDF, 
to the single-particle properties, to other 
missing correlations at the SCMF level, and to 
the forms of the IBM-2 and IBFFM-2 Hamiltonians.

\appendix

\section{Parameters for the IBFFM-2 Hamiltonian\label{sec:paraoo}}

The strength parameters adopted for the 
boson-fermion interactions $\hbf^{\nu}$ and $\hbf^{\pi}$, 
and the residual neutron-proton 
interaction $\hff$ in the IBFFM-2 Hamiltonian 
in the three cases of the separable 
pairing strength in the RHB-SCMF calculations are 
listed in Tables~\ref{tab:parabfn}, \ref{tab:parabfp}, 
and \ref{tab:paraff}, respectively. 
Some updates have been made 
to the adopted IBFFM-2 parameters since 
the previous study of Ref.~\cite{nomura2022bb} 
concerning the parameters $\Lambda_\nu$, $A_\nu$, 
for both parities, and the tensor strength $\vt$, 
for the nucleus $^{76}$As, when the 
default pairing strength $V$ is used. 
New parameters are here also employed for 
some other nuclei: 
$\vt$ for $^{82}$Br, boson-fermion interactions 
for $^{136}$Cs and $^{150}$Pm. 
For details, compare entries in Table~\ref{tab:parabfn}, 
\ref{tab:parabfp}, and \ref{tab:parabfp}, 
with those in Table XVI of Ref.~\cite{nomura2022bb}.

\begin{table*}
\caption{\label{tab:parabfn}
Strength parameters for the 
boson-fermion interaction $\hbf^\nu$ (\ref{eq:hbf}) 
for the 
odd-odd nuclei obtained for the single-particle 
spaces corresponding to the positive- 
and negative-parity states in the cases of 
the reduced ($0.9V$), 
default ($V$), and increased ($1.15V$) 
pairing strengths in the RHB-SCMF calculations. 
Note that the single-neutron space for $^{48}$Sc 
does not include orbitals of positive parity, 
so the strength parameters 
corresponding to positive parity are taken 
to be the same as those 
determined for the negative-parity 
($2p_{1/2,3/2},1f_{5/2,7/2}$) configuration. 
}
 \begin{center}
 \begin{ruledtabular}
  \begin{tabular}{llccccccccc}
\multirow{2}{*}{Nucleus} &
\multirow{2}{*}{Single-particle space} &
\multicolumn{3}{c}{$\Gamma_\nu$ (MeV)} &
\multicolumn{3}{c}{$\Lambda_\nu$ (MeV)} &
\multicolumn{3}{c}{$A_\nu$ (MeV)} \\
\cline{3-5}
\cline{6-8}
\cline{9-11}
 & & 
$0.9V$  & $V$ & $1.15V$ &
$0.9V$  & $V$ & $1.15V$ &
$0.9V$  & $V$ & $1.15V$ \\
\hline
$^{48}$Sc & $2p_{1/2,3/2},1f_{5/2,7/2}$ & $0.30$ & $0.30$ & $0.30$ & $1.00$ & $1.00$ & $1.00$ & ${}$ & ${}$ & ${}$ \\ 
[1.0ex]
\multirow{2}{*}{$^{76}$As} & $1g_{9/2}$ & $0.30$ & $0.30$ & $0.60$ & $1.20$ & $1.00$ & $1.00$ & $-0.60$ & $-0.50$ & $-0.50$ \\ 
 & $2p_{1/2,3/2},1f_{5/2}$ & $0.30$ & $0.30$ & $0.60$ & $0.80$ & $0.80$ & $1.00$ & $-0.50$ & $-0.50$ & $-0.40$ \\ 
[1.0ex]
\multirow{2}{*}{$^{82}$Br} & $1g_{9/2}$ & $0.30$ & $0.30$ & $0.30$ & $2.10$ & $2.10$ & $1.70$ & ${}$ & ${}$ & ${}$ \\ 
 & $2p_{1/2,3/2},1f_{5/2}$ & $0.30$ & $0.30$ & $0.30$ & $0.80$ & $0.80$ & $0.80$ & ${}$ & ${}$ & ${}$ \\ 
[1.0ex]
\multirow{2}{*}{$^{96}$Nb} & $3s_{1/2},2d_{3/2,5/2},1g_{7/2}$ & $0.30$ & $0.30$ & $0.30$ & $0.40$ & $0.40$ & $0.40$ & ${}$ & ${}$ & ${}$ \\ 
 & $1h_{11/2}$ & $0.30$ & $0.30$ & $0.30$ & ${}$ & ${}$ & ${}$ & $-1.50$ & $-1.50$ & $-0.80$ \\ 
[1.0ex]
\multirow{2}{*}{$^{100}$Tc} & $3s_{1/2},2d_{3/2,5/2},1g_{7/2}$ & $0.30$ & $0.30$ & $0.30$ & $0.35$ & $0.35$ & $0.35$ & ${}$ & ${}$ & ${}$ \\ 
 & $1h_{11/2}$ & $0.30$ & $0.30$ & $0.30$ & ${}$ & ${}$ & ${}$ & ${}$ & ${}$ & ${}$ \\ 
[1.0ex]
\multirow{2}{*}{$^{116}$In} & $3s_{1/2},2d_{3/2,5/2},1g_{7/2}$ & $0.30$ & $0.30$ & $0.30$ & $0.20$ & $0.20$ & $0.20$ & $-0.15$ & $-0.15$ & $-0.15$ \\ 
 & $1h_{11/2}$ & $0.30$ & $0.30$ & $0.30$ & $0.20$ & $0.20$ & $0.20$ & $-0.15$ & $-0.15$ & $-0.15$ \\ 
[1.0ex]
\multirow{2}{*}{$^{128}$I} & $3s_{1/2},2d_{3/2,5/2},1g_{7/2}$ & $0.30$ & $0.30$ & $0.30$ & $6.50$ & $6.50$ & $6.50$ & ${}$ & ${}$ & ${}$ \\ 
 & $1h_{11/2}$ & $0.30$ & $0.30$ & $0.30$ & $0.90$ & $0.90$ & $0.90$ & $-0.20$ & $-0.20$ & $-0.20$ \\ 
[1.0ex]
\multirow{2}{*}{$^{130}$I} & $3s_{1/2},2d_{3/2,5/2},1g_{7/2}$ & $0.30$ & $0.30$ & $0.30$ & $7.60$ & $7.60$ & $5.00$ & ${}$ & ${}$ & ${}$ \\ 
 & $1h_{11/2}$ & $0.30$ & $0.30$ & $0.30$ & $0.90$ & $0.90$ & $0.90$ & $-0.50$ & $-0.50$ & $-0.50$ \\ 
[1.0ex]
\multirow{2}{*}{$^{136}$Cs} & $3s_{1/2},2d_{3/2,5/2},1g_{7/2}$ & $0.30$ & $0.30$ & $0.30$ & $0.20$ & $0.20$ & $0.20$ & $-0.15$ & $-0.15$ & $-0.15$ \\ 
 & $1h_{11/2}$ & $0.30$ & $0.30$ & $0.30$ & $0.20$ & $0.20$ & $0.20$ & $-0.15$ & $-0.15$ & $-0.15$ \\ 
[1.0ex]
\multirow{2}{*}{$^{150}$Pm} & $1i_{13/2}$ & $0.30$ & $0.30$ & $0.30$ & $16.00$ & $10.00$ & $6.00$ & ${}$ & $-0.80$ & $-0.80$ \\ 
 & $3p_{1/2,3/2},2f_{5/2,7/2},1h_{9/2}$ & $0.30$ & $0.30$ & $0.30$ & $0.50$ & $0.60$ & $0.50$ & $-0.80$ & ${}$ & $-0.80$ \\
  \end{tabular}
 \end{ruledtabular}
 \end{center}
\end{table*}

\begin{table*}
\caption{\label{tab:parabfp}
Same as Table~\ref{tab:parabfn}, but for $\hbf^\pi$.
}
 \begin{center}
 \begin{ruledtabular}
  \begin{tabular}{llccccccccc}
\multirow{2}{*}{Nucleus} &
\multirow{2}{*}{Single-particle space} &
\multicolumn{3}{c}{$\Gamma_\pi$ (MeV)} &
\multicolumn{3}{c}{$\Lambda_\pi$ (MeV)} &
\multicolumn{3}{c}{$A_\pi$ (MeV)} \\
\cline{3-5}
\cline{6-8}
\cline{9-11}
 & & 
$0.9V$  & $V$ & $1.15V$ &
$0.9V$  & $V$ & $1.15V$ &
$0.9V$  & $V$ & $1.15V$ \\
\hline
$^{48}$Sc & $2p_{1/2,3/2},1f_{5/2,7/2}$ & $0.30$ & $0.30$ & $0.30$ & ${}$ & ${}$ & ${}$ & ${}$ & ${}$ & ${}$ \\ 
[1.0ex]
\multirow{2}{*}{$^{76}$As} & $1g_{9/2}$ & $1.00$ & $1.00$ & $1.00$ & ${}$ & ${}$ & ${}$ & ${}$ & ${}$ & ${}$ \\ 
 & $2p_{1/2,3/2},1f_{5/2}$ & $0.30$ & $0.30$ & $0.60$ & $1.60$ & $1.60$ & $0.35$ & $-0.80$ & $-0.80$ & ${}$ \\ 
[1.0ex]
\multirow{2}{*}{$^{82}$Br} & $1g_{9/2}$ & $2.50$ & $2.50$ & $2.00$ & ${}$ & ${}$ & $1.30$ & ${}$ & ${}$ & ${}$ \\ 
 & $2p_{1/2,3/2},1f_{5/2}$ & $0.30$ & $0.30$ & $0.30$ & $0.80$ & $0.80$ & $0.80$ & ${}$ & ${}$ & ${}$ \\ 
[1.0ex]
\multirow{2}{*}{$^{96}$Nb} & $1g_{9/2}$ & $0.30$ & $0.30$ & $0.30$ & $0.90$ & $0.90$ & $0.90$ & $-0.50$ & $-0.50$ & ${}$ \\ 
 & $2p_{1/2,3/2},1f_{5/2}$ & $0.30$ & $0.30$ & $0.30$ & $1.60$ & $1.60$ & $1.60$ & $-0.30$ & $-0.30$ & $-0.30$ \\ 
[1.0ex]
\multirow{2}{*}{$^{100}$Tc} & $1g_{9/2}$ & $0.30$ & $0.30$ & $0.30$ & $0.90$ & $0.90$ & $0.60$ & ${}$ & ${}$ & ${}$ \\ 
 & $2p_{1/2,3/2},1f_{5/2}$ & $0.30$ & $0.30$ & $0.30$ & $5.00$ & $5.00$ & $5.00$ & $-1.00$ & $-1.00$ & $-1.00$ \\ 
[1.0ex]
\multirow{2}{*}{$^{116}$In} & $1g_{9/2}$ & $0.30$ & $0.30$ & $0.30$ & ${}$ & ${}$ & ${}$ & ${}$ & ${}$ & ${}$ \\ 
 & $2p_{1/2,3/2},1f_{5/2}$ & $1.00$ & $1.00$ & $1.00$ & ${}$ & ${}$ & ${}$ & ${}$ & ${}$ & ${}$ \\ 
[1.0ex]
\multirow{2}{*}{$^{128}$I} & $3s_{1/2},2d_{3/2,5/2},1g_{7/2}$ & $0.30$ & $0.30$ & $0.30$ & $0.60$ & $0.60$ & $0.60$ & $-1.00$ & $-1.00$ & $-1.00$ \\ 
 & $1h_{11/2}$ & $0.30$ & $0.30$ & $0.30$ & ${}$ & ${}$ & ${}$ & $-1.05$ & $-1.05$ & $-1.05$ \\ 
[1.0ex]
\multirow{2}{*}{$^{130}$I} & $3s_{1/2},2d_{3/2,5/2},1g_{7/2}$ & $0.30$ & $0.30$ & $0.30$ & $0.80$ & $0.80$ & $0.80$ & $-0.75$ & $-0.75$ & $-0.50$ \\ 
 & $1h_{11/2}$ & $0.30$ & $0.30$ & $0.30$ & ${}$ & ${}$ & ${}$ & $-1.05$ & $-1.05$ & $-1.05$ \\ 
[1.0ex]
\multirow{2}{*}{$^{136}$Cs} & $3s_{1/2},2d_{3/2,5/2},1g_{7/2}$ & $0.30$ & $0.30$ & $0.30$ & ${}$ & ${}$ & ${}$ & ${}$ & ${}$ & ${}$ \\ 
 & $1h_{11/2}$ & $1.00$ & $1.00$ & $1.00$ & ${}$ & ${}$ & ${}$ & ${}$ & ${}$ & ${}$ \\ 
[1.0ex]
\multirow{2}{*}{$^{150}$Pm} & $3s_{1/2},2d_{3/2,5/2},1g_{7/2}$ & $0.30$ & $0.30$ & $0.30$ & $0.40$ & $0.40$ & $0.40$ & $-0.70$ & $-1.00$ & $-0.70$ \\ 
 & $1h_{9/2,11/2}$ & $1.00$ & $1.00$ & $1.00$ & $2.80$ & $3.00$ & $2.80$ & ${}$ & ${}$ & ${}$ \\
  \end{tabular}
 \end{ruledtabular}
 \end{center}
\end{table*}

\begin{table}
\caption{\label{tab:paraff}
Strength parameters used for the residual 
neutron-proton interaction $\hff$ (\ref{eq:hff}) 
in the IBFFM-2 
Hamiltonian for the intermediate odd-odd nuclei 
for the calculations with the three different 
pairing strengths. Note that the spin-spin interaction 
strength $\vsss=0$ MeV for all the nuclei.}
 \begin{center}
 \begin{ruledtabular}
  \begin{tabular}{lcccccccc}
Nucleus & Pairing & 
$\vd$ (MeV) & $\vssd$ (MeV) & $\vt$ (MeV) \\
\hline
\multirow{3}{*}{$^{48}$Sc} & $0.9V$ & $0.60$ & ${}$ & ${}$ \\ 
 & $V$ & $0.60$ & ${}$ & ${}$ \\ 
 & $1.15V$ & $0.60$ & ${}$ & $-0.02$ \\ 
[1.0ex]
\multirow{3}{*}{$^{76}$As} & $0.9V$ & $0.80$ & ${}$ & $0.15$ \\ 
 & $V$ & $0.80$ & ${}$ & $0.15$ \\ 
 & $1.15V$ & $0.80$ & ${}$ & $0.02$ \\ 
[1.0ex]
\multirow{3}{*}{$^{82}$Br} & $0.9V$ & ${}$ & $-0.23$ & $0.10$ \\ 
 & $V$ & ${}$ & $-0.23$ & $0.10$ \\ 
 & $1.15V$ & ${}$ & $-0.23$ & $0.10$ \\ 
[1.0ex]
\multirow{3}{*}{$^{96}$Nb} & $0.9V$ & $0.80$ & $0.25$ & ${}$ \\ 
 & $V$ & $0.80$ & ${}$ & ${}$ \\ 
 & $1.15V$ & $0.40$ & $0.10$ & ${}$ \\ 
[1.0ex]
\multirow{3}{*}{$^{100}$Tc} & $0.9V$ & $-0.08$ & ${}$ & $0.12$ \\ 
 & $V$ & $-0.08$ & ${}$ & $0.05$ \\ 
 & $1.15V$ & $-0.08$ & ${}$ & $0.20$ \\ 
[1.0ex]
\multirow{3}{*}{$^{116}$In} & $0.9V$ & $-0.80$ & ${}$ & $0.40$ \\ 
 & $V$ & $-0.80$ & ${}$ & $0.40$ \\ 
 & $1.15V$ & $-0.80$ & ${}$ & $0.43$ \\ 
[1.0ex]
\multirow{3}{*}{$^{128}$I} & $0.9V$ & ${}$ & $-0.05$ & ${}$ \\ 
 & $V$ & ${}$ & $-0.51$ & ${}$ \\ 
 & $1.15V$ & ${}$ & $-0.30$ & ${}$ \\ 
[1.0ex]
\multirow{3}{*}{$^{130}$I} & $0.9V$ & $-0.02$ & ${}$ & $0.01$ \\ 
 & $V$ & $-0.08$ & ${}$ & $0.01$ \\ 
 & $1.15V$ & $0.01$ & ${}$ & $0.01$ \\ 
[1.0ex]
\multirow{3}{*}{$^{136}$Cs} & $0.9V$ & $-0.08$ & ${}$ & $0.15$ \\ 
 & $V$ & $-0.08$ & ${}$ & $0.09$ \\ 
 & $1.15V$ & $-0.08$ & ${}$ & $0.04$ \\ 
[1.0ex]
\multirow{3}{*}{$^{150}$Pm} & $0.9V$ & $-0.08$ & ${}$ & $0.20$ \\ 
 & $V$ & $-0.08$ & ${}$ & $0.14$ \\ 
 & $1.15V$ & $-0.08$ & ${}$ & $0.18$ \\ 
  \end{tabular}
 \end{ruledtabular}
 \end{center}
\end{table}

\bibliography{refs}

\begin{thebibliography}{71}%
\makeatletter
\providecommand \@ifxundefined [1]{%
 \@ifx{#1\undefined}
}%
\providecommand \@ifnum [1]{%
 \ifnum #1\expandafter \@firstoftwo
 \else \expandafter \@secondoftwo
 \fi
}%
\providecommand \@ifx [1]{%
 \ifx #1\expandafter \@firstoftwo
 \else \expandafter \@secondoftwo
 \fi
}%
\providecommand \natexlab [1]{#1}%
\providecommand \enquote  [1]{``#1''}%
\providecommand \bibnamefont  [1]{#1}%
\providecommand \bibfnamefont [1]{#1}%
\providecommand \citenamefont [1]{#1}%
\providecommand \href@noop [0]{\@secondoftwo}%
\providecommand \href [0]{\begingroup \@sanitize@url \@href}%
\providecommand \@href[1]{\@@startlink{#1}\@@href}%
\providecommand \@@href[1]{\endgroup#1\@@endlink}%
\providecommand \@sanitize@url [0]{\catcode `\\12\catcode `\$12\catcode
  `\&12\catcode `\#12\catcode `\^12\catcode `\_12\catcode `\%12\relax}%
\providecommand \@@startlink[1]{}%
\providecommand \@@endlink[0]{}%
\providecommand \url  [0]{\begingroup\@sanitize@url \@url }%
\providecommand \@url [1]{\endgroup\@href {#1}{\urlprefix }}%
\providecommand \urlprefix  [0]{URL }%
\providecommand \Eprint [0]{\href }%
\providecommand \doibase [0]{https://doi.org/}%
\providecommand \selectlanguage [0]{\@gobble}%
\providecommand \bibinfo  [0]{\@secondoftwo}%
\providecommand \bibfield  [0]{\@secondoftwo}%
\providecommand \translation [1]{[#1]}%
\providecommand \BibitemOpen [0]{}%
\providecommand \bibitemStop [0]{}%
\providecommand \bibitemNoStop [0]{.\EOS\space}%
\providecommand \EOS [0]{\spacefactor3000\relax}%
\providecommand \BibitemShut  [1]{\csname bibitem#1\endcsname}%
\let\auto@bib@innerbib\@empty
\bibitem [{\citenamefont {Primakoff}\ and\ \citenamefont
  {Rosen}(1959)}]{primakoff1959}%
  \BibitemOpen
  \bibfield  {author} {\bibinfo {author} {\bibfnamefont {H.}~\bibnamefont
  {Primakoff}}\ and\ \bibinfo {author} {\bibfnamefont {S.~P.}\ \bibnamefont
  {Rosen}},\ }\href {https://doi.org/10.1088/0034-4885/22/1/305} {\bibfield
  {journal} {\bibinfo  {journal} {Rep. Prog. Phys.}\ }\textbf {\bibinfo
  {volume} {22}},\ \bibinfo {pages} {121} (\bibinfo {year} {1959})}\BibitemShut
  {NoStop}%
\bibitem [{\citenamefont {Haxton}\ and\ \citenamefont
  {Stephenson}(1984)}]{haxton1984}%
  \BibitemOpen
  \bibfield  {author} {\bibinfo {author} {\bibfnamefont {W.}~\bibnamefont
  {Haxton}}\ and\ \bibinfo {author} {\bibfnamefont {G.}~\bibnamefont
  {Stephenson}},\ }\href
  {https://doi.org/https://doi.org/10.1016/0146-6410(84)90006-1} {\bibfield
  {journal} {\bibinfo  {journal} {Prog. Part. Nucl. Phys.}\ }\textbf {\bibinfo
  {volume} {12}},\ \bibinfo {pages} {409} (\bibinfo {year} {1984})}\BibitemShut
  {NoStop}%
\bibitem [{\citenamefont {Doi}\ \emph {et~al.}(1985)\citenamefont {Doi},
  \citenamefont {Kotani},\ and\ \citenamefont {Takasugi}}]{doi1985}%
  \BibitemOpen
  \bibfield  {author} {\bibinfo {author} {\bibfnamefont {M.}~\bibnamefont
  {Doi}}, \bibinfo {author} {\bibfnamefont {T.}~\bibnamefont {Kotani}},\ and\
  \bibinfo {author} {\bibfnamefont {E.}~\bibnamefont {Takasugi}},\ }\href
  {https://doi.org/10.1143/PTPS.83.1} {\bibfield  {journal} {\bibinfo
  {journal} {Prog. Theor. Phys. Suppl.}\ }\textbf {\bibinfo {volume} {83}},\
  \bibinfo {pages} {1} (\bibinfo {year} {1985})}\BibitemShut {NoStop}%
\bibitem [{\citenamefont {Tomoda}(1991)}]{tomoda1991}%
  \BibitemOpen
  \bibfield  {author} {\bibinfo {author} {\bibfnamefont {T.}~\bibnamefont
  {Tomoda}},\ }\href {https://doi.org/10.1088/0034-4885/54/1/002} {\bibfield
  {journal} {\bibinfo  {journal} {Rep. Prog. Phys.}\ }\textbf {\bibinfo
  {volume} {54}},\ \bibinfo {pages} {53} (\bibinfo {year} {1991})}\BibitemShut
  {NoStop}%
\bibitem [{\citenamefont {Suhonen}\ and\ \citenamefont
  {Civitarese}(1998)}]{suhonen1998}%
  \BibitemOpen
  \bibfield  {author} {\bibinfo {author} {\bibfnamefont {J.}~\bibnamefont
  {Suhonen}}\ and\ \bibinfo {author} {\bibfnamefont {O.}~\bibnamefont
  {Civitarese}},\ }\href
  {https://doi.org/https://doi.org/10.1016/S0370-1573(97)00087-2} {\bibfield
  {journal} {\bibinfo  {journal} {Phys. Rep.}\ }\textbf {\bibinfo {volume}
  {300}},\ \bibinfo {pages} {123} (\bibinfo {year} {1998})}\BibitemShut
  {NoStop}%
\bibitem [{\citenamefont {Faessler}\ and\ \citenamefont
  {Simkovic}(1998)}]{faessler1998}%
  \BibitemOpen
  \bibfield  {author} {\bibinfo {author} {\bibfnamefont {A.}~\bibnamefont
  {Faessler}}\ and\ \bibinfo {author} {\bibfnamefont {F.}~\bibnamefont
  {Simkovic}},\ }\href {https://doi.org/10.1088/0954-3899/24/12/001} {\bibfield
   {journal} {\bibinfo  {journal} {J. Phys. G: Nucl. Part. Phys.}\ }\textbf
  {\bibinfo {volume} {24}},\ \bibinfo {pages} {2139} (\bibinfo {year}
  {1998})}\BibitemShut {NoStop}%
\bibitem [{\citenamefont {Vogel}(2012)}]{vogel2012}%
  \BibitemOpen
  \bibfield  {author} {\bibinfo {author} {\bibfnamefont {P.}~\bibnamefont
  {Vogel}},\ }\href {https://doi.org/10.1088/0954-3899/39/12/124002} {\bibfield
   {journal} {\bibinfo  {journal} {J. Phys. G: Nucl. Part. Phys.}\ }\textbf
  {\bibinfo {volume} {39}},\ \bibinfo {pages} {124002} (\bibinfo {year}
  {2012})}\BibitemShut {NoStop}%
\bibitem [{\citenamefont {Vergados}\ \emph {et~al.}(2012)\citenamefont
  {Vergados}, \citenamefont {Ejiri},\ and\ \citenamefont
  {{\v{S}}imkovic}}]{vergados2012}%
  \BibitemOpen
  \bibfield  {author} {\bibinfo {author} {\bibfnamefont {J.~D.}\ \bibnamefont
  {Vergados}}, \bibinfo {author} {\bibfnamefont {H.}~\bibnamefont {Ejiri}},\
  and\ \bibinfo {author} {\bibfnamefont {F.}~\bibnamefont {{\v{S}}imkovic}},\
  }\href {https://doi.org/10.1088/0034-4885/75/10/106301} {\bibfield  {journal}
  {\bibinfo  {journal} {Rep. Prog. Phys.}\ }\textbf {\bibinfo {volume} {75}},\
  \bibinfo {pages} {106301} (\bibinfo {year} {2012})}\BibitemShut {NoStop}%
\bibitem [{\citenamefont {Engel}\ and\ \citenamefont
  {Men{\'{e}}ndez}(2017)}]{engel2017}%
  \BibitemOpen
  \bibfield  {author} {\bibinfo {author} {\bibfnamefont {J.}~\bibnamefont
  {Engel}}\ and\ \bibinfo {author} {\bibfnamefont {J.}~\bibnamefont
  {Men{\'{e}}ndez}},\ }\href {https://doi.org/10.1088/1361-6633/aa5bc5}
  {\bibfield  {journal} {\bibinfo  {journal} {Rep. Prog. Phys.}\ }\textbf
  {\bibinfo {volume} {80}},\ \bibinfo {pages} {046301} (\bibinfo {year}
  {2017})}\BibitemShut {NoStop}%
\bibitem [{\citenamefont {Avignone}\ \emph {et~al.}(2008)\citenamefont
  {Avignone}, \citenamefont {Elliott},\ and\ \citenamefont
  {Engel}}]{avignone2008}%
  \BibitemOpen
  \bibfield  {author} {\bibinfo {author} {\bibfnamefont {F.~T.}\ \bibnamefont
  {Avignone}}, \bibinfo {author} {\bibfnamefont {S.~R.}\ \bibnamefont
  {Elliott}},\ and\ \bibinfo {author} {\bibfnamefont {J.}~\bibnamefont
  {Engel}},\ }\href {https://doi.org/10.1103/RevModPhys.80.481} {\bibfield
  {journal} {\bibinfo  {journal} {Rev. Mod. Phys.}\ }\textbf {\bibinfo {volume}
  {80}},\ \bibinfo {pages} {481} (\bibinfo {year} {2008})}\BibitemShut
  {NoStop}%
\bibitem [{\citenamefont {Ejiri}\ \emph {et~al.}(2019)\citenamefont {Ejiri},
  \citenamefont {Suhonen},\ and\ \citenamefont {Zuber}}]{ejiri2019}%
  \BibitemOpen
  \bibfield  {author} {\bibinfo {author} {\bibfnamefont {H.}~\bibnamefont
  {Ejiri}}, \bibinfo {author} {\bibfnamefont {J.}~\bibnamefont {Suhonen}},\
  and\ \bibinfo {author} {\bibfnamefont {K.}~\bibnamefont {Zuber}},\ }\href
  {https://doi.org/https://doi.org/10.1016/j.physrep.2018.12.001} {\bibfield
  {journal} {\bibinfo  {journal} {Phys. Rep.}\ }\textbf {\bibinfo {volume}
  {797}},\ \bibinfo {pages} {1} (\bibinfo {year} {2019})}\BibitemShut {NoStop}%
\bibitem [{\citenamefont {Agostini}\ \emph
  {et~al.}(2023{\natexlab{a}})\citenamefont {Agostini}, \citenamefont {Benato},
  \citenamefont {Detwiler}, \citenamefont {Men\'endez},\ and\ \citenamefont
  {Vissani}}]{agostini2023}%
  \BibitemOpen
  \bibfield  {author} {\bibinfo {author} {\bibfnamefont {M.}~\bibnamefont
  {Agostini}}, \bibinfo {author} {\bibfnamefont {G.}~\bibnamefont {Benato}},
  \bibinfo {author} {\bibfnamefont {J.~A.}\ \bibnamefont {Detwiler}}, \bibinfo
  {author} {\bibfnamefont {J.}~\bibnamefont {Men\'endez}},\ and\ \bibinfo
  {author} {\bibfnamefont {F.}~\bibnamefont {Vissani}},\ }\href
  {https://doi.org/10.1103/RevModPhys.95.025002} {\bibfield  {journal}
  {\bibinfo  {journal} {Rev. Mod. Phys.}\ }\textbf {\bibinfo {volume} {95}},\
  \bibinfo {pages} {025002} (\bibinfo {year} {2023}{\natexlab{a}})}\BibitemShut
  {NoStop}%
\bibitem [{\citenamefont {Barabash}(2020)}]{barabash2020}%
  \BibitemOpen
  \bibfield  {author} {\bibinfo {author} {\bibfnamefont {A.}~\bibnamefont
  {Barabash}},\ }\href {https://www.mdpi.com/2218-1997/6/10/159} {\bibfield
  {journal} {\bibinfo  {journal} {Universe}\ }\textbf {\bibinfo {volume} {6}}
  (\bibinfo {year} {2020})}\BibitemShut {NoStop}%
\bibitem [{\citenamefont {Agostini}\ \emph
  {et~al.}(2023{\natexlab{b}})\citenamefont {Agostini}, \citenamefont
  {Alexander}, \citenamefont {Araujo}, \citenamefont {Bakalyarov},
  \citenamefont {Balata}, \citenamefont {Barabanov}, \citenamefont {Baudis},
  \citenamefont {Bauer}, \citenamefont {Belogurov}, \citenamefont {Bettini},
  \citenamefont {Bezrukov}, \citenamefont {Biancacci}, \citenamefont {Bossio},
  \citenamefont {Bothe}, \citenamefont {Brugnera}, \citenamefont {Caldwell},
  \citenamefont {Calgaro}, \citenamefont {Cattadori}, \citenamefont
  {Chernogorov}, \citenamefont {Chiu}, \citenamefont {Comellato}, \citenamefont
  {D'Andrea}, \citenamefont {Demidova}, \citenamefont {Di~Giacinto},
  \citenamefont {Di~Marco}, \citenamefont {Doroshkevich}, \citenamefont
  {Fischer}, \citenamefont {Fomina}, \citenamefont {Gangapshev}, \citenamefont
  {Garfagnini}, \citenamefont {Gooch}, \citenamefont {Grabmayr}, \citenamefont
  {Gurentsov}, \citenamefont {Gusev}, \citenamefont {Hackenm\"uller},
  \citenamefont {Hemmer}, \citenamefont {Hofmann}, \citenamefont {Huang},
  \citenamefont {Hult}, \citenamefont {Inzhechik}, \citenamefont
  {Janicsk\'o~Cs\'athy}, \citenamefont {Jochum}, \citenamefont {Junker},
  \citenamefont {Kazalov}, \citenamefont {Kerma\"{\i}dic}, \citenamefont
  {Khushbakht}, \citenamefont {Kihm}, \citenamefont {Kilgus}, \citenamefont
  {Kirpichnikov}, \citenamefont {Klimenko}, \citenamefont {Kn\"opfle},
  \citenamefont {Kochetov}, \citenamefont {Kornoukhov}, \citenamefont {Krause},
  \citenamefont {Kuzminov}, \citenamefont {Laubenstein}, \citenamefont
  {Lehnert}, \citenamefont {Lindner}, \citenamefont {Lippi}, \citenamefont
  {Lubashevskiy}, \citenamefont {Lubsandorzhiev}, \citenamefont {Lutter},
  \citenamefont {Macolino}, \citenamefont {Majorovits}, \citenamefont
  {Maneschg}, \citenamefont {Manzanillas}, \citenamefont {Marshall},
  \citenamefont {Miloradovic}, \citenamefont {Mingazheva}, \citenamefont
  {Misiaszek}, \citenamefont {Morella}, \citenamefont {M\"uller}, \citenamefont
  {Nemchenok}, \citenamefont {Neuberger}, \citenamefont {Pandola},
  \citenamefont {Pelczar}, \citenamefont {Pertoldi}, \citenamefont {Piseri},
  \citenamefont {Pullia}, \citenamefont {Ransom}, \citenamefont {Rauscher},
  \citenamefont {Redchuk}, \citenamefont {Riboldi}, \citenamefont
  {Rumyantseva}, \citenamefont {Sada}, \citenamefont {Sailer}, \citenamefont
  {Salamida}, \citenamefont {Sch\"onert}, \citenamefont {Schreiner},
  \citenamefont {Sch\"utt}, \citenamefont {Sch\"utz}, \citenamefont {Schulz},
  \citenamefont {Schwarz}, \citenamefont {Schwingenheuer}, \citenamefont
  {Selivanenko}, \citenamefont {Shevchik}, \citenamefont {Shirchenko},
  \citenamefont {Shtembari}, \citenamefont {Simgen}, \citenamefont {Smolnikov},
  \citenamefont {Stukov}, \citenamefont {Sullivan}, \citenamefont {Vasenko},
  \citenamefont {Veresnikova}, \citenamefont {Vignoli}, \citenamefont {von
  Sturm}, \citenamefont {Wester}, \citenamefont {Wiesinger}, \citenamefont
  {Wojcik}, \citenamefont {Yanovich}, \citenamefont {Zatschler}, \citenamefont
  {Zhitnikov}, \citenamefont {Zhukov}, \citenamefont {Zinatulina},
  \citenamefont {Zschocke}, \citenamefont {Zsigmond}, \citenamefont {Zuber},\
  and\ \citenamefont {Zuzel}}]{agostini2023-76Ge}%
  \BibitemOpen
  \bibfield  {author} {\bibinfo {author} {\bibfnamefont {M.}~\bibnamefont
  {Agostini}}, \bibinfo {author} {\bibfnamefont {A.}~\bibnamefont {Alexander}},
  \bibinfo {author} {\bibfnamefont {G.~R.}\ \bibnamefont {Araujo}}, \bibinfo
  {author} {\bibfnamefont {A.~M.}\ \bibnamefont {Bakalyarov}}, \bibinfo
  {author} {\bibfnamefont {M.}~\bibnamefont {Balata}}, \bibinfo {author}
  {\bibfnamefont {I.}~\bibnamefont {Barabanov}}, \bibinfo {author}
  {\bibfnamefont {L.}~\bibnamefont {Baudis}}, \bibinfo {author} {\bibfnamefont
  {C.}~\bibnamefont {Bauer}}, \bibinfo {author} {\bibfnamefont
  {S.}~\bibnamefont {Belogurov}}, \bibinfo {author} {\bibfnamefont
  {A.}~\bibnamefont {Bettini}}, \bibinfo {author} {\bibfnamefont
  {L.}~\bibnamefont {Bezrukov}}, \bibinfo {author} {\bibfnamefont
  {V.}~\bibnamefont {Biancacci}}, \bibinfo {author} {\bibfnamefont
  {E.}~\bibnamefont {Bossio}}, \bibinfo {author} {\bibfnamefont
  {V.}~\bibnamefont {Bothe}}, \bibinfo {author} {\bibfnamefont
  {R.}~\bibnamefont {Brugnera}}, \bibinfo {author} {\bibfnamefont
  {A.}~\bibnamefont {Caldwell}}, \bibinfo {author} {\bibfnamefont
  {S.}~\bibnamefont {Calgaro}}, \bibinfo {author} {\bibfnamefont
  {C.}~\bibnamefont {Cattadori}}, \bibinfo {author} {\bibfnamefont
  {A.}~\bibnamefont {Chernogorov}}, \bibinfo {author} {\bibfnamefont {P.-J.}\
  \bibnamefont {Chiu}}, \bibinfo {author} {\bibfnamefont {T.}~\bibnamefont
  {Comellato}}, \bibinfo {author} {\bibfnamefont {V.}~\bibnamefont {D'Andrea}},
  \bibinfo {author} {\bibfnamefont {E.~V.}\ \bibnamefont {Demidova}}, \bibinfo
  {author} {\bibfnamefont {A.}~\bibnamefont {Di~Giacinto}}, \bibinfo {author}
  {\bibfnamefont {N.}~\bibnamefont {Di~Marco}}, \bibinfo {author}
  {\bibfnamefont {E.}~\bibnamefont {Doroshkevich}}, \bibinfo {author}
  {\bibfnamefont {F.}~\bibnamefont {Fischer}}, \bibinfo {author} {\bibfnamefont
  {M.}~\bibnamefont {Fomina}}, \bibinfo {author} {\bibfnamefont
  {A.}~\bibnamefont {Gangapshev}}, \bibinfo {author} {\bibfnamefont
  {A.}~\bibnamefont {Garfagnini}}, \bibinfo {author} {\bibfnamefont
  {C.}~\bibnamefont {Gooch}}, \bibinfo {author} {\bibfnamefont
  {P.}~\bibnamefont {Grabmayr}}, \bibinfo {author} {\bibfnamefont
  {V.}~\bibnamefont {Gurentsov}}, \bibinfo {author} {\bibfnamefont
  {K.}~\bibnamefont {Gusev}}, \bibinfo {author} {\bibfnamefont
  {S.}~\bibnamefont {Hackenm\"uller}}, \bibinfo {author} {\bibfnamefont
  {S.}~\bibnamefont {Hemmer}}, \bibinfo {author} {\bibfnamefont
  {W.}~\bibnamefont {Hofmann}}, \bibinfo {author} {\bibfnamefont
  {J.}~\bibnamefont {Huang}}, \bibinfo {author} {\bibfnamefont
  {M.}~\bibnamefont {Hult}}, \bibinfo {author} {\bibfnamefont {L.~V.}\
  \bibnamefont {Inzhechik}}, \bibinfo {author} {\bibfnamefont {J.}~\bibnamefont
  {Janicsk\'o~Cs\'athy}}, \bibinfo {author} {\bibfnamefont {J.}~\bibnamefont
  {Jochum}}, \bibinfo {author} {\bibfnamefont {M.}~\bibnamefont {Junker}},
  \bibinfo {author} {\bibfnamefont {V.}~\bibnamefont {Kazalov}}, \bibinfo
  {author} {\bibfnamefont {Y.}~\bibnamefont {Kerma\"{\i}dic}}, \bibinfo
  {author} {\bibfnamefont {H.}~\bibnamefont {Khushbakht}}, \bibinfo {author}
  {\bibfnamefont {T.}~\bibnamefont {Kihm}}, \bibinfo {author} {\bibfnamefont
  {K.}~\bibnamefont {Kilgus}}, \bibinfo {author} {\bibfnamefont {I.~V.}\
  \bibnamefont {Kirpichnikov}}, \bibinfo {author} {\bibfnamefont
  {A.}~\bibnamefont {Klimenko}}, \bibinfo {author} {\bibfnamefont {K.~T.}\
  \bibnamefont {Kn\"opfle}}, \bibinfo {author} {\bibfnamefont {O.}~\bibnamefont
  {Kochetov}}, \bibinfo {author} {\bibfnamefont {V.~N.}\ \bibnamefont
  {Kornoukhov}}, \bibinfo {author} {\bibfnamefont {P.}~\bibnamefont {Krause}},
  \bibinfo {author} {\bibfnamefont {V.~V.}\ \bibnamefont {Kuzminov}}, \bibinfo
  {author} {\bibfnamefont {M.}~\bibnamefont {Laubenstein}}, \bibinfo {author}
  {\bibfnamefont {B.}~\bibnamefont {Lehnert}}, \bibinfo {author} {\bibfnamefont
  {M.}~\bibnamefont {Lindner}}, \bibinfo {author} {\bibfnamefont
  {I.}~\bibnamefont {Lippi}}, \bibinfo {author} {\bibfnamefont
  {A.}~\bibnamefont {Lubashevskiy}}, \bibinfo {author} {\bibfnamefont
  {B.}~\bibnamefont {Lubsandorzhiev}}, \bibinfo {author} {\bibfnamefont
  {G.}~\bibnamefont {Lutter}}, \bibinfo {author} {\bibfnamefont
  {C.}~\bibnamefont {Macolino}}, \bibinfo {author} {\bibfnamefont
  {B.}~\bibnamefont {Majorovits}}, \bibinfo {author} {\bibfnamefont
  {W.}~\bibnamefont {Maneschg}}, \bibinfo {author} {\bibfnamefont
  {L.}~\bibnamefont {Manzanillas}}, \bibinfo {author} {\bibfnamefont
  {G.}~\bibnamefont {Marshall}}, \bibinfo {author} {\bibfnamefont
  {M.}~\bibnamefont {Miloradovic}}, \bibinfo {author} {\bibfnamefont
  {R.}~\bibnamefont {Mingazheva}}, \bibinfo {author} {\bibfnamefont
  {M.}~\bibnamefont {Misiaszek}}, \bibinfo {author} {\bibfnamefont
  {M.}~\bibnamefont {Morella}}, \bibinfo {author} {\bibfnamefont
  {Y.}~\bibnamefont {M\"uller}}, \bibinfo {author} {\bibfnamefont
  {I.}~\bibnamefont {Nemchenok}}, \bibinfo {author} {\bibfnamefont
  {M.}~\bibnamefont {Neuberger}}, \bibinfo {author} {\bibfnamefont
  {L.}~\bibnamefont {Pandola}}, \bibinfo {author} {\bibfnamefont
  {K.}~\bibnamefont {Pelczar}}, \bibinfo {author} {\bibfnamefont
  {L.}~\bibnamefont {Pertoldi}}, \bibinfo {author} {\bibfnamefont
  {P.}~\bibnamefont {Piseri}}, \bibinfo {author} {\bibfnamefont
  {A.}~\bibnamefont {Pullia}}, \bibinfo {author} {\bibfnamefont
  {C.}~\bibnamefont {Ransom}}, \bibinfo {author} {\bibfnamefont
  {L.}~\bibnamefont {Rauscher}}, \bibinfo {author} {\bibfnamefont
  {M.}~\bibnamefont {Redchuk}}, \bibinfo {author} {\bibfnamefont
  {S.}~\bibnamefont {Riboldi}}, \bibinfo {author} {\bibfnamefont
  {N.}~\bibnamefont {Rumyantseva}}, \bibinfo {author} {\bibfnamefont
  {C.}~\bibnamefont {Sada}}, \bibinfo {author} {\bibfnamefont {S.}~\bibnamefont
  {Sailer}}, \bibinfo {author} {\bibfnamefont {F.}~\bibnamefont {Salamida}},
  \bibinfo {author} {\bibfnamefont {S.}~\bibnamefont {Sch\"onert}}, \bibinfo
  {author} {\bibfnamefont {J.}~\bibnamefont {Schreiner}}, \bibinfo {author}
  {\bibfnamefont {M.}~\bibnamefont {Sch\"utt}}, \bibinfo {author}
  {\bibfnamefont {A.-K.}\ \bibnamefont {Sch\"utz}}, \bibinfo {author}
  {\bibfnamefont {O.}~\bibnamefont {Schulz}}, \bibinfo {author} {\bibfnamefont
  {M.}~\bibnamefont {Schwarz}}, \bibinfo {author} {\bibfnamefont
  {B.}~\bibnamefont {Schwingenheuer}}, \bibinfo {author} {\bibfnamefont
  {O.}~\bibnamefont {Selivanenko}}, \bibinfo {author} {\bibfnamefont
  {E.}~\bibnamefont {Shevchik}}, \bibinfo {author} {\bibfnamefont
  {M.}~\bibnamefont {Shirchenko}}, \bibinfo {author} {\bibfnamefont
  {L.}~\bibnamefont {Shtembari}}, \bibinfo {author} {\bibfnamefont
  {H.}~\bibnamefont {Simgen}}, \bibinfo {author} {\bibfnamefont
  {A.}~\bibnamefont {Smolnikov}}, \bibinfo {author} {\bibfnamefont
  {D.}~\bibnamefont {Stukov}}, \bibinfo {author} {\bibfnamefont
  {S.}~\bibnamefont {Sullivan}}, \bibinfo {author} {\bibfnamefont {A.~A.}\
  \bibnamefont {Vasenko}}, \bibinfo {author} {\bibfnamefont {A.}~\bibnamefont
  {Veresnikova}}, \bibinfo {author} {\bibfnamefont {C.}~\bibnamefont
  {Vignoli}}, \bibinfo {author} {\bibfnamefont {K.}~\bibnamefont {von Sturm}},
  \bibinfo {author} {\bibfnamefont {T.}~\bibnamefont {Wester}}, \bibinfo
  {author} {\bibfnamefont {C.}~\bibnamefont {Wiesinger}}, \bibinfo {author}
  {\bibfnamefont {M.}~\bibnamefont {Wojcik}}, \bibinfo {author} {\bibfnamefont
  {E.}~\bibnamefont {Yanovich}}, \bibinfo {author} {\bibfnamefont
  {B.}~\bibnamefont {Zatschler}}, \bibinfo {author} {\bibfnamefont
  {I.}~\bibnamefont {Zhitnikov}}, \bibinfo {author} {\bibfnamefont {S.~V.}\
  \bibnamefont {Zhukov}}, \bibinfo {author} {\bibfnamefont {D.}~\bibnamefont
  {Zinatulina}}, \bibinfo {author} {\bibfnamefont {A.}~\bibnamefont
  {Zschocke}}, \bibinfo {author} {\bibfnamefont {A.~J.}\ \bibnamefont
  {Zsigmond}}, \bibinfo {author} {\bibfnamefont {K.}~\bibnamefont {Zuber}},\
  and\ \bibinfo {author} {\bibfnamefont {G.}~\bibnamefont {Zuzel}} (\bibinfo
  {collaboration} {GERDA Collaboration}),\ }\href
  {https://doi.org/10.1103/PhysRevLett.131.142501} {\bibfield  {journal}
  {\bibinfo  {journal} {Phys. Rev. Lett.}\ }\textbf {\bibinfo {volume} {131}},\
  \bibinfo {pages} {142501} (\bibinfo {year} {2023}{\natexlab{b}})}\BibitemShut
  {NoStop}%
\bibitem [{\citenamefont {Augier}\ \emph
  {et~al.}(2023{\natexlab{a}})\citenamefont {Augier}, \citenamefont {Barabash},
  \citenamefont {Bellini}, \citenamefont {Benato}, \citenamefont {Beretta},
  \citenamefont {Berg\'e}, \citenamefont {Billard}, \citenamefont {Borovlev},
  \citenamefont {Cardani}, \citenamefont {Casali}, \citenamefont {Cazes},
  \citenamefont {Celi}, \citenamefont {Chapellier}, \citenamefont {Chiesa},
  \citenamefont {Dafinei}, \citenamefont {Danevich}, \citenamefont {De~Jesus},
  \citenamefont {Dixon}, \citenamefont {Dumoulin}, \citenamefont {Eitel},
  \citenamefont {Ferri}, \citenamefont {Fujikawa}, \citenamefont {Gascon},
  \citenamefont {Gironi}, \citenamefont {Giuliani}, \citenamefont {Grigorieva},
  \citenamefont {Gros}, \citenamefont {Helis}, \citenamefont {Huang},
  \citenamefont {Huang}, \citenamefont {Imbert}, \citenamefont {Johnston},
  \citenamefont {Juillard}, \citenamefont {Khalife}, \citenamefont {Kleifges},
  \citenamefont {Kobychev}, \citenamefont {Kolomensky}, \citenamefont
  {Konovalov}, \citenamefont {Kotila}, \citenamefont {Loaiza}, \citenamefont
  {Ma}, \citenamefont {Makarov}, \citenamefont {de~Marcillac}, \citenamefont
  {Mariam}, \citenamefont {Marini}, \citenamefont {Marnieros}, \citenamefont
  {Navick}, \citenamefont {Nones}, \citenamefont {Norman}, \citenamefont
  {Olivieri}, \citenamefont {Ouellet}, \citenamefont {Pagnanini}, \citenamefont
  {Pattavina}, \citenamefont {Paul}, \citenamefont {Pavan}, \citenamefont
  {Peng}, \citenamefont {Pessina}, \citenamefont {Pirro}, \citenamefont {Poda},
  \citenamefont {Polischuk}, \citenamefont {Pozzi}, \citenamefont {Previtali},
  \citenamefont {Redon}, \citenamefont {Rojas}, \citenamefont {Rozov},
  \citenamefont {Sanglard}, \citenamefont {Scarpaci}, \citenamefont {Schmidt},
  \citenamefont {Shen}, \citenamefont {Shlegel}, \citenamefont
  {\ifmmode~\check{S}\else \v{S}\fi{}imkovic}, \citenamefont {Singh},
  \citenamefont {Tomei}, \citenamefont {Tretyak}, \citenamefont {Umatov},
  \citenamefont {Vagneron}, \citenamefont {Vel\'azquez}, \citenamefont {Ware},
  \citenamefont {Welliver}, \citenamefont {Winslow}, \citenamefont {Xue},
  \citenamefont {Yakushev}, \citenamefont {Zarytskyy},\ and\ \citenamefont
  {Zolotarova}}]{augier2023-100Mo-Letter}%
  \BibitemOpen
  \bibfield  {author} {\bibinfo {author} {\bibfnamefont {C.}~\bibnamefont
  {Augier}}, \bibinfo {author} {\bibfnamefont {A.~S.}\ \bibnamefont
  {Barabash}}, \bibinfo {author} {\bibfnamefont {F.}~\bibnamefont {Bellini}},
  \bibinfo {author} {\bibfnamefont {G.}~\bibnamefont {Benato}}, \bibinfo
  {author} {\bibfnamefont {M.}~\bibnamefont {Beretta}}, \bibinfo {author}
  {\bibfnamefont {L.}~\bibnamefont {Berg\'e}}, \bibinfo {author} {\bibfnamefont
  {J.}~\bibnamefont {Billard}}, \bibinfo {author} {\bibfnamefont {Y.~A.}\
  \bibnamefont {Borovlev}}, \bibinfo {author} {\bibfnamefont {L.}~\bibnamefont
  {Cardani}}, \bibinfo {author} {\bibfnamefont {N.}~\bibnamefont {Casali}},
  \bibinfo {author} {\bibfnamefont {A.}~\bibnamefont {Cazes}}, \bibinfo
  {author} {\bibfnamefont {E.}~\bibnamefont {Celi}}, \bibinfo {author}
  {\bibfnamefont {M.}~\bibnamefont {Chapellier}}, \bibinfo {author}
  {\bibfnamefont {D.}~\bibnamefont {Chiesa}}, \bibinfo {author} {\bibfnamefont
  {I.}~\bibnamefont {Dafinei}}, \bibinfo {author} {\bibfnamefont {F.~A.}\
  \bibnamefont {Danevich}}, \bibinfo {author} {\bibfnamefont {M.}~\bibnamefont
  {De~Jesus}}, \bibinfo {author} {\bibfnamefont {T.}~\bibnamefont {Dixon}},
  \bibinfo {author} {\bibfnamefont {L.}~\bibnamefont {Dumoulin}}, \bibinfo
  {author} {\bibfnamefont {K.}~\bibnamefont {Eitel}}, \bibinfo {author}
  {\bibfnamefont {F.}~\bibnamefont {Ferri}}, \bibinfo {author} {\bibfnamefont
  {B.~K.}\ \bibnamefont {Fujikawa}}, \bibinfo {author} {\bibfnamefont
  {J.}~\bibnamefont {Gascon}}, \bibinfo {author} {\bibfnamefont
  {L.}~\bibnamefont {Gironi}}, \bibinfo {author} {\bibfnamefont
  {A.}~\bibnamefont {Giuliani}}, \bibinfo {author} {\bibfnamefont {V.~D.}\
  \bibnamefont {Grigorieva}}, \bibinfo {author} {\bibfnamefont
  {M.}~\bibnamefont {Gros}}, \bibinfo {author} {\bibfnamefont {D.~L.}\
  \bibnamefont {Helis}}, \bibinfo {author} {\bibfnamefont {H.~Z.}\ \bibnamefont
  {Huang}}, \bibinfo {author} {\bibfnamefont {R.}~\bibnamefont {Huang}},
  \bibinfo {author} {\bibfnamefont {L.}~\bibnamefont {Imbert}}, \bibinfo
  {author} {\bibfnamefont {J.}~\bibnamefont {Johnston}}, \bibinfo {author}
  {\bibfnamefont {A.}~\bibnamefont {Juillard}}, \bibinfo {author}
  {\bibfnamefont {H.}~\bibnamefont {Khalife}}, \bibinfo {author} {\bibfnamefont
  {M.}~\bibnamefont {Kleifges}}, \bibinfo {author} {\bibfnamefont {V.~V.}\
  \bibnamefont {Kobychev}}, \bibinfo {author} {\bibfnamefont {Y.~G.}\
  \bibnamefont {Kolomensky}}, \bibinfo {author} {\bibfnamefont {S.~I.}\
  \bibnamefont {Konovalov}}, \bibinfo {author} {\bibfnamefont {J.}~\bibnamefont
  {Kotila}}, \bibinfo {author} {\bibfnamefont {P.}~\bibnamefont {Loaiza}},
  \bibinfo {author} {\bibfnamefont {L.}~\bibnamefont {Ma}}, \bibinfo {author}
  {\bibfnamefont {E.~P.}\ \bibnamefont {Makarov}}, \bibinfo {author}
  {\bibfnamefont {P.}~\bibnamefont {de~Marcillac}}, \bibinfo {author}
  {\bibfnamefont {R.}~\bibnamefont {Mariam}}, \bibinfo {author} {\bibfnamefont
  {L.}~\bibnamefont {Marini}}, \bibinfo {author} {\bibfnamefont
  {S.}~\bibnamefont {Marnieros}}, \bibinfo {author} {\bibfnamefont {X.-F.}\
  \bibnamefont {Navick}}, \bibinfo {author} {\bibfnamefont {C.}~\bibnamefont
  {Nones}}, \bibinfo {author} {\bibfnamefont {E.~B.}\ \bibnamefont {Norman}},
  \bibinfo {author} {\bibfnamefont {E.}~\bibnamefont {Olivieri}}, \bibinfo
  {author} {\bibfnamefont {J.~L.}\ \bibnamefont {Ouellet}}, \bibinfo {author}
  {\bibfnamefont {L.}~\bibnamefont {Pagnanini}}, \bibinfo {author}
  {\bibfnamefont {L.}~\bibnamefont {Pattavina}}, \bibinfo {author}
  {\bibfnamefont {B.}~\bibnamefont {Paul}}, \bibinfo {author} {\bibfnamefont
  {M.}~\bibnamefont {Pavan}}, \bibinfo {author} {\bibfnamefont
  {H.}~\bibnamefont {Peng}}, \bibinfo {author} {\bibfnamefont {G.}~\bibnamefont
  {Pessina}}, \bibinfo {author} {\bibfnamefont {S.}~\bibnamefont {Pirro}},
  \bibinfo {author} {\bibfnamefont {D.~V.}\ \bibnamefont {Poda}}, \bibinfo
  {author} {\bibfnamefont {O.~G.}\ \bibnamefont {Polischuk}}, \bibinfo {author}
  {\bibfnamefont {S.}~\bibnamefont {Pozzi}}, \bibinfo {author} {\bibfnamefont
  {E.}~\bibnamefont {Previtali}}, \bibinfo {author} {\bibfnamefont
  {T.}~\bibnamefont {Redon}}, \bibinfo {author} {\bibfnamefont
  {A.}~\bibnamefont {Rojas}}, \bibinfo {author} {\bibfnamefont
  {S.}~\bibnamefont {Rozov}}, \bibinfo {author} {\bibfnamefont
  {V.}~\bibnamefont {Sanglard}}, \bibinfo {author} {\bibfnamefont {J.~A.}\
  \bibnamefont {Scarpaci}}, \bibinfo {author} {\bibfnamefont {B.}~\bibnamefont
  {Schmidt}}, \bibinfo {author} {\bibfnamefont {Y.}~\bibnamefont {Shen}},
  \bibinfo {author} {\bibfnamefont {V.~N.}\ \bibnamefont {Shlegel}}, \bibinfo
  {author} {\bibfnamefont {F.}~\bibnamefont {\ifmmode~\check{S}\else
  \v{S}\fi{}imkovic}}, \bibinfo {author} {\bibfnamefont {V.}~\bibnamefont
  {Singh}}, \bibinfo {author} {\bibfnamefont {C.}~\bibnamefont {Tomei}},
  \bibinfo {author} {\bibfnamefont {V.~I.}\ \bibnamefont {Tretyak}}, \bibinfo
  {author} {\bibfnamefont {V.~I.}\ \bibnamefont {Umatov}}, \bibinfo {author}
  {\bibfnamefont {L.}~\bibnamefont {Vagneron}}, \bibinfo {author}
  {\bibfnamefont {M.}~\bibnamefont {Vel\'azquez}}, \bibinfo {author}
  {\bibfnamefont {B.}~\bibnamefont {Ware}}, \bibinfo {author} {\bibfnamefont
  {B.}~\bibnamefont {Welliver}}, \bibinfo {author} {\bibfnamefont
  {L.}~\bibnamefont {Winslow}}, \bibinfo {author} {\bibfnamefont
  {M.}~\bibnamefont {Xue}}, \bibinfo {author} {\bibfnamefont {E.}~\bibnamefont
  {Yakushev}}, \bibinfo {author} {\bibfnamefont {M.}~\bibnamefont
  {Zarytskyy}},\ and\ \bibinfo {author} {\bibfnamefont {A.~S.}\ \bibnamefont
  {Zolotarova}} (\bibinfo {collaboration} {CUPID-Mo Collaboration}),\ }\href
  {https://doi.org/10.1103/PhysRevLett.131.162501} {\bibfield  {journal}
  {\bibinfo  {journal} {Phys. Rev. Lett.}\ }\textbf {\bibinfo {volume} {131}},\
  \bibinfo {pages} {162501} (\bibinfo {year} {2023}{\natexlab{a}})}\BibitemShut
  {NoStop}%
\bibitem [{\citenamefont {Augier}\ \emph
  {et~al.}(2023{\natexlab{b}})\citenamefont {Augier}, \citenamefont {Barabash},
  \citenamefont {Bellini}, \citenamefont {Benato}, \citenamefont {Beretta},
  \citenamefont {Berg\'e}, \citenamefont {Billard}, \citenamefont {Borovlev},
  \citenamefont {Cardani}, \citenamefont {Casali}, \citenamefont {Cazes},
  \citenamefont {Chapellier}, \citenamefont {Chiesa}, \citenamefont {Dafinei},
  \citenamefont {Danevich}, \citenamefont {De~Jesus}, \citenamefont {Dixon},
  \citenamefont {Dumoulin}, \citenamefont {Eitel}, \citenamefont {Ferri},
  \citenamefont {Fujikawa}, \citenamefont {Gascon}, \citenamefont {Gironi},
  \citenamefont {Giuliani}, \citenamefont {Grigorieva}, \citenamefont {Gros},
  \citenamefont {Helis}, \citenamefont {Huang}, \citenamefont {Huang},
  \citenamefont {Imbert}, \citenamefont {Johnston}, \citenamefont {Juillard},
  \citenamefont {Khalife}, \citenamefont {Kleifges}, \citenamefont {Kobychev},
  \citenamefont {Kolomensky}, \citenamefont {Konovalov}, \citenamefont
  {Kotila}, \citenamefont {Loaiza}, \citenamefont {Ma}, \citenamefont
  {Makarov}, \citenamefont {de~Marcillac}, \citenamefont {Mariam},
  \citenamefont {Marini}, \citenamefont {Marnieros}, \citenamefont {Navick},
  \citenamefont {Nones}, \citenamefont {Norman}, \citenamefont {Olivieri},
  \citenamefont {Ouellet}, \citenamefont {Pagnanini}, \citenamefont
  {Pattavina}, \citenamefont {Paul}, \citenamefont {Pavan}, \citenamefont
  {Peng}, \citenamefont {Pessina}, \citenamefont {Pirro}, \citenamefont {Poda},
  \citenamefont {Polischuk}, \citenamefont {Pozzi}, \citenamefont {Previtali},
  \citenamefont {Redon}, \citenamefont {Rojas}, \citenamefont {Rozov},
  \citenamefont {Sanglard}, \citenamefont {Scarpaci}, \citenamefont {Schmidt},
  \citenamefont {Shen}, \citenamefont {Shlegel}, \citenamefont {Singh},
  \citenamefont {Tomei}, \citenamefont {Tretyak}, \citenamefont {Umatov},
  \citenamefont {Vagneron}, \citenamefont {Vel\'azquez}, \citenamefont
  {Welliver}, \citenamefont {Winslow}, \citenamefont {Xue}, \citenamefont
  {Yakushev}, \citenamefont {Zarytskyy},\ and\ \citenamefont
  {Zolotarova}}]{augier2023-100Mo}%
  \BibitemOpen
  \bibfield  {author} {\bibinfo {author} {\bibfnamefont {C.}~\bibnamefont
  {Augier}}, \bibinfo {author} {\bibfnamefont {A.~S.}\ \bibnamefont
  {Barabash}}, \bibinfo {author} {\bibfnamefont {F.}~\bibnamefont {Bellini}},
  \bibinfo {author} {\bibfnamefont {G.}~\bibnamefont {Benato}}, \bibinfo
  {author} {\bibfnamefont {M.}~\bibnamefont {Beretta}}, \bibinfo {author}
  {\bibfnamefont {L.}~\bibnamefont {Berg\'e}}, \bibinfo {author} {\bibfnamefont
  {J.}~\bibnamefont {Billard}}, \bibinfo {author} {\bibfnamefont {Y.~A.}\
  \bibnamefont {Borovlev}}, \bibinfo {author} {\bibfnamefont {L.}~\bibnamefont
  {Cardani}}, \bibinfo {author} {\bibfnamefont {N.}~\bibnamefont {Casali}},
  \bibinfo {author} {\bibfnamefont {A.}~\bibnamefont {Cazes}}, \bibinfo
  {author} {\bibfnamefont {M.}~\bibnamefont {Chapellier}}, \bibinfo {author}
  {\bibfnamefont {D.}~\bibnamefont {Chiesa}}, \bibinfo {author} {\bibfnamefont
  {I.}~\bibnamefont {Dafinei}}, \bibinfo {author} {\bibfnamefont {F.~A.}\
  \bibnamefont {Danevich}}, \bibinfo {author} {\bibfnamefont {M.}~\bibnamefont
  {De~Jesus}}, \bibinfo {author} {\bibfnamefont {T.}~\bibnamefont {Dixon}},
  \bibinfo {author} {\bibfnamefont {L.}~\bibnamefont {Dumoulin}}, \bibinfo
  {author} {\bibfnamefont {K.}~\bibnamefont {Eitel}}, \bibinfo {author}
  {\bibfnamefont {F.}~\bibnamefont {Ferri}}, \bibinfo {author} {\bibfnamefont
  {B.~K.}\ \bibnamefont {Fujikawa}}, \bibinfo {author} {\bibfnamefont
  {J.}~\bibnamefont {Gascon}}, \bibinfo {author} {\bibfnamefont
  {L.}~\bibnamefont {Gironi}}, \bibinfo {author} {\bibfnamefont
  {A.}~\bibnamefont {Giuliani}}, \bibinfo {author} {\bibfnamefont {V.~D.}\
  \bibnamefont {Grigorieva}}, \bibinfo {author} {\bibfnamefont
  {M.}~\bibnamefont {Gros}}, \bibinfo {author} {\bibfnamefont {D.~L.}\
  \bibnamefont {Helis}}, \bibinfo {author} {\bibfnamefont {H.~Z.}\ \bibnamefont
  {Huang}}, \bibinfo {author} {\bibfnamefont {R.}~\bibnamefont {Huang}},
  \bibinfo {author} {\bibfnamefont {L.}~\bibnamefont {Imbert}}, \bibinfo
  {author} {\bibfnamefont {J.}~\bibnamefont {Johnston}}, \bibinfo {author}
  {\bibfnamefont {A.}~\bibnamefont {Juillard}}, \bibinfo {author}
  {\bibfnamefont {H.}~\bibnamefont {Khalife}}, \bibinfo {author} {\bibfnamefont
  {M.}~\bibnamefont {Kleifges}}, \bibinfo {author} {\bibfnamefont {V.~V.}\
  \bibnamefont {Kobychev}}, \bibinfo {author} {\bibfnamefont {Y.~G.}\
  \bibnamefont {Kolomensky}}, \bibinfo {author} {\bibfnamefont {S.~I.}\
  \bibnamefont {Konovalov}}, \bibinfo {author} {\bibfnamefont {J.}~\bibnamefont
  {Kotila}}, \bibinfo {author} {\bibfnamefont {P.}~\bibnamefont {Loaiza}},
  \bibinfo {author} {\bibfnamefont {L.}~\bibnamefont {Ma}}, \bibinfo {author}
  {\bibfnamefont {E.~P.}\ \bibnamefont {Makarov}}, \bibinfo {author}
  {\bibfnamefont {P.}~\bibnamefont {de~Marcillac}}, \bibinfo {author}
  {\bibfnamefont {R.}~\bibnamefont {Mariam}}, \bibinfo {author} {\bibfnamefont
  {L.}~\bibnamefont {Marini}}, \bibinfo {author} {\bibfnamefont
  {S.}~\bibnamefont {Marnieros}}, \bibinfo {author} {\bibfnamefont {X.-F.}\
  \bibnamefont {Navick}}, \bibinfo {author} {\bibfnamefont {C.}~\bibnamefont
  {Nones}}, \bibinfo {author} {\bibfnamefont {E.~B.}\ \bibnamefont {Norman}},
  \bibinfo {author} {\bibfnamefont {E.}~\bibnamefont {Olivieri}}, \bibinfo
  {author} {\bibfnamefont {J.~L.}\ \bibnamefont {Ouellet}}, \bibinfo {author}
  {\bibfnamefont {L.}~\bibnamefont {Pagnanini}}, \bibinfo {author}
  {\bibfnamefont {L.}~\bibnamefont {Pattavina}}, \bibinfo {author}
  {\bibfnamefont {B.}~\bibnamefont {Paul}}, \bibinfo {author} {\bibfnamefont
  {M.}~\bibnamefont {Pavan}}, \bibinfo {author} {\bibfnamefont
  {H.}~\bibnamefont {Peng}}, \bibinfo {author} {\bibfnamefont {G.}~\bibnamefont
  {Pessina}}, \bibinfo {author} {\bibfnamefont {S.}~\bibnamefont {Pirro}},
  \bibinfo {author} {\bibfnamefont {D.~V.}\ \bibnamefont {Poda}}, \bibinfo
  {author} {\bibfnamefont {O.~G.}\ \bibnamefont {Polischuk}}, \bibinfo {author}
  {\bibfnamefont {S.}~\bibnamefont {Pozzi}}, \bibinfo {author} {\bibfnamefont
  {E.}~\bibnamefont {Previtali}}, \bibinfo {author} {\bibfnamefont
  {T.}~\bibnamefont {Redon}}, \bibinfo {author} {\bibfnamefont
  {A.}~\bibnamefont {Rojas}}, \bibinfo {author} {\bibfnamefont
  {S.}~\bibnamefont {Rozov}}, \bibinfo {author} {\bibfnamefont
  {V.}~\bibnamefont {Sanglard}}, \bibinfo {author} {\bibfnamefont {J.~A.}\
  \bibnamefont {Scarpaci}}, \bibinfo {author} {\bibfnamefont {B.}~\bibnamefont
  {Schmidt}}, \bibinfo {author} {\bibfnamefont {Y.}~\bibnamefont {Shen}},
  \bibinfo {author} {\bibfnamefont {V.~N.}\ \bibnamefont {Shlegel}}, \bibinfo
  {author} {\bibfnamefont {V.}~\bibnamefont {Singh}}, \bibinfo {author}
  {\bibfnamefont {C.}~\bibnamefont {Tomei}}, \bibinfo {author} {\bibfnamefont
  {V.~I.}\ \bibnamefont {Tretyak}}, \bibinfo {author} {\bibfnamefont {V.~I.}\
  \bibnamefont {Umatov}}, \bibinfo {author} {\bibfnamefont {L.}~\bibnamefont
  {Vagneron}}, \bibinfo {author} {\bibfnamefont {M.}~\bibnamefont
  {Vel\'azquez}}, \bibinfo {author} {\bibfnamefont {B.}~\bibnamefont
  {Welliver}}, \bibinfo {author} {\bibfnamefont {L.}~\bibnamefont {Winslow}},
  \bibinfo {author} {\bibfnamefont {M.}~\bibnamefont {Xue}}, \bibinfo {author}
  {\bibfnamefont {E.}~\bibnamefont {Yakushev}}, \bibinfo {author}
  {\bibfnamefont {M.}~\bibnamefont {Zarytskyy}},\ and\ \bibinfo {author}
  {\bibfnamefont {A.~S.}\ \bibnamefont {Zolotarova}} (\bibinfo {collaboration}
  {CUPID-Mo Collaboration}),\ }\href
  {https://doi.org/10.1103/PhysRevC.107.025503} {\bibfield  {journal} {\bibinfo
   {journal} {Phys. Rev. C}\ }\textbf {\bibinfo {volume} {107}},\ \bibinfo
  {pages} {025503} (\bibinfo {year} {2023}{\natexlab{b}})}\BibitemShut
  {NoStop}%
\bibitem [{\citenamefont {Adams}\ \emph {et~al.}(2021)\citenamefont {Adams},
  \citenamefont {Alduino}, \citenamefont {Alfonso}, \citenamefont {Avignone},
  \citenamefont {Azzolini}, \citenamefont {Bari}, \citenamefont {Bellini},
  \citenamefont {Benato}, \citenamefont {Biassoni}, \citenamefont {Branca},
  \citenamefont {Brofferio}, \citenamefont {Bucci}, \citenamefont {Camilleri},
  \citenamefont {Caminata}, \citenamefont {Campani}, \citenamefont {Canonica},
  \citenamefont {Cao}, \citenamefont {Capelli}, \citenamefont {Cappelli},
  \citenamefont {Cardani}, \citenamefont {Carniti}, \citenamefont {Casali},
  \citenamefont {Chiesa}, \citenamefont {Clemenza}, \citenamefont {Copello},
  \citenamefont {Cosmelli}, \citenamefont {Cremonesi}, \citenamefont
  {Creswick}, \citenamefont {D'Addabbo}, \citenamefont {Dafinei}, \citenamefont
  {Davis}, \citenamefont {Dell'Oro}, \citenamefont {Di~Domizio}, \citenamefont
  {Domp\`e}, \citenamefont {Fang}, \citenamefont {Fantini}, \citenamefont
  {Faverzani}, \citenamefont {Ferri}, \citenamefont {Ferroni}, \citenamefont
  {Fiorini}, \citenamefont {Franceschi}, \citenamefont {Freedman},
  \citenamefont {Fu}, \citenamefont {Fujikawa}, \citenamefont {Giachero},
  \citenamefont {Gironi}, \citenamefont {Giuliani}, \citenamefont {Gorla},
  \citenamefont {Gotti}, \citenamefont {Gutierrez}, \citenamefont {Han},
  \citenamefont {Heeger}, \citenamefont {Huang}, \citenamefont {Huang},
  \citenamefont {Johnston}, \citenamefont {Keppel}, \citenamefont {Kolomensky},
  \citenamefont {Ligi}, \citenamefont {Ma}, \citenamefont {Ma}, \citenamefont
  {Marini}, \citenamefont {Maruyama}, \citenamefont {Mayer}, \citenamefont
  {Mei}, \citenamefont {Moggi}, \citenamefont {Morganti}, \citenamefont
  {Napolitano}, \citenamefont {Nastasi}, \citenamefont {Nikkel}, \citenamefont
  {Nones}, \citenamefont {Norman}, \citenamefont {Nucciotti}, \citenamefont
  {Nutini}, \citenamefont {O'Donnell}, \citenamefont {Ouellet}, \citenamefont
  {Pagan}, \citenamefont {Pagliarone}, \citenamefont {Pagnanini}, \citenamefont
  {Pallavicini}, \citenamefont {Pattavina}, \citenamefont {Pavan},
  \citenamefont {Pessina}, \citenamefont {Pettinacci}, \citenamefont {Pira},
  \citenamefont {Pirro}, \citenamefont {Pozzi}, \citenamefont {Previtali},
  \citenamefont {Puiu}, \citenamefont {Rosenfeld}, \citenamefont {Rusconi},
  \citenamefont {Sakai}, \citenamefont {Sangiorgio}, \citenamefont {Schmidt},
  \citenamefont {Scielzo}, \citenamefont {Sharma}, \citenamefont {Singh},
  \citenamefont {Sisti}, \citenamefont {Speller}, \citenamefont {Surukuchi},
  \citenamefont {Taffarello}, \citenamefont {Terranova}, \citenamefont {Tomei},
  \citenamefont {Vetter}, \citenamefont {Vignati}, \citenamefont
  {Wagaarachchi}, \citenamefont {Wang}, \citenamefont {Welliver}, \citenamefont
  {Wilson}, \citenamefont {Wilson}, \citenamefont {Winslow}, \citenamefont
  {Zimmermann},\ and\ \citenamefont {Zucchelli}}]{adams2021-130Te}%
  \BibitemOpen
  \bibfield  {author} {\bibinfo {author} {\bibfnamefont {D.~Q.}\ \bibnamefont
  {Adams}}, \bibinfo {author} {\bibfnamefont {C.}~\bibnamefont {Alduino}},
  \bibinfo {author} {\bibfnamefont {K.}~\bibnamefont {Alfonso}}, \bibinfo
  {author} {\bibfnamefont {F.~T.}\ \bibnamefont {Avignone}}, \bibinfo {author}
  {\bibfnamefont {O.}~\bibnamefont {Azzolini}}, \bibinfo {author}
  {\bibfnamefont {G.}~\bibnamefont {Bari}}, \bibinfo {author} {\bibfnamefont
  {F.}~\bibnamefont {Bellini}}, \bibinfo {author} {\bibfnamefont
  {G.}~\bibnamefont {Benato}}, \bibinfo {author} {\bibfnamefont
  {M.}~\bibnamefont {Biassoni}}, \bibinfo {author} {\bibfnamefont
  {A.}~\bibnamefont {Branca}}, \bibinfo {author} {\bibfnamefont
  {C.}~\bibnamefont {Brofferio}}, \bibinfo {author} {\bibfnamefont
  {C.}~\bibnamefont {Bucci}}, \bibinfo {author} {\bibfnamefont
  {J.}~\bibnamefont {Camilleri}}, \bibinfo {author} {\bibfnamefont
  {A.}~\bibnamefont {Caminata}}, \bibinfo {author} {\bibfnamefont
  {A.}~\bibnamefont {Campani}}, \bibinfo {author} {\bibfnamefont
  {L.}~\bibnamefont {Canonica}}, \bibinfo {author} {\bibfnamefont {X.~G.}\
  \bibnamefont {Cao}}, \bibinfo {author} {\bibfnamefont {S.}~\bibnamefont
  {Capelli}}, \bibinfo {author} {\bibfnamefont {L.}~\bibnamefont {Cappelli}},
  \bibinfo {author} {\bibfnamefont {L.}~\bibnamefont {Cardani}}, \bibinfo
  {author} {\bibfnamefont {P.}~\bibnamefont {Carniti}}, \bibinfo {author}
  {\bibfnamefont {N.}~\bibnamefont {Casali}}, \bibinfo {author} {\bibfnamefont
  {D.}~\bibnamefont {Chiesa}}, \bibinfo {author} {\bibfnamefont
  {M.}~\bibnamefont {Clemenza}}, \bibinfo {author} {\bibfnamefont
  {S.}~\bibnamefont {Copello}}, \bibinfo {author} {\bibfnamefont
  {C.}~\bibnamefont {Cosmelli}}, \bibinfo {author} {\bibfnamefont
  {O.}~\bibnamefont {Cremonesi}}, \bibinfo {author} {\bibfnamefont {R.~J.}\
  \bibnamefont {Creswick}}, \bibinfo {author} {\bibfnamefont {A.}~\bibnamefont
  {D'Addabbo}}, \bibinfo {author} {\bibfnamefont {I.}~\bibnamefont {Dafinei}},
  \bibinfo {author} {\bibfnamefont {C.~J.}\ \bibnamefont {Davis}}, \bibinfo
  {author} {\bibfnamefont {S.}~\bibnamefont {Dell'Oro}}, \bibinfo {author}
  {\bibfnamefont {S.}~\bibnamefont {Di~Domizio}}, \bibinfo {author}
  {\bibfnamefont {V.}~\bibnamefont {Domp\`e}}, \bibinfo {author} {\bibfnamefont
  {D.~Q.}\ \bibnamefont {Fang}}, \bibinfo {author} {\bibfnamefont
  {G.}~\bibnamefont {Fantini}}, \bibinfo {author} {\bibfnamefont
  {M.}~\bibnamefont {Faverzani}}, \bibinfo {author} {\bibfnamefont
  {E.}~\bibnamefont {Ferri}}, \bibinfo {author} {\bibfnamefont
  {F.}~\bibnamefont {Ferroni}}, \bibinfo {author} {\bibfnamefont
  {E.}~\bibnamefont {Fiorini}}, \bibinfo {author} {\bibfnamefont {M.~A.}\
  \bibnamefont {Franceschi}}, \bibinfo {author} {\bibfnamefont {S.~J.}\
  \bibnamefont {Freedman}}, \bibinfo {author} {\bibfnamefont {S.~H.}\
  \bibnamefont {Fu}}, \bibinfo {author} {\bibfnamefont {B.~K.}\ \bibnamefont
  {Fujikawa}}, \bibinfo {author} {\bibfnamefont {A.}~\bibnamefont {Giachero}},
  \bibinfo {author} {\bibfnamefont {L.}~\bibnamefont {Gironi}}, \bibinfo
  {author} {\bibfnamefont {A.}~\bibnamefont {Giuliani}}, \bibinfo {author}
  {\bibfnamefont {P.}~\bibnamefont {Gorla}}, \bibinfo {author} {\bibfnamefont
  {C.}~\bibnamefont {Gotti}}, \bibinfo {author} {\bibfnamefont {T.~D.}\
  \bibnamefont {Gutierrez}}, \bibinfo {author} {\bibfnamefont {K.}~\bibnamefont
  {Han}}, \bibinfo {author} {\bibfnamefont {K.~M.}\ \bibnamefont {Heeger}},
  \bibinfo {author} {\bibfnamefont {R.~G.}\ \bibnamefont {Huang}}, \bibinfo
  {author} {\bibfnamefont {H.~Z.}\ \bibnamefont {Huang}}, \bibinfo {author}
  {\bibfnamefont {J.}~\bibnamefont {Johnston}}, \bibinfo {author}
  {\bibfnamefont {G.}~\bibnamefont {Keppel}}, \bibinfo {author} {\bibfnamefont
  {Y.~G.}\ \bibnamefont {Kolomensky}}, \bibinfo {author} {\bibfnamefont
  {C.}~\bibnamefont {Ligi}}, \bibinfo {author} {\bibfnamefont {L.}~\bibnamefont
  {Ma}}, \bibinfo {author} {\bibfnamefont {Y.~G.}\ \bibnamefont {Ma}}, \bibinfo
  {author} {\bibfnamefont {L.}~\bibnamefont {Marini}}, \bibinfo {author}
  {\bibfnamefont {R.~H.}\ \bibnamefont {Maruyama}}, \bibinfo {author}
  {\bibfnamefont {D.}~\bibnamefont {Mayer}}, \bibinfo {author} {\bibfnamefont
  {Y.}~\bibnamefont {Mei}}, \bibinfo {author} {\bibfnamefont {N.}~\bibnamefont
  {Moggi}}, \bibinfo {author} {\bibfnamefont {S.}~\bibnamefont {Morganti}},
  \bibinfo {author} {\bibfnamefont {T.}~\bibnamefont {Napolitano}}, \bibinfo
  {author} {\bibfnamefont {M.}~\bibnamefont {Nastasi}}, \bibinfo {author}
  {\bibfnamefont {J.}~\bibnamefont {Nikkel}}, \bibinfo {author} {\bibfnamefont
  {C.}~\bibnamefont {Nones}}, \bibinfo {author} {\bibfnamefont {E.~B.}\
  \bibnamefont {Norman}}, \bibinfo {author} {\bibfnamefont {A.}~\bibnamefont
  {Nucciotti}}, \bibinfo {author} {\bibfnamefont {I.}~\bibnamefont {Nutini}},
  \bibinfo {author} {\bibfnamefont {T.}~\bibnamefont {O'Donnell}}, \bibinfo
  {author} {\bibfnamefont {J.~L.}\ \bibnamefont {Ouellet}}, \bibinfo {author}
  {\bibfnamefont {S.}~\bibnamefont {Pagan}}, \bibinfo {author} {\bibfnamefont
  {C.~E.}\ \bibnamefont {Pagliarone}}, \bibinfo {author} {\bibfnamefont
  {L.}~\bibnamefont {Pagnanini}}, \bibinfo {author} {\bibfnamefont
  {M.}~\bibnamefont {Pallavicini}}, \bibinfo {author} {\bibfnamefont
  {L.}~\bibnamefont {Pattavina}}, \bibinfo {author} {\bibfnamefont
  {M.}~\bibnamefont {Pavan}}, \bibinfo {author} {\bibfnamefont
  {G.}~\bibnamefont {Pessina}}, \bibinfo {author} {\bibfnamefont
  {V.}~\bibnamefont {Pettinacci}}, \bibinfo {author} {\bibfnamefont
  {C.}~\bibnamefont {Pira}}, \bibinfo {author} {\bibfnamefont {S.}~\bibnamefont
  {Pirro}}, \bibinfo {author} {\bibfnamefont {S.}~\bibnamefont {Pozzi}},
  \bibinfo {author} {\bibfnamefont {E.}~\bibnamefont {Previtali}}, \bibinfo
  {author} {\bibfnamefont {A.}~\bibnamefont {Puiu}}, \bibinfo {author}
  {\bibfnamefont {C.}~\bibnamefont {Rosenfeld}}, \bibinfo {author}
  {\bibfnamefont {C.}~\bibnamefont {Rusconi}}, \bibinfo {author} {\bibfnamefont
  {M.}~\bibnamefont {Sakai}}, \bibinfo {author} {\bibfnamefont
  {S.}~\bibnamefont {Sangiorgio}}, \bibinfo {author} {\bibfnamefont
  {B.}~\bibnamefont {Schmidt}}, \bibinfo {author} {\bibfnamefont {N.~D.}\
  \bibnamefont {Scielzo}}, \bibinfo {author} {\bibfnamefont {V.}~\bibnamefont
  {Sharma}}, \bibinfo {author} {\bibfnamefont {V.}~\bibnamefont {Singh}},
  \bibinfo {author} {\bibfnamefont {M.}~\bibnamefont {Sisti}}, \bibinfo
  {author} {\bibfnamefont {D.}~\bibnamefont {Speller}}, \bibinfo {author}
  {\bibfnamefont {P.~T.}\ \bibnamefont {Surukuchi}}, \bibinfo {author}
  {\bibfnamefont {L.}~\bibnamefont {Taffarello}}, \bibinfo {author}
  {\bibfnamefont {F.}~\bibnamefont {Terranova}}, \bibinfo {author}
  {\bibfnamefont {C.}~\bibnamefont {Tomei}}, \bibinfo {author} {\bibfnamefont
  {K.~J.}\ \bibnamefont {Vetter}}, \bibinfo {author} {\bibfnamefont
  {M.}~\bibnamefont {Vignati}}, \bibinfo {author} {\bibfnamefont {S.~L.}\
  \bibnamefont {Wagaarachchi}}, \bibinfo {author} {\bibfnamefont {B.~S.}\
  \bibnamefont {Wang}}, \bibinfo {author} {\bibfnamefont {B.}~\bibnamefont
  {Welliver}}, \bibinfo {author} {\bibfnamefont {J.}~\bibnamefont {Wilson}},
  \bibinfo {author} {\bibfnamefont {K.}~\bibnamefont {Wilson}}, \bibinfo
  {author} {\bibfnamefont {L.~A.}\ \bibnamefont {Winslow}}, \bibinfo {author}
  {\bibfnamefont {S.}~\bibnamefont {Zimmermann}},\ and\ \bibinfo {author}
  {\bibfnamefont {S.}~\bibnamefont {Zucchelli}},\ }\href
  {https://doi.org/10.1103/PhysRevLett.126.171801} {\bibfield  {journal}
  {\bibinfo  {journal} {Phys. Rev. Lett.}\ }\textbf {\bibinfo {volume} {126}},\
  \bibinfo {pages} {171801} (\bibinfo {year} {2021})}\BibitemShut {NoStop}%
\bibitem [{\citenamefont {Pirinen}\ and\ \citenamefont
  {Suhonen}(2015)}]{pirinen2015}%
  \BibitemOpen
  \bibfield  {author} {\bibinfo {author} {\bibfnamefont {P.}~\bibnamefont
  {Pirinen}}\ and\ \bibinfo {author} {\bibfnamefont {J.}~\bibnamefont
  {Suhonen}},\ }\href {https://doi.org/10.1103/PhysRevC.91.054309} {\bibfield
  {journal} {\bibinfo  {journal} {Phys. Rev. C}\ }\textbf {\bibinfo {volume}
  {91}},\ \bibinfo {pages} {054309} (\bibinfo {year} {2015})}\BibitemShut
  {NoStop}%
\bibitem [{\citenamefont {\ifmmode~\check{S}\else \v{S}\fi{}imkovic}\ \emph
  {et~al.}(2018)\citenamefont {\ifmmode~\check{S}\else \v{S}\fi{}imkovic},
  \citenamefont {Smetana},\ and\ \citenamefont {Vogel}}]{simkovic2018}%
  \BibitemOpen
  \bibfield  {author} {\bibinfo {author} {\bibfnamefont {F.}~\bibnamefont
  {\ifmmode~\check{S}\else \v{S}\fi{}imkovic}}, \bibinfo {author}
  {\bibfnamefont {A.}~\bibnamefont {Smetana}},\ and\ \bibinfo {author}
  {\bibfnamefont {P.}~\bibnamefont {Vogel}},\ }\href
  {https://doi.org/10.1103/PhysRevC.98.064325} {\bibfield  {journal} {\bibinfo
  {journal} {Phys. Rev. C}\ }\textbf {\bibinfo {volume} {98}},\ \bibinfo
  {pages} {064325} (\bibinfo {year} {2018})}\BibitemShut {NoStop}%
\bibitem [{\citenamefont {Caurier}\ \emph {et~al.}(2007)\citenamefont
  {Caurier}, \citenamefont {Nowacki},\ and\ \citenamefont
  {Poves}}]{caurier2007}%
  \BibitemOpen
  \bibfield  {author} {\bibinfo {author} {\bibfnamefont {E.}~\bibnamefont
  {Caurier}}, \bibinfo {author} {\bibfnamefont {F.}~\bibnamefont {Nowacki}},\
  and\ \bibinfo {author} {\bibfnamefont {A.}~\bibnamefont {Poves}},\ }\href
  {https://doi.org/10.1142/S0218301307005983} {\bibfield  {journal} {\bibinfo
  {journal} {Int. J. Mod. Phys. E}\ }\textbf {\bibinfo {volume} {16}},\
  \bibinfo {pages} {552} (\bibinfo {year} {2007})}\BibitemShut {NoStop}%
\bibitem [{\citenamefont {Yoshinaga}\ \emph {et~al.}(2018)\citenamefont
  {Yoshinaga}, \citenamefont {Yanase}, \citenamefont {Higashiyama},
  \citenamefont {Teruya},\ and\ \citenamefont {Taguchi}}]{yoshinaga2018}%
  \BibitemOpen
  \bibfield  {author} {\bibinfo {author} {\bibfnamefont {N.}~\bibnamefont
  {Yoshinaga}}, \bibinfo {author} {\bibfnamefont {K.}~\bibnamefont {Yanase}},
  \bibinfo {author} {\bibfnamefont {K.}~\bibnamefont {Higashiyama}}, \bibinfo
  {author} {\bibfnamefont {E.}~\bibnamefont {Teruya}},\ and\ \bibinfo {author}
  {\bibfnamefont {D.}~\bibnamefont {Taguchi}},\ }\href
  {https://doi.org/10.1093/ptep/ptx174} {\bibfield  {journal} {\bibinfo
  {journal} {Prog. Theor. Exp. Phys.}\ }\textbf {\bibinfo {volume} {2018}},\
  \bibinfo {pages} {023D02} (\bibinfo {year} {2018})}\BibitemShut {NoStop}%
\bibitem [{\citenamefont {Caurier}\ \emph {et~al.}(1990)\citenamefont
  {Caurier}, \citenamefont {Poves},\ and\ \citenamefont {Zuker}}]{caurier1990}%
  \BibitemOpen
  \bibfield  {author} {\bibinfo {author} {\bibfnamefont {E.}~\bibnamefont
  {Caurier}}, \bibinfo {author} {\bibfnamefont {A.}~\bibnamefont {Poves}},\
  and\ \bibinfo {author} {\bibfnamefont {A.}~\bibnamefont {Zuker}},\ }\href
  {https://doi.org/https://doi.org/10.1016/0370-2693(90)91071-I} {\bibfield
  {journal} {\bibinfo  {journal} {Phys. Lett. B}\ }\textbf {\bibinfo {volume}
  {252}},\ \bibinfo {pages} {13} (\bibinfo {year} {1990})}\BibitemShut
  {NoStop}%
\bibitem [{\citenamefont {Caurier}\ \emph {et~al.}(2012)\citenamefont
  {Caurier}, \citenamefont {Nowacki},\ and\ \citenamefont
  {Poves}}]{caurier2012}%
  \BibitemOpen
  \bibfield  {author} {\bibinfo {author} {\bibfnamefont {E.}~\bibnamefont
  {Caurier}}, \bibinfo {author} {\bibfnamefont {F.}~\bibnamefont {Nowacki}},\
  and\ \bibinfo {author} {\bibfnamefont {A.}~\bibnamefont {Poves}},\ }\href
  {https://doi.org/https://doi.org/10.1016/j.physletb.2012.03.076} {\bibfield
  {journal} {\bibinfo  {journal} {Phys. Lett. B}\ }\textbf {\bibinfo {volume}
  {711}},\ \bibinfo {pages} {62} (\bibinfo {year} {2012})}\BibitemShut
  {NoStop}%
\bibitem [{\citenamefont {Sen'kov}\ and\ \citenamefont
  {Horoi}(2016)}]{senkov2016}%
  \BibitemOpen
  \bibfield  {author} {\bibinfo {author} {\bibfnamefont {R.~A.}\ \bibnamefont
  {Sen'kov}}\ and\ \bibinfo {author} {\bibfnamefont {M.}~\bibnamefont
  {Horoi}},\ }\href {https://doi.org/10.1103/PhysRevC.93.044334} {\bibfield
  {journal} {\bibinfo  {journal} {Phys. Rev. C}\ }\textbf {\bibinfo {volume}
  {93}},\ \bibinfo {pages} {044334} (\bibinfo {year} {2016})}\BibitemShut
  {NoStop}%
\bibitem [{\citenamefont {Coraggio}\ \emph {et~al.}(2019)\citenamefont
  {Coraggio}, \citenamefont {De~Angelis}, \citenamefont {Fukui}, \citenamefont
  {Gargano}, \citenamefont {Itaco},\ and\ \citenamefont
  {Nowacki}}]{coraggio2019}%
  \BibitemOpen
  \bibfield  {author} {\bibinfo {author} {\bibfnamefont {L.}~\bibnamefont
  {Coraggio}}, \bibinfo {author} {\bibfnamefont {L.}~\bibnamefont
  {De~Angelis}}, \bibinfo {author} {\bibfnamefont {T.}~\bibnamefont {Fukui}},
  \bibinfo {author} {\bibfnamefont {A.}~\bibnamefont {Gargano}}, \bibinfo
  {author} {\bibfnamefont {N.}~\bibnamefont {Itaco}},\ and\ \bibinfo {author}
  {\bibfnamefont {F.}~\bibnamefont {Nowacki}},\ }\href
  {https://doi.org/10.1103/PhysRevC.100.014316} {\bibfield  {journal} {\bibinfo
   {journal} {Phys. Rev. C}\ }\textbf {\bibinfo {volume} {100}},\ \bibinfo
  {pages} {014316} (\bibinfo {year} {2019})}\BibitemShut {NoStop}%
\bibitem [{\citenamefont {Yoshida}\ and\ \citenamefont
  {Iachello}(2013)}]{yoshida2013}%
  \BibitemOpen
  \bibfield  {author} {\bibinfo {author} {\bibfnamefont {N.}~\bibnamefont
  {Yoshida}}\ and\ \bibinfo {author} {\bibfnamefont {F.}~\bibnamefont
  {Iachello}},\ }\href {https://doi.org/10.1093/ptep/ptt007} {\bibfield
  {journal} {\bibinfo  {journal} {Prog. Theor. Exp. Phys.}\ }\textbf {\bibinfo
  {volume} {2013}},\ \bibinfo {pages} {043D01} (\bibinfo {year}
  {2013})}\BibitemShut {NoStop}%
\bibitem [{\citenamefont {Nomura}(2022{\natexlab{a}})}]{nomura2022bb}%
  \BibitemOpen
  \bibfield  {author} {\bibinfo {author} {\bibfnamefont {K.}~\bibnamefont
  {Nomura}},\ }\href {https://doi.org/10.1103/PhysRevC.105.044301} {\bibfield
  {journal} {\bibinfo  {journal} {Phys. Rev. C}\ }\textbf {\bibinfo {volume}
  {105}},\ \bibinfo {pages} {044301} (\bibinfo {year}
  {2022}{\natexlab{a}})}\BibitemShut {NoStop}%
\bibitem [{\citenamefont {Otsuka}\ \emph
  {et~al.}(1978{\natexlab{a}})\citenamefont {Otsuka}, \citenamefont {Arima},
  \citenamefont {Iachello},\ and\ \citenamefont {Talmi}}]{OAIT}%
  \BibitemOpen
  \bibfield  {author} {\bibinfo {author} {\bibfnamefont {T.}~\bibnamefont
  {Otsuka}}, \bibinfo {author} {\bibfnamefont {A.}~\bibnamefont {Arima}},
  \bibinfo {author} {\bibfnamefont {F.}~\bibnamefont {Iachello}},\ and\
  \bibinfo {author} {\bibfnamefont {I.}~\bibnamefont {Talmi}},\ }\href
  {https://doi.org/10.1016/0370-2693(78)90260-5} {\bibfield  {journal}
  {\bibinfo  {journal} {Phys. Lett. B}\ }\textbf {\bibinfo {volume} {76}},\
  \bibinfo {pages} {139 } (\bibinfo {year} {1978}{\natexlab{a}})}\BibitemShut
  {NoStop}%
\bibitem [{\citenamefont {Otsuka}\ \emph
  {et~al.}(1978{\natexlab{b}})\citenamefont {Otsuka}, \citenamefont {Arima},\
  and\ \citenamefont {Iachello}}]{OAI}%
  \BibitemOpen
  \bibfield  {author} {\bibinfo {author} {\bibfnamefont {T.}~\bibnamefont
  {Otsuka}}, \bibinfo {author} {\bibfnamefont {A.}~\bibnamefont {Arima}},\ and\
  \bibinfo {author} {\bibfnamefont {F.}~\bibnamefont {Iachello}},\ }\href
  {https://doi.org/10.1016/0375-9474(78)90532-8} {\bibfield  {journal}
  {\bibinfo  {journal} {Nucl. Phys. A}\ }\textbf {\bibinfo {volume} {309}},\
  \bibinfo {pages} {1} (\bibinfo {year} {1978}{\natexlab{b}})}\BibitemShut
  {NoStop}%
\bibitem [{\citenamefont {Ring}\ and\ \citenamefont {Schuck}(1980)}]{RS}%
  \BibitemOpen
  \bibfield  {author} {\bibinfo {author} {\bibfnamefont {P.}~\bibnamefont
  {Ring}}\ and\ \bibinfo {author} {\bibfnamefont {P.}~\bibnamefont {Schuck}},\
  }\href@noop {} {\emph {\bibinfo {title} {The nuclear many-body problem}}}\
  (\bibinfo  {publisher} {Springer, Berlin},\ \bibinfo {year}
  {1980})\BibitemShut {NoStop}%
\bibitem [{\citenamefont {Bender}\ \emph {et~al.}(2003)\citenamefont {Bender},
  \citenamefont {Heenen},\ and\ \citenamefont {Reinhard}}]{bender2003}%
  \BibitemOpen
  \bibfield  {author} {\bibinfo {author} {\bibfnamefont {M.}~\bibnamefont
  {Bender}}, \bibinfo {author} {\bibfnamefont {P.-H.}\ \bibnamefont {Heenen}},\
  and\ \bibinfo {author} {\bibfnamefont {P.-G.}\ \bibnamefont {Reinhard}},\
  }\href {https://doi.org/10.1103/RevModPhys.75.121} {\bibfield  {journal}
  {\bibinfo  {journal} {Rev. Mod. Phys.}\ }\textbf {\bibinfo {volume} {75}},\
  \bibinfo {pages} {121} (\bibinfo {year} {2003})}\BibitemShut {NoStop}%
\bibitem [{\citenamefont {Vretenar}\ \emph {et~al.}(2005)\citenamefont
  {Vretenar}, \citenamefont {Afanasjev}, \citenamefont {Lalazissis},\ and\
  \citenamefont {Ring}}]{vretenar2005}%
  \BibitemOpen
  \bibfield  {author} {\bibinfo {author} {\bibfnamefont {D.}~\bibnamefont
  {Vretenar}}, \bibinfo {author} {\bibfnamefont {A.~V.}\ \bibnamefont
  {Afanasjev}}, \bibinfo {author} {\bibfnamefont {G.~A.}\ \bibnamefont
  {Lalazissis}},\ and\ \bibinfo {author} {\bibfnamefont {P.}~\bibnamefont
  {Ring}},\ }\href {https://doi.org/10.1016/j.physrep.2004.10.001} {\bibfield
  {journal} {\bibinfo  {journal} {Phys. Rep.}\ }\textbf {\bibinfo {volume}
  {409}},\ \bibinfo {pages} {101 } (\bibinfo {year} {2005})}\BibitemShut
  {NoStop}%
\bibitem [{\citenamefont {Nik\ifmmode \check{s}\else
  \v{s}\fi{}i\ifmmode~\acute{c}\else \'{c}\fi{}}\ \emph
  {et~al.}(2011)\citenamefont {Nik\ifmmode \check{s}\else
  \v{s}\fi{}i\ifmmode~\acute{c}\else \'{c}\fi{}}, \citenamefont {Vretenar},\
  and\ \citenamefont {Ring}}]{niksic2011}%
  \BibitemOpen
  \bibfield  {author} {\bibinfo {author} {\bibfnamefont {T.}~\bibnamefont
  {Nik\ifmmode \check{s}\else \v{s}\fi{}i\ifmmode~\acute{c}\else \'{c}\fi{}}},
  \bibinfo {author} {\bibfnamefont {D.}~\bibnamefont {Vretenar}},\ and\
  \bibinfo {author} {\bibfnamefont {P.}~\bibnamefont {Ring}},\ }\href
  {https://doi.org/10.1016/j.ppnp.2011.01.055} {\bibfield  {journal} {\bibinfo
  {journal} {Prog. Part. Nucl. Phys.}\ }\textbf {\bibinfo {volume} {66}},\
  \bibinfo {pages} {519} (\bibinfo {year} {2011})}\BibitemShut {NoStop}%
\bibitem [{\citenamefont {Robledo}\ \emph {et~al.}(2019)\citenamefont
  {Robledo}, \citenamefont {Rodríguez},\ and\ \citenamefont
  {Rodríguez-Guzmán}}]{robledo2019}%
  \BibitemOpen
  \bibfield  {author} {\bibinfo {author} {\bibfnamefont {L.~M.}\ \bibnamefont
  {Robledo}}, \bibinfo {author} {\bibfnamefont {T.~R.}\ \bibnamefont
  {Rodríguez}},\ and\ \bibinfo {author} {\bibfnamefont {R.~R.}\ \bibnamefont
  {Rodríguez-Guzmán}},\ }\href
  {http://stacks.iop.org/0954-3899/46/i=1/a=013001} {\bibfield  {journal}
  {\bibinfo  {journal} {J. Phys. G: Nucl. Part. Phys.}\ }\textbf {\bibinfo
  {volume} {46}},\ \bibinfo {pages} {013001} (\bibinfo {year}
  {2019})}\BibitemShut {NoStop}%
\bibitem [{\citenamefont {Nik\ifmmode \check{s}\else
  \v{s}\fi{}i\ifmmode~\acute{c}\else \'{c}\fi{}}\ \emph
  {et~al.}(2008)\citenamefont {Nik\ifmmode \check{s}\else
  \v{s}\fi{}i\ifmmode~\acute{c}\else \'{c}\fi{}}, \citenamefont {Vretenar},\
  and\ \citenamefont {Ring}}]{DDPC1}%
  \BibitemOpen
  \bibfield  {author} {\bibinfo {author} {\bibfnamefont {T.}~\bibnamefont
  {Nik\ifmmode \check{s}\else \v{s}\fi{}i\ifmmode~\acute{c}\else \'{c}\fi{}}},
  \bibinfo {author} {\bibfnamefont {D.}~\bibnamefont {Vretenar}},\ and\
  \bibinfo {author} {\bibfnamefont {P.}~\bibnamefont {Ring}},\ }\href
  {https://doi.org/10.1103/PhysRevC.78.034318} {\bibfield  {journal} {\bibinfo
  {journal} {Phys. Rev. C}\ }\textbf {\bibinfo {volume} {78}},\ \bibinfo
  {pages} {034318} (\bibinfo {year} {2008})}\BibitemShut {NoStop}%
\bibitem [{\citenamefont {Tian}\ \emph {et~al.}(2009)\citenamefont {Tian},
  \citenamefont {Ma},\ and\ \citenamefont {Ring}}]{tian2009}%
  \BibitemOpen
  \bibfield  {author} {\bibinfo {author} {\bibfnamefont {Y.}~\bibnamefont
  {Tian}}, \bibinfo {author} {\bibfnamefont {Z.~Y.}\ \bibnamefont {Ma}},\ and\
  \bibinfo {author} {\bibfnamefont {P.}~\bibnamefont {Ring}},\ }\href
  {https://doi.org/10.1016/j.physletb.2009.04.067} {\bibfield  {journal}
  {\bibinfo  {journal} {Phys. Lett. B}\ }\textbf {\bibinfo {volume} {676}},\
  \bibinfo {pages} {44 } (\bibinfo {year} {2009})}\BibitemShut {NoStop}%
\bibitem [{\citenamefont {Brant}\ \emph {et~al.}(1984)\citenamefont {Brant},
  \citenamefont {Paar},\ and\ \citenamefont {Vretenar}}]{brant1984}%
  \BibitemOpen
  \bibfield  {author} {\bibinfo {author} {\bibfnamefont {S.}~\bibnamefont
  {Brant}}, \bibinfo {author} {\bibfnamefont {V.}~\bibnamefont {Paar}},\ and\
  \bibinfo {author} {\bibfnamefont {D.}~\bibnamefont {Vretenar}},\ }\href
  {https://doi.org/10.1007/BF01412551} {\bibfield  {journal} {\bibinfo
  {journal} {Z. Phys. A}\ }\textbf {\bibinfo {volume} {319}},\ \bibinfo {pages}
  {355} (\bibinfo {year} {1984})}\BibitemShut {NoStop}%
\bibitem [{\citenamefont {Iachello}\ and\ \citenamefont {{Van
  Isacker}}(1991)}]{IBFM}%
  \BibitemOpen
  \bibfield  {author} {\bibinfo {author} {\bibfnamefont {F.}~\bibnamefont
  {Iachello}}\ and\ \bibinfo {author} {\bibfnamefont {P.}~\bibnamefont {{Van
  Isacker}}},\ }\href@noop {} {\emph {\bibinfo {title} {The interacting
  boson-fermion model}}}\ (\bibinfo  {publisher} {Cambridge University Press,
  Cambridge},\ \bibinfo {year} {1991})\BibitemShut {NoStop}%
\bibitem [{\citenamefont {Duval}\ and\ \citenamefont
  {Barrett}(1981)}]{duval1981}%
  \BibitemOpen
  \bibfield  {author} {\bibinfo {author} {\bibfnamefont {P.~D.}\ \bibnamefont
  {Duval}}\ and\ \bibinfo {author} {\bibfnamefont {B.~R.}\ \bibnamefont
  {Barrett}},\ }\href {https://doi.org/10.1016/0370-2693(81)90321-X} {\bibfield
   {journal} {\bibinfo  {journal} {Phys. Lett. B}\ }\textbf {\bibinfo {volume}
  {100}},\ \bibinfo {pages} {223} (\bibinfo {year} {1981})}\BibitemShut
  {NoStop}%
\bibitem [{\citenamefont {Nomura}\ \emph
  {et~al.}(2016{\natexlab{a}})\citenamefont {Nomura}, \citenamefont {Otsuka},\
  and\ \citenamefont {{Van Isacker}}}]{nomura2016sc}%
  \BibitemOpen
  \bibfield  {author} {\bibinfo {author} {\bibfnamefont {K.}~\bibnamefont
  {Nomura}}, \bibinfo {author} {\bibfnamefont {T.}~\bibnamefont {Otsuka}},\
  and\ \bibinfo {author} {\bibfnamefont {P.}~\bibnamefont {{Van Isacker}}},\
  }\href {https://doi.org/10.1088/0954-3899/43/2/024008} {\bibfield  {journal}
  {\bibinfo  {journal} {J. Phys. G: Nucl. Part. Phys.}\ }\textbf {\bibinfo
  {volume} {43}},\ \bibinfo {pages} {024008} (\bibinfo {year}
  {2016}{\natexlab{a}})}\BibitemShut {NoStop}%
\bibitem [{\citenamefont {Nomura}\ \emph
  {et~al.}(2016{\natexlab{b}})\citenamefont {Nomura}, \citenamefont
  {Rodr\'{\i}guez-Guzm\'an},\ and\ \citenamefont {Robledo}}]{nomura2016zr}%
  \BibitemOpen
  \bibfield  {author} {\bibinfo {author} {\bibfnamefont {K.}~\bibnamefont
  {Nomura}}, \bibinfo {author} {\bibfnamefont {R.}~\bibnamefont
  {Rodr\'{\i}guez-Guzm\'an}},\ and\ \bibinfo {author} {\bibfnamefont {L.~M.}\
  \bibnamefont {Robledo}},\ }\href {https://doi.org/10.1103/PhysRevC.94.044314}
  {\bibfield  {journal} {\bibinfo  {journal} {Phys. Rev. C}\ }\textbf {\bibinfo
  {volume} {94}},\ \bibinfo {pages} {044314} (\bibinfo {year}
  {2016}{\natexlab{b}})}\BibitemShut {NoStop}%
\bibitem [{\citenamefont {Nomura}(2022{\natexlab{b}})}]{nomura2022octcm}%
  \BibitemOpen
  \bibfield  {author} {\bibinfo {author} {\bibfnamefont {K.}~\bibnamefont
  {Nomura}},\ }\href {https://doi.org/10.1103/PhysRevC.106.024330} {\bibfield
  {journal} {\bibinfo  {journal} {Phys. Rev. C}\ }\textbf {\bibinfo {volume}
  {106}},\ \bibinfo {pages} {024330} (\bibinfo {year}
  {2022}{\natexlab{b}})}\BibitemShut {NoStop}%
\bibitem [{\citenamefont {Nomura}\ \emph {et~al.}(2020)\citenamefont {Nomura},
  \citenamefont {Vretenar}, \citenamefont {Li},\ and\ \citenamefont
  {Xiang}}]{nomura2020pv}%
  \BibitemOpen
  \bibfield  {author} {\bibinfo {author} {\bibfnamefont {K.}~\bibnamefont
  {Nomura}}, \bibinfo {author} {\bibfnamefont {D.}~\bibnamefont {Vretenar}},
  \bibinfo {author} {\bibfnamefont {Z.~P.}\ \bibnamefont {Li}},\ and\ \bibinfo
  {author} {\bibfnamefont {J.}~\bibnamefont {Xiang}},\ }\href
  {https://doi.org/10.1103/PhysRevC.102.054313} {\bibfield  {journal} {\bibinfo
   {journal} {Phys. Rev. C}\ }\textbf {\bibinfo {volume} {102}},\ \bibinfo
  {pages} {054313} (\bibinfo {year} {2020})}\BibitemShut {NoStop}%
\bibitem [{\citenamefont {Nomura}\ \emph
  {et~al.}(2021{\natexlab{a}})\citenamefont {Nomura}, \citenamefont {Vretenar},
  \citenamefont {Li},\ and\ \citenamefont {Xiang}}]{nomura2021pv}%
  \BibitemOpen
  \bibfield  {author} {\bibinfo {author} {\bibfnamefont {K.}~\bibnamefont
  {Nomura}}, \bibinfo {author} {\bibfnamefont {D.}~\bibnamefont {Vretenar}},
  \bibinfo {author} {\bibfnamefont {Z.~P.}\ \bibnamefont {Li}},\ and\ \bibinfo
  {author} {\bibfnamefont {J.}~\bibnamefont {Xiang}},\ }\href
  {https://doi.org/10.1103/PhysRevC.103.054322} {\bibfield  {journal} {\bibinfo
   {journal} {Phys. Rev. C}\ }\textbf {\bibinfo {volume} {103}},\ \bibinfo
  {pages} {054322} (\bibinfo {year} {2021}{\natexlab{a}})}\BibitemShut
  {NoStop}%
\bibitem [{\citenamefont {Nomura}\ \emph
  {et~al.}(2021{\natexlab{b}})\citenamefont {Nomura}, \citenamefont
  {Gavrielov},\ and\ \citenamefont {Leviatan}}]{nomura2021pds}%
  \BibitemOpen
  \bibfield  {author} {\bibinfo {author} {\bibfnamefont {K.}~\bibnamefont
  {Nomura}}, \bibinfo {author} {\bibfnamefont {N.}~\bibnamefont {Gavrielov}},\
  and\ \bibinfo {author} {\bibfnamefont {A.}~\bibnamefont {Leviatan}},\ }\href
  {https://doi.org/10.1103/PhysRevC.104.044317} {\bibfield  {journal} {\bibinfo
   {journal} {Phys. Rev. C}\ }\textbf {\bibinfo {volume} {104}},\ \bibinfo
  {pages} {044317} (\bibinfo {year} {2021}{\natexlab{b}})}\BibitemShut
  {NoStop}%
\bibitem [{\citenamefont {Homma}\ and\ \citenamefont
  {Nomura}(2024)}]{homma2024}%
  \BibitemOpen
  \bibfield  {author} {\bibinfo {author} {\bibfnamefont {M.}~\bibnamefont
  {Homma}}\ and\ \bibinfo {author} {\bibfnamefont {K.}~\bibnamefont {Nomura}},\
  }\href {https://doi.org/10.1103/PhysRevC.110.014303} {\bibfield  {journal}
  {\bibinfo  {journal} {Phys. Rev. C}\ }\textbf {\bibinfo {volume} {110}},\
  \bibinfo {pages} {014303} (\bibinfo {year} {2024})}\BibitemShut {NoStop}%
\bibitem [{\citenamefont {Berger}\ \emph {et~al.}(1984)\citenamefont {Berger},
  \citenamefont {Girod},\ and\ \citenamefont {Gogny}}]{D1S}%
  \BibitemOpen
  \bibfield  {author} {\bibinfo {author} {\bibfnamefont {J.~F.}\ \bibnamefont
  {Berger}}, \bibinfo {author} {\bibfnamefont {M.}~\bibnamefont {Girod}},\ and\
  \bibinfo {author} {\bibfnamefont {D.}~\bibnamefont {Gogny}},\ }\href
  {https://doi.org/10.1016/0375-9474(84)90240-9} {\bibfield  {journal}
  {\bibinfo  {journal} {Nucl. Phys. A}\ }\textbf {\bibinfo {volume} {428}},\
  \bibinfo {pages} {23 } (\bibinfo {year} {1984})}\BibitemShut {NoStop}%
\bibitem [{\citenamefont {Teeti}\ and\ \citenamefont
  {Afanasjev}(2021)}]{teeti2021}%
  \BibitemOpen
  \bibfield  {author} {\bibinfo {author} {\bibfnamefont {S.}~\bibnamefont
  {Teeti}}\ and\ \bibinfo {author} {\bibfnamefont {A.~V.}\ \bibnamefont
  {Afanasjev}},\ }\href {https://doi.org/10.1103/PhysRevC.103.034310}
  {\bibfield  {journal} {\bibinfo  {journal} {Phys. Rev. C}\ }\textbf {\bibinfo
  {volume} {103}},\ \bibinfo {pages} {034310} (\bibinfo {year}
  {2021})}\BibitemShut {NoStop}%
\bibitem [{\citenamefont {Bohr}\ and\ \citenamefont {Mottelson}(1975)}]{BM}%
  \BibitemOpen
  \bibfield  {author} {\bibinfo {author} {\bibfnamefont {A.}~\bibnamefont
  {Bohr}}\ and\ \bibinfo {author} {\bibfnamefont {B.~R.}\ \bibnamefont
  {Mottelson}},\ }\href@noop {} {\emph {\bibinfo {title} {Nuclear Structure}}}\
  (\bibinfo  {publisher} {Benjamin, New York},\ \bibinfo {year}
  {1975})\BibitemShut {NoStop}%
\bibitem [{\citenamefont {Nomura}\ \emph
  {et~al.}(2016{\natexlab{c}})\citenamefont {Nomura}, \citenamefont
  {Nik\ifmmode \check{s}\else \v{s}\fi{}i\ifmmode~\acute{c}\else \'{c}\fi{}},\
  and\ \citenamefont {Vretenar}}]{nomura2016odd}%
  \BibitemOpen
  \bibfield  {author} {\bibinfo {author} {\bibfnamefont {K.}~\bibnamefont
  {Nomura}}, \bibinfo {author} {\bibfnamefont {T.}~\bibnamefont {Nik\ifmmode
  \check{s}\else \v{s}\fi{}i\ifmmode~\acute{c}\else \'{c}\fi{}}},\ and\
  \bibinfo {author} {\bibfnamefont {D.}~\bibnamefont {Vretenar}},\ }\href
  {https://doi.org/10.1103/PhysRevC.93.054305} {\bibfield  {journal} {\bibinfo
  {journal} {Phys. Rev. C}\ }\textbf {\bibinfo {volume} {93}},\ \bibinfo
  {pages} {054305} (\bibinfo {year} {2016}{\natexlab{c}})}\BibitemShut
  {NoStop}%
\bibitem [{\citenamefont {Nomura}\ \emph {et~al.}(2017)\citenamefont {Nomura},
  \citenamefont {Rodr\'{\i}guez-Guzm\'an},\ and\ \citenamefont
  {Robledo}}]{nomura2017odd-2}%
  \BibitemOpen
  \bibfield  {author} {\bibinfo {author} {\bibfnamefont {K.}~\bibnamefont
  {Nomura}}, \bibinfo {author} {\bibfnamefont {R.}~\bibnamefont
  {Rodr\'{\i}guez-Guzm\'an}},\ and\ \bibinfo {author} {\bibfnamefont {L.~M.}\
  \bibnamefont {Robledo}},\ }\href {https://doi.org/10.1103/PhysRevC.96.014314}
  {\bibfield  {journal} {\bibinfo  {journal} {Phys. Rev. C}\ }\textbf {\bibinfo
  {volume} {96}},\ \bibinfo {pages} {014314} (\bibinfo {year}
  {2017})}\BibitemShut {NoStop}%
\bibitem [{\citenamefont {Nomura}\ \emph {et~al.}(2008)\citenamefont {Nomura},
  \citenamefont {Shimizu},\ and\ \citenamefont {Otsuka}}]{nomura2008}%
  \BibitemOpen
  \bibfield  {author} {\bibinfo {author} {\bibfnamefont {K.}~\bibnamefont
  {Nomura}}, \bibinfo {author} {\bibfnamefont {N.}~\bibnamefont {Shimizu}},\
  and\ \bibinfo {author} {\bibfnamefont {T.}~\bibnamefont {Otsuka}},\ }\href
  {https://doi.org/10.1103/PhysRevLett.101.142501} {\bibfield  {journal}
  {\bibinfo  {journal} {Phys. Rev. Lett.}\ }\textbf {\bibinfo {volume} {101}},\
  \bibinfo {pages} {142501} (\bibinfo {year} {2008})}\BibitemShut {NoStop}%
\bibitem [{\citenamefont {Nomura}\ \emph {et~al.}(2010)\citenamefont {Nomura},
  \citenamefont {Shimizu},\ and\ \citenamefont {Otsuka}}]{nomura2010}%
  \BibitemOpen
  \bibfield  {author} {\bibinfo {author} {\bibfnamefont {K.}~\bibnamefont
  {Nomura}}, \bibinfo {author} {\bibfnamefont {N.}~\bibnamefont {Shimizu}},\
  and\ \bibinfo {author} {\bibfnamefont {T.}~\bibnamefont {Otsuka}},\ }\href
  {https://doi.org/10.1103/PhysRevC.81.044307} {\bibfield  {journal} {\bibinfo
  {journal} {Phys. Rev. C}\ }\textbf {\bibinfo {volume} {81}},\ \bibinfo
  {pages} {044307} (\bibinfo {year} {2010})}\BibitemShut {NoStop}%
\bibitem [{\citenamefont {Dieperink}\ \emph {et~al.}(1980)\citenamefont
  {Dieperink}, \citenamefont {Scholten},\ and\ \citenamefont
  {Iachello}}]{dieperink1980}%
  \BibitemOpen
  \bibfield  {author} {\bibinfo {author} {\bibfnamefont {A.~E.~L.}\
  \bibnamefont {Dieperink}}, \bibinfo {author} {\bibfnamefont {O.}~\bibnamefont
  {Scholten}},\ and\ \bibinfo {author} {\bibfnamefont {F.}~\bibnamefont
  {Iachello}},\ }\href {https://doi.org/10.1103/PhysRevLett.44.1747} {\bibfield
   {journal} {\bibinfo  {journal} {Phys. Rev. Lett.}\ }\textbf {\bibinfo
  {volume} {44}},\ \bibinfo {pages} {1747} (\bibinfo {year}
  {1980})}\BibitemShut {NoStop}%
\bibitem [{\citenamefont {Ginocchio}\ and\ \citenamefont
  {Kirson}(1980)}]{ginocchio1980}%
  \BibitemOpen
  \bibfield  {author} {\bibinfo {author} {\bibfnamefont {J.~N.}\ \bibnamefont
  {Ginocchio}}\ and\ \bibinfo {author} {\bibfnamefont {M.~W.}\ \bibnamefont
  {Kirson}},\ }\href {https://doi.org/10.1016/0375-9474(80)90387-5} {\bibfield
  {journal} {\bibinfo  {journal} {Nucl. Phys. A}\ }\textbf {\bibinfo {volume}
  {350}},\ \bibinfo {pages} {31} (\bibinfo {year} {1980})}\BibitemShut
  {NoStop}%
\bibitem [{\citenamefont {Nomura}\ \emph {et~al.}(2011)\citenamefont {Nomura},
  \citenamefont {Otsuka}, \citenamefont {Shimizu},\ and\ \citenamefont
  {Guo}}]{nomura2011rot}%
  \BibitemOpen
  \bibfield  {author} {\bibinfo {author} {\bibfnamefont {K.}~\bibnamefont
  {Nomura}}, \bibinfo {author} {\bibfnamefont {T.}~\bibnamefont {Otsuka}},
  \bibinfo {author} {\bibfnamefont {N.}~\bibnamefont {Shimizu}},\ and\ \bibinfo
  {author} {\bibfnamefont {L.}~\bibnamefont {Guo}},\ }\href
  {https://doi.org/10.1103/PhysRevC.83.041302} {\bibfield  {journal} {\bibinfo
  {journal} {Phys. Rev. C}\ }\textbf {\bibinfo {volume} {83}},\ \bibinfo
  {pages} {041302} (\bibinfo {year} {2011})}\BibitemShut {NoStop}%
\bibitem [{\citenamefont {Schaaser}\ and\ \citenamefont
  {Brink}(1986)}]{Schaaser86}%
  \BibitemOpen
  \bibfield  {author} {\bibinfo {author} {\bibfnamefont {H.}~\bibnamefont
  {Schaaser}}\ and\ \bibinfo {author} {\bibfnamefont {D.~M.}\ \bibnamefont
  {Brink}},\ }\href@noop {} {\bibfield  {journal} {\bibinfo  {journal} {Nucl.
  Phys. A}\ }\textbf {\bibinfo {volume} {452}},\ \bibinfo {pages} {1 }
  (\bibinfo {year} {1986})}\BibitemShut {NoStop}%
\bibitem [{\citenamefont {Inglis}(1956)}]{inglis1956}%
  \BibitemOpen
  \bibfield  {author} {\bibinfo {author} {\bibfnamefont {D.~R.}\ \bibnamefont
  {Inglis}},\ }\href {https://doi.org/10.1103/PhysRev.103.1786} {\bibfield
  {journal} {\bibinfo  {journal} {Phys. Rev.}\ }\textbf {\bibinfo {volume}
  {103}},\ \bibinfo {pages} {1786 } (\bibinfo {year} {1956})}\BibitemShut
  {NoStop}%
\bibitem [{\citenamefont {Beliaev}(1961)}]{belyaev1961}%
  \BibitemOpen
  \bibfield  {author} {\bibinfo {author} {\bibfnamefont {S.~T.}\ \bibnamefont
  {Beliaev}},\ }\href {https://doi.org/10.1016/0029-5582(61)90384-4} {\bibfield
   {journal} {\bibinfo  {journal} {Nucl. Phys.}\ }\textbf {\bibinfo {volume}
  {24}},\ \bibinfo {pages} {322 } (\bibinfo {year} {1961})}\BibitemShut
  {NoStop}%
\bibitem [{\citenamefont {Scholten}(1985)}]{scholten1985}%
  \BibitemOpen
  \bibfield  {author} {\bibinfo {author} {\bibfnamefont {O.}~\bibnamefont
  {Scholten}},\ }\href
  {https://doi.org/https://doi.org/10.1016/0146-6410(85)90054-7} {\bibfield
  {journal} {\bibinfo  {journal} {Prog. Part. Nucl. Phys.}\ }\textbf {\bibinfo
  {volume} {14}},\ \bibinfo {pages} {189} (\bibinfo {year} {1985})}\BibitemShut
  {NoStop}%
\bibitem [{\citenamefont {Nomura}\ \emph {et~al.}(2019)\citenamefont {Nomura},
  \citenamefont {Rodr\'{\i}guez-Guzm\'an},\ and\ \citenamefont
  {Robledo}}]{nomura2019dodd}%
  \BibitemOpen
  \bibfield  {author} {\bibinfo {author} {\bibfnamefont {K.}~\bibnamefont
  {Nomura}}, \bibinfo {author} {\bibfnamefont {R.}~\bibnamefont
  {Rodr\'{\i}guez-Guzm\'an}},\ and\ \bibinfo {author} {\bibfnamefont {L.~M.}\
  \bibnamefont {Robledo}},\ }\href {https://doi.org/10.1103/PhysRevC.99.034308}
  {\bibfield  {journal} {\bibinfo  {journal} {Phys. Rev. C}\ }\textbf {\bibinfo
  {volume} {99}},\ \bibinfo {pages} {034308} (\bibinfo {year}
  {2019})}\BibitemShut {NoStop}%
\bibitem [{\citenamefont {Dellagiacoma}(1988)}]{dellagiacoma1988phdthesis}%
  \BibitemOpen
  \bibfield  {author} {\bibinfo {author} {\bibfnamefont {F.}~\bibnamefont
  {Dellagiacoma}},\ }\emph {\bibinfo {title} {Beta decay of odd mass nuclei in
  the interacting boson-fermion model}},\ \href@noop {} {Ph.D. thesis},\
  \bibinfo  {school} {Yale University} (\bibinfo {year} {1988})\BibitemShut
  {NoStop}%
\bibitem [{\citenamefont {Dellagiacoma}\ and\ \citenamefont
  {Iachello}(1989)}]{DELLAGIACOMA1989}%
  \BibitemOpen
  \bibfield  {author} {\bibinfo {author} {\bibfnamefont {F.}~\bibnamefont
  {Dellagiacoma}}\ and\ \bibinfo {author} {\bibfnamefont {F.}~\bibnamefont
  {Iachello}},\ }\href
  {https://doi.org/https://doi.org/10.1016/0370-2693(89)91434-2} {\bibfield
  {journal} {\bibinfo  {journal} {Phys. Lett. B}\ }\textbf {\bibinfo {volume}
  {218}},\ \bibinfo {pages} {399 } (\bibinfo {year} {1989})}\BibitemShut
  {NoStop}%
\bibitem [{\citenamefont {{Brookhaven National Nuclear Data Center}}()}]{data}%
  \BibitemOpen
  \bibfield  {author} {\bibinfo {author} {\bibnamefont {{Brookhaven National
  Nuclear Data Center}}},\ }\href@noop {} {}\bibinfo {howpublished}
  {{http://www.nndc.bnl.gov}}\BibitemShut {NoStop}%
\bibitem [{\citenamefont {Kotila}\ and\ \citenamefont
  {Iachello}(2012)}]{kotila2012}%
  \BibitemOpen
  \bibfield  {author} {\bibinfo {author} {\bibfnamefont {J.}~\bibnamefont
  {Kotila}}\ and\ \bibinfo {author} {\bibfnamefont {F.}~\bibnamefont
  {Iachello}},\ }\href {https://doi.org/10.1103/PhysRevC.85.034316} {\bibfield
  {journal} {\bibinfo  {journal} {Phys. Rev. C}\ }\textbf {\bibinfo {volume}
  {85}},\ \bibinfo {pages} {034316} (\bibinfo {year} {2012})}\BibitemShut
  {NoStop}%
\bibitem [{\citenamefont {Stone}(2005)}]{stone2005}%
  \BibitemOpen
  \bibfield  {author} {\bibinfo {author} {\bibfnamefont {N.}~\bibnamefont
  {Stone}},\ }\href@noop {} {\bibfield  {journal} {\bibinfo  {journal} {At.
  Data Nucl. Data Tables}\ }\textbf {\bibinfo {volume} {90}},\ \bibinfo {pages}
  {75} (\bibinfo {year} {2005})}\BibitemShut {NoStop}%
\bibitem [{\citenamefont {Barea}\ \emph {et~al.}(2013)\citenamefont {Barea},
  \citenamefont {Kotila},\ and\ \citenamefont {Iachello}}]{barea2013}%
  \BibitemOpen
  \bibfield  {author} {\bibinfo {author} {\bibfnamefont {J.}~\bibnamefont
  {Barea}}, \bibinfo {author} {\bibfnamefont {J.}~\bibnamefont {Kotila}},\ and\
  \bibinfo {author} {\bibfnamefont {F.}~\bibnamefont {Iachello}},\ }\href
  {https://doi.org/10.1103/PhysRevC.87.014315} {\bibfield  {journal} {\bibinfo
  {journal} {Phys. Rev. C}\ }\textbf {\bibinfo {volume} {87}},\ \bibinfo
  {pages} {014315} (\bibinfo {year} {2013})}\BibitemShut {NoStop}%
\bibitem [{\citenamefont {Barea}\ \emph {et~al.}(2015)\citenamefont {Barea},
  \citenamefont {Kotila},\ and\ \citenamefont {Iachello}}]{barea2015}%
  \BibitemOpen
  \bibfield  {author} {\bibinfo {author} {\bibfnamefont {J.}~\bibnamefont
  {Barea}}, \bibinfo {author} {\bibfnamefont {J.}~\bibnamefont {Kotila}},\ and\
  \bibinfo {author} {\bibfnamefont {F.}~\bibnamefont {Iachello}},\ }\href
  {https://doi.org/10.1103/PhysRevC.91.034304} {\bibfield  {journal} {\bibinfo
  {journal} {Phys. Rev. C}\ }\textbf {\bibinfo {volume} {91}},\ \bibinfo
  {pages} {034304} (\bibinfo {year} {2015})}\BibitemShut {NoStop}%
\bibitem [{\citenamefont {Griffiths}\ and\ \citenamefont
  {Vogel}(1992)}]{griffiths1992}%
  \BibitemOpen
  \bibfield  {author} {\bibinfo {author} {\bibfnamefont {A.}~\bibnamefont
  {Griffiths}}\ and\ \bibinfo {author} {\bibfnamefont {P.}~\bibnamefont
  {Vogel}},\ }\href {https://doi.org/10.1103/PhysRevC.46.181} {\bibfield
  {journal} {\bibinfo  {journal} {Phys. Rev. C}\ }\textbf {\bibinfo {volume}
  {46}},\ \bibinfo {pages} {181} (\bibinfo {year} {1992})}\BibitemShut
  {NoStop}%
\bibitem [{\citenamefont {Civitarese}\ and\ \citenamefont
  {Suhonen}(1998)}]{civitarese1998}%
  \BibitemOpen
  \bibfield  {author} {\bibinfo {author} {\bibfnamefont {O.}~\bibnamefont
  {Civitarese}}\ and\ \bibinfo {author} {\bibfnamefont {J.}~\bibnamefont
  {Suhonen}},\ }\href {https://doi.org/10.1103/PhysRevC.58.1535} {\bibfield
  {journal} {\bibinfo  {journal} {Phys. Rev. C}\ }\textbf {\bibinfo {volume}
  {58}},\ \bibinfo {pages} {1535} (\bibinfo {year} {1998})}\BibitemShut
  {NoStop}%
\bibitem [{\citenamefont {Moreno}\ \emph {et~al.}(2008)\citenamefont {Moreno},
  \citenamefont {{\'{A}}lvarez-Rodr{\'{\i}}guez}, \citenamefont {Sarriguren},
  \citenamefont {de~Guerra}, \citenamefont {{\v{S}}imkovic},\ and\
  \citenamefont {Faessler}}]{moreno2008}%
  \BibitemOpen
  \bibfield  {author} {\bibinfo {author} {\bibfnamefont {O.}~\bibnamefont
  {Moreno}}, \bibinfo {author} {\bibfnamefont {R.}~\bibnamefont
  {{\'{A}}lvarez-Rodr{\'{\i}}guez}}, \bibinfo {author} {\bibfnamefont
  {P.}~\bibnamefont {Sarriguren}}, \bibinfo {author} {\bibfnamefont {E.~M.}\
  \bibnamefont {de~Guerra}}, \bibinfo {author} {\bibfnamefont {F.}~\bibnamefont
  {{\v{S}}imkovic}},\ and\ \bibinfo {author} {\bibfnamefont {A.}~\bibnamefont
  {Faessler}},\ }\href {https://doi.org/10.1088/0954-3899/36/1/015106}
  {\bibfield  {journal} {\bibinfo  {journal} {J. Phys. G: Nucl. Part. Phys.}\
  }\textbf {\bibinfo {volume} {36}},\ \bibinfo {pages} {015106} (\bibinfo
  {year} {2008})}\BibitemShut {NoStop}%
\end{thebibliography}%

\end{document}